\documentclass[aps,prl,twocolumn,preprintnumbers,amsmath,amssymb,amsfonts,superscriptaddress,footinbib,10pt]{revtex4-2}
\pdfoutput=1 

\usepackage[colorlinks=true,breaklinks=true]{hyperref}
\usepackage[normalem]{ulem}
\usepackage[utf8]{inputenc}
\hypersetup{allcolors=[rgb]{0.0 0.0 0.6},linkcolor=[rgb]{0.75 0.05 0.05}}
\usepackage{amsmath,amssymb}

\usepackage{graphicx}   
\usepackage{slashed}       
\usepackage{tikz}
\usepackage[compat=1.1.0]{tikz-feynman}
\usepackage{url}
\usepackage{color}
\usepackage{multirow}
\usepackage{comment}
\usepackage{mathtools}
\usepackage{bm}
\usepackage{braket}
\usepackage{amsmath, amsthm, epsfig, fancyhdr}
\usepackage{physics}
\tikzfeynmanset{compat=1.1.0}
\usepackage{tikzsymbols}
\usepackage{cleveref}

\usepackage{booktabs}
\usepackage{adjustbox}
\usepackage{array}
\usepackage{tikz,xcolor}

\newcommand{\be}{\begin{equation}}
\newcommand{\ee}{\end{equation}}
\newcommand{\bea}{\begin{eqnarray}}
\newcommand{\eea}{\end{eqnarray}}

\graphicspath{{./fig/}{./}{./figs/}}

\begin{document}


\title{\texorpdfstring{Primordial Black Hole from Tensor-induced Density Fluctuation: \\ \textit{First-order Phase Transitions and Domain Walls}}{Primordial Black Hole from Tensor-induced Density Fluctuation: First-order Phase Transitions and Domain Walls}}

\author{Utkarsh Kumar}
\email{utkarshkumar.physics@gmail.com}
\affiliation{Department of Physics, University of Ottawa, Ottawa, ON K1N6N5, Canada}

\author{Anish Ghoshal}
\email{A.Ghoshal@sussex.ac.uk}
\affiliation{Department of Physics and Astronomy, University of Sussex,
Brighton, BN1 9RH, United Kingdom}

\smallskip

\begin{abstract}
Gravitational waves generated by violent processes in the early Universe necessarily couple to scalar perturbations beyond linear order. We show that tensor perturbations produced by first-order phase transitions and by annihilating domain-wall networks source density fluctuations at second order, thereby opening a distinct channel for primordial-black-hole (PBH) formation. We compute the tensor-induced density spectrum for both sources and map the resulting PBH abundance onto the macroscopic parameters $(\alpha,\beta/H,T_\star)$ of a first-order phase transition and $(\alpha_{\rm ann},V_{\rm bias},\sigma)$ of a domain-wall network. Existing PBH limits therefore impose complementary constraints on early-Universe sources of stochastic gravitational waves. We find viable regions where one can have an observable gravitational-wave background and an appreciable abundance of asteroid-mass PBHs, including benchmark points that saturate the dark-matter abundance. Our results establish a direct, testable correlation among the source scale, the gravitational-wave spectrum, and the PBH mass function, distinguishing this tensor-induced channel from PBHs formed through delayed vacuum decay or direct defect collapse. We also discuss viable particle physics origin of such FOPT and DW, and therefore, constraints on such microphysics, either in the visible, or in dark sector models.
\end{abstract}

\maketitle

\paragraph{\textbf{Introduction:-} }
Primordial black holes (PBHs) probe fluctuations and dynamics on scales far beyond those accessible to the cosmic microwave background, with consequences for dark matter, gravitational-wave astronomy, and physics beyond the Standard Model~\cite{Carr:1975qj,Carr:2020gox}. In the conventional picture, PBHs originate from unusually large primordial scalar perturbations, motivating a broad range of inflationary and post-inflationary mechanisms that enhance the curvature spectrum on small scales. The early Universe, however, also contains intrinsically tensorial sources. In particular, first-order phase transitions (FOPTs) and annihilating domain-wall (DW) networks can generate intense stochastic gravitational-wave (GW) backgrounds through bubble collisions, acoustic motion, and defect dynamics~\cite{Witten:1984rs,Hogan:1986qda,Caprini:2015zlo,Caprini:2019egz,LISACosmologyWorkingGroup:2022jok,Hiramatsu:2013qaa,Saikawa:2017hiv}. This raises a basic question: can the tensor perturbations generated by these causal sources themselves seed PBHs?

In this Letter we demonstrate that they can. First-order tensor perturbations inevitably source scalar density fluctuations at second order in cosmological perturbation theory. Once the source has switched off, these induced overdensities evolve in the radiation era and may collapse into PBHs at horizon re-entry. The mechanism is distinct from PBH production by false-vacuum bubbles, delayed transition dynamics, or direct wall collapse~\cite{Deng:2017uwc,Deng:2020mds,Kusenko:2020pcg,Wang:2025lti,Ferrer:2018uiu,Gouttenoire:2023gbn,Ferreira:2023jbu,Wang:2025hwc,Ning:2026nfs}: it isolates the contribution generated by the tensor sector itself and therefore applies across otherwise different microscopic realizations of FOPTs and DWs.

We calculate the tensor-induced density spectrum for finite-time, sub-Hubble GW sources and determine the associated PBH mass function. For FOPTs, the result maps PBH limits onto the transition strength $\alpha$, inverse duration $\beta/H$, and temperature $T_\star$; for DWs, it constrains the wall energy fraction at annihilation $\alpha_{\rm ann}$, the bias $V_{\rm bias}$, and the surface tension $\sigma$. We further derive semi-analytical relations that expose how the induced spectrum, characteristic PBH mass, and relic abundance inherit the amplitude and scale of the source GW spectrum.

The central outcome is a joint observational target. Regions in which tensor-induced PBHs are cosmologically relevant also predict stochastic GW signals in the sensitivity ranges of pulsar-timing arrays and future space- and ground-based interferometers. Conversely, PBH non-observation constrains phase transitions and topological defects even when their GW backgrounds are not directly detected. The resulting correlation between the GW peak, the PBH mass scale, and the source parameters provides a discriminant between this second-order tensor channel and source-specific PBH formation mechanisms.

\paragraph*{\textbf{GW sources from phase transitions and domain walls.-}}
We work in radiation domination. For FOPTs\footnote{For a review  on strong FOPT related to supercooled phase transition, see Ref. \cite{Athron:2023xlk}.}, $\alpha\equiv\rho_{\rm vac}/\rho_r$ is the transition strength, $\beta^{-1}$ the characteristic duration, $H_\star$ the Hubble rate, $v_w$ the wall speed, $\kappa_i$ the efficiency of source $i$, and $\Upsilon_i$ its finite-lifetime suppression. A general component scales as $\Omega_{{\rm GW},\star}^{i}\propto C_i(v_w)(H_\star/\beta)^{p_i}[\kappa_i\alpha/(1+\alpha)]^2\Upsilon_i{\cal S}_i$, with $p_{\rm col}=2$ for collisions and $p_{\rm sw}=1$ for a long-lived acoustic source before lifetime suppression~\cite{Caprini:2015zlo,Caprini:2019egz}. The numerical results below use the collision-dominated benchmark $p_i=2$, $\kappa_i=\Upsilon_i=1$; the resulting limits are therefore conditional on that source prescription. For $\alpha\gg1$ we also require percolation and completion without prolonged vacuum domination.

For the collision-dominated benchmark, the GW background can be written as a broken power law
\begin{equation}
\Omega_{\rm GW}^{\rm FOPT}(f)
=
\left(\frac{\alpha}{1+\alpha}\right)^2
\left(\frac{\beta}{H}\right)^{-2}
{\cal S}_{\rm FOPT}(f/f_p),
\label{eq:OmegaFOPT}
\end{equation}
where $\alpha$ is the transition strength, $\beta/H$ is the inverse duration in Hubble units, and $T_\star$ sets the peak frequency through
\begin{equation}
f_p \simeq 0.7\,f_H(T_\star)\,\beta/H,
\qquad
f_H(T)\simeq 2.6\times10^{-8}\,{\rm Hz}
\left(\frac{T}{\rm GeV}\right),
\label{eq:fpeakFOPT}
\end{equation}
up to the usual dependence on relativistic degrees of freedom~\cite{Caprini:2015zlo,Caprini:2019egz,LISACosmologyWorkingGroup:2022jok}. The ${\cal S}_{\rm FOPT}$ is defined as 
\begin{eqnarray}
  \mathcal{S}_{\rm FOPT} &=&  \frac{A (a+b)^c S_H(f,f_H(T_{\star}))}{\left(b \left[\frac{f}{f_p}\right]^{\!-\frac{a}{c}} + a \left[\frac{f}{f_p}\right]^{\frac{b}{c}}\right)^{\!c}}\,.
\end{eqnarray}
Here $A$ is the peak normalization, $a$ and $b$ are the infrared and ultraviolet spectral indices, $c$ controls the turnover, and $S_H$ imposes the causal Hubble-scale suppression.  In the infrared regime, causality enforces $\Omega_{\rm GW}\propto f^3$~\cite{Caprini:2009fx,Cai:2019cdl,Domenech:2020kqm}.

For annihilating domain walls, the GW background is similarly characterized by the wall energy fraction at annihilation, $\alpha_{\rm ann}\equiv \rho_{\rm DW}/\rho_r$, and the annihilation scale $T_{\rm ann}$:
\begin{equation}
\Omega_{\rm GW}^{\rm DW}(f)
=
\frac{3\epsilon_{\rm GW} \alpha_{\rm ann}^2}{8\pi} \!\left( \frac14\! \left[ \frac{\Omega_{\rm CT}(f_p)}{\Omega_{\rm CT}(f)}\right]^{\frac{1}{\delta}} \!\!+ \frac34\! \left[\frac{f}{f_p}\right]^{\!\frac{1}{\delta}}\right)^{\!\!-\delta},
\label{eq:OmegaDW}
\end{equation}
where $\epsilon_{\rm GW}$ is the simulation-calibrated GW efficiency, $\delta$ controls the turnover, and $\Omega_{\rm CT}$ is the causal-turnover factor normalized at the peak with peak frequency $f_p=f_H(T_{\rm ann})$ and a high-frequency tail scaling approximately as $f^{-1}$~\cite{Hiramatsu:2013qaa,Saikawa:2017hiv,Ferreira:2022zzo}. The annihilation temperature is controlled by the competition between the bias energy density $(V_{\rm bias})$ and wall tension $(\sigma)$,
\begin{equation}
T_{\rm ann}
\simeq
\frac{5\,{\rm MeV}}{\sqrt{A}}
\left(\frac{V_{\rm bias}^{1/4}}{10\,{\rm MeV}}\right)^2
\left(\frac{10^5\,{\rm GeV}}{\sigma^{1/3}}\right)^{3/2},
\label{eq:Tann}
\end{equation}
where $A\simeq0.8$ is fixed by simulations~\cite{Hiramatsu:2013qaa,Ferreira:2022zzo}. To avoid cosmological problems, annihilation must occur before DW domination and before BBN~\cite{Gelmini:1988sf,Fields:2019pfx,Planck:2018vyg}. For more details regarding the GW spectra considered in this Letter, see the Supplemental Material.\footnote{The Supplemental Material gives the source spectra, first-order scalar comparison, kernel derivation and validation, tensor-to-scalar transfer, spectral enhancement, PBH abundance prescription, parameter scalings, inflationary-source caveat, particle-model examples, and uncertainty summary.} 

\paragraph*{\textbf{Tensor-induced density fluctuations.-}}
We consider a radiation-dominated Universe with first-order tensor perturbations $\chi_{ij}$ and compute the induced scalar perturbations at second order. Combining the fluid conservation equations with Einstein’s equations yields a sourced evolution equation for the second-order velocity perturbation, whose source is quadratic in the first-order tensors~\cite{Bari:2022grh}. The resulting second-order comoving-gauge density contrast $\Delta^{(2)}$ can be written in Fourier space as a convolution of the primordial tensor modes with a radiation-era kernel. Its power spectrum is \cite{Kumar:2025jfi}\footnote{Refer to SM Section 3 for details}
\begin{equation}
\begin{split}
{\cal P}_{\Delta^{(2)}}(\eta,k) &= \frac12\int_0^\infty dv \int_{|1-v|}^{1+v} du
  \frac{f(u,v)}{(uv)^3} \overline{{\cal I}^2(u,v)} \\
&\quad \times {\cal P}_{\chi_{\rm ini}}(ku) {\cal P}_{\chi_{\rm ini}}(kv),
\end{split}
\label{eq:Pdelta2}
\end{equation}
where $x\equiv k\eta$, $v\equiv p/k$, $u\equiv|\mathbf{k}-\mathbf p|/k$, $f(u,v)$ is the polarization contraction, and ${\cal I}(u,v,x;x_{\rm ini})$ is the radiation-era response integrated from $x_{\rm ini}=k\eta_{\rm ini}$~\cite{Bari:2022grh,Bari:2021xvf}. The quantity $x_{\rm ini}$ is a lower limit of the time integral defining ${\cal I}$, not of the momentum convolution. After the active source has switched off, freely propagating subhorizon tensors satisfy
\begin{equation}
\Omega_{\rm GW}(k,\eta)=\frac{1}{3}\left(\frac{k}{\mathcal H}\right)^2\overline{{\cal P}_{\chi}(k,\eta)},
\label{eq:PhOmega}
\end{equation}
where ${\cal H}=aH$ and ${\cal P}_{\chi}$ includes both helicities. This matching assumes a finite active stage followed by free propagation; a long-lived incoherent source requires its tensor unequal-time correlator and can modify the normalization. \footnote{We therefore treat the conversion from the source GW spectrum to ${\cal P}_{\chi,\rm ini}$ as a free-propagation matching approximation. A residual normalization uncertainty in this matching rescales ${\cal P}_{\Delta^{(2)}}$ quadratically and is part of the systematic error budget discussed in the Supplemental Material.} For transient sub-horizon sources such as FOPTs and DWs, the initial tensor spectrum ${\cal P}_{\chi_{\rm ini}}(k)$ is related to the GW energy density at horizon crossing via $ {\cal P}_{\chi_{\rm ini}}(k) \simeq 3\,\Omega_{\rm GW}(x_{\rm ini} k)$. \footnote{For the detailed derivation refer to Eq. (S50) to S51 in Supplemental Material.}

A key feature of \cref{eq:Pdelta2} is that
$\mathcal{P}_{\delta^{(2)}}\propto\mathcal{P}_{\chi_\text{ini}}^2$ meaning that even a moderate GW amplitude can drive a substantial density perturbation through
the convolution.
For sub-Hubble FOPT and DW sources, the lower integration limit is
$x_\text{ini}$, parameterizing the time of Hubble crossing\footnote{For the case of FOPT, the $x_{\rm ini}$ is simply related to $\beta / H$ while for the domain wall case we use the results extracted from lattice simulations. ~\cite{Notari:2025kqq} See SM for details}. The resulting $\mathcal{P}_{\delta^{(2)}}$ admits the following scaling laws for FOPT and DW respectively:

\begin{align}
\mathcal{P}_{\delta^{(2)}}^\text{FOPT}
&\propto
\left(\frac{\alpha}{1+\alpha}\right)^4
\left(\frac{\beta}{H}\right)^{-4}
\begin{cases}
(k/k_p)^{2a}, & k\ll k_p \\[2pt]
(k/k_p)^{a-b}, & k_p\leq k\leq k_p^\text{TIS}\\[2pt]
(k/k_p)^{-2b}, & k\gg k_p^\text{TIS}
\end{cases}
\label{eq:ind_PT_ana} \\
\mathcal{P}_{\delta^{(2)}}^\text{DW}
&\propto
\frac{9\epsilon^2\alpha_\text{ann}^4}{64\pi^2}
\begin{cases}
(k/k_p)^6, & k\ll k_p \\[2pt]
(k/k_p)^2, & k_p\leq k\leq k_p^\text{TIS}\\[2pt]
(k/k_p)^{-2}, & k\gg k_p^\text{TIS}
\end{cases}
\label{eq:indDW_ana}
\end{align}
The tensor-induced scalar (TIS) peak sits well above the GW peak:
$k_p^\text{TIS}\approx3.88\times10^2\,k_p^\text{GW}$ for FOPT and
$\approx10^3\,k_p^\text{GW}$ for DW\footnote{These factors, $3.88 \times 10^2$ and $10^3$, do not depend upon the source parameters.}. The quoted peak ratios are benchmark dependent rather than universal: they are not the acoustic resonance factors themselves. The acoustic resonance is an order-unity condition, while the larger separation also depends on $x_{\rm ini}$ and on the hierarchy between Hubble and microscopic source scales. We therefore evaluate $k_p^{\rm TIS}$ for each numerical parameter point. The kernel enhancement follows from the resonance condition $c_s(u+v)=1$, with $c_s=1/\sqrt{3}$, i.e. $u+v=\sqrt{3}$
that maximizes the kernel $\overline{\mathcal{I}^2}$ in \cref{eq:Pdelta2}. The induced scalar spectrum can dominate over the corresponding first-order scalar perturbations associated with the source, making the tensor-induced contribution the leading effect for PBH formation in the relevant parameter space. We have validated the scale dependence of \cref{eq:ind_PT_ana,eq:indDW_ana} via numerical integration of  \cref{eq:Pdelta2}. The numerical spectra also reproduce the expected causal large-scale scaling.\footnote{We refer the reader to SM for detailed discussion of parameter dependence of GW spectrum and causal behavior of induced density spectrum.}

\paragraph*{\textbf{PBH abundance.-}}
PBHs form when the smoothed comoving density contrast at horizon entry exceeds $\Delta_c$. We use a real-space top-hat $W(y)=3(\sin y-y\cos y)/y^3$ and evaluate the variance directly from the induced density spectrum,
\begin{equation}
\sigma_R^2(\eta_R)=\int d\ln k\,W^2(kR){\cal P}_{\Delta^{(2)}}(k,\eta_R),\qquad aH|_{\eta_R}=R^{-1}.
\label{eq:variance}
\end{equation}
We adopt $\Delta_c=0.51$ \footnote{The threshold value $\Delta_c=0.51$ should be understood as a benchmark collapse prescription. It is calibrated in the standard long-wavelength treatment of PBH formation from radiation-era overdensities and is widely used for comparison with conventional PBH calculations. The density perturbations studied here, however, are generated by causal tensor sources and can have profiles that differ from the adiabatic super-horizon profiles used in numerical-collapse calibrations. A dedicated compaction-function analysis, or numerical-relativity collapse simulation, for tensor-induced causal profiles would be required to determine the source-specific threshold. We therefore interpret the abundance contours as conditional on this benchmark threshold, while the tensor-induced spectrum, its parameter scaling, and the GW-PBH mass correlation do not rely on the precise numerical value of $\Delta_c$. } and critical collapse,
\begin{equation}
M={\cal K}M_H(\Delta-\Delta_c)^\gamma,\qquad {\cal K}=4.36,\quad\gamma=0.38.
\label{eq:MPBH}
\end{equation}
Since $\Delta^{(2)}$ is quadratic in the tensor field, it is intrinsically non-Gaussian even for Gaussian tensor modes~\cite{Abdelaziz:2025qpn}. In this first study we nevertheless retain the Gaussian one-point approximation $P_G(\Delta_R)=\exp[-\Delta_R^2/(2\sigma_R^2)]/(\sqrt{2\pi}\sigma_R)$ throughout \footnote{This supplies a transparent benchmark and allows direct comparison with conventional PBH calculations, but PBHs probe the far tail and connected higher cumulants can change the abundance exponentially.}. The contours below must therefore be read as Gaussian-benchmark constraints; deriving the full quadratic-field probability distribution and compaction-function statistics is left for future work. The complete PBH prescription, including window choice, critical collapse, extended-mass-function recasting, mass mapping, and uncertainty budget, is collected in the Supplemental Material (SM) \footnote{Refer to section 7 in SM.}.

A comment on the collapse threshold is important. The threshold $\Delta_c=0.51$ used in the Letter is the conventional radiation-era benchmark for PBH formation from long-wavelength overdensities. The FOPT and DW sources considered in the main analysis are causal and can generate density profiles whose detailed compaction functions differ from those of standard super-horizon curvature perturbations. Since PBH collapse is known to depend on the profile shape and on the maximum of the compaction function, the numerical value of $\Delta_c$ is not universal. In this work we therefore use $\Delta_c=0.51$ only as a common benchmark that allows comparison with the standard PBH literature. A source-specific determination of $\Delta_c$ for tensor-induced causal profiles is left for future numerical-relativity or compaction-function studies.

The formalism itself is not restricted to sub-Hubble sources. The $x_{\rm ini}\to0$ kernels derived in \cite{Kumar:2025jfi} apply to super-horizon tensor initial conditions, such as inflationary tensor spectra. We include the super-horizon limit to show that the tensor-induced density formalism connects smoothly to the standard long-wavelength setup in which the usual PBH collapse threshold is most directly motivated. The FOPT and DW results in the Letter, by contrast, use the finite-$x_{\rm ini}$ kernels appropriate to causal post-inflationary sources and should be read with the benchmark-threshold caveat above.

Critical collapse produces an extended mass function,
\begin{equation} 
\frac{d\beta}{d\ln M}=\frac{M}{M_H}P_G[\Delta(M)]\left|\frac{d\Delta}{d\ln M}\right|,
\label{eq:massfunction} 
\end{equation}
where $\Delta(M)=\Delta_c+\left(\frac{M}{{\cal K}M_H}\right)^{1/\gamma}$. After red-shifting this spectrum as non-relativistic matter, bounds that are linear in abundance are recast as
\begin{equation}
\int d\ln M\,\frac{f_{\rm PBH}(M)}{f_{\max}^{\rm mono}(M)}\leq1.
\label{eq:extendedconstraint}
\end{equation}
Accretion, merger-rate, and dynamical limits have additional nonlinear and astrophysical dependence and are displayed separately. The characteristic mass is obtained from the numerical maximum $k_p^{\rm TIS}$ using $M_H(k)\propto k^{-2}g_*^{-1/6}$. For FOPTs,
\begin{equation}
k_p^{\rm TIS}=C_{\rm TIS}(x_{\rm ini},a,b)k_p^{\rm GW}\propto C_{\rm TIS}T_\star\frac{\beta}{H},
\label{eq:MPBHFOPT}
\end{equation}
whereas for DWs
\begin{equation}
k_p^{\rm TIS}=C_{\rm TIS}^{\rm DW}k_H(T_{\rm ann}).
\label{eq:MPBHDW}
\end{equation}
Thus $M_p\propto C_{\rm TIS}^{-2}T^{-2}$, with an additional $(\beta/H)^{-2}$ dependence for FOPTs unless it is compensated by the matching-time dependence of $C_{\rm TIS}$.
The peak variance scales as $\sigma_{R,p}^2\propto[\kappa_i\alpha/(1+\alpha)]^4(H_\star/\beta)^{2p_i}$ for FOPTs and as $\sigma_{R,p}^2\propto\epsilon_{\rm GW}^2\alpha_{\rm ann}^4$ for DWs.

\begin{figure*}[ht]
\includegraphics[width=8.5 cm,height=6.0 cm]{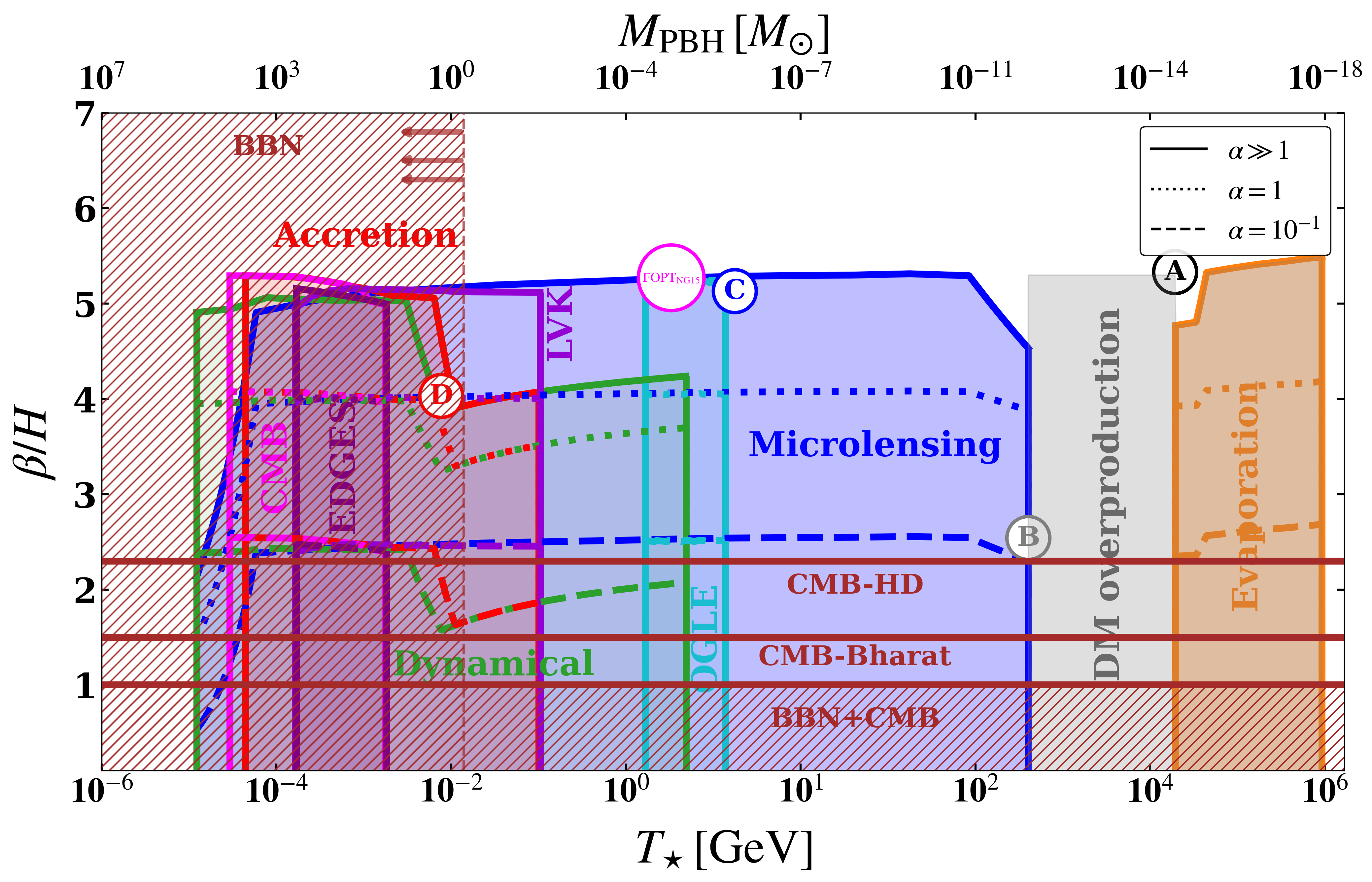}
\includegraphics[width=8.5 cm,height=6.0 cm]{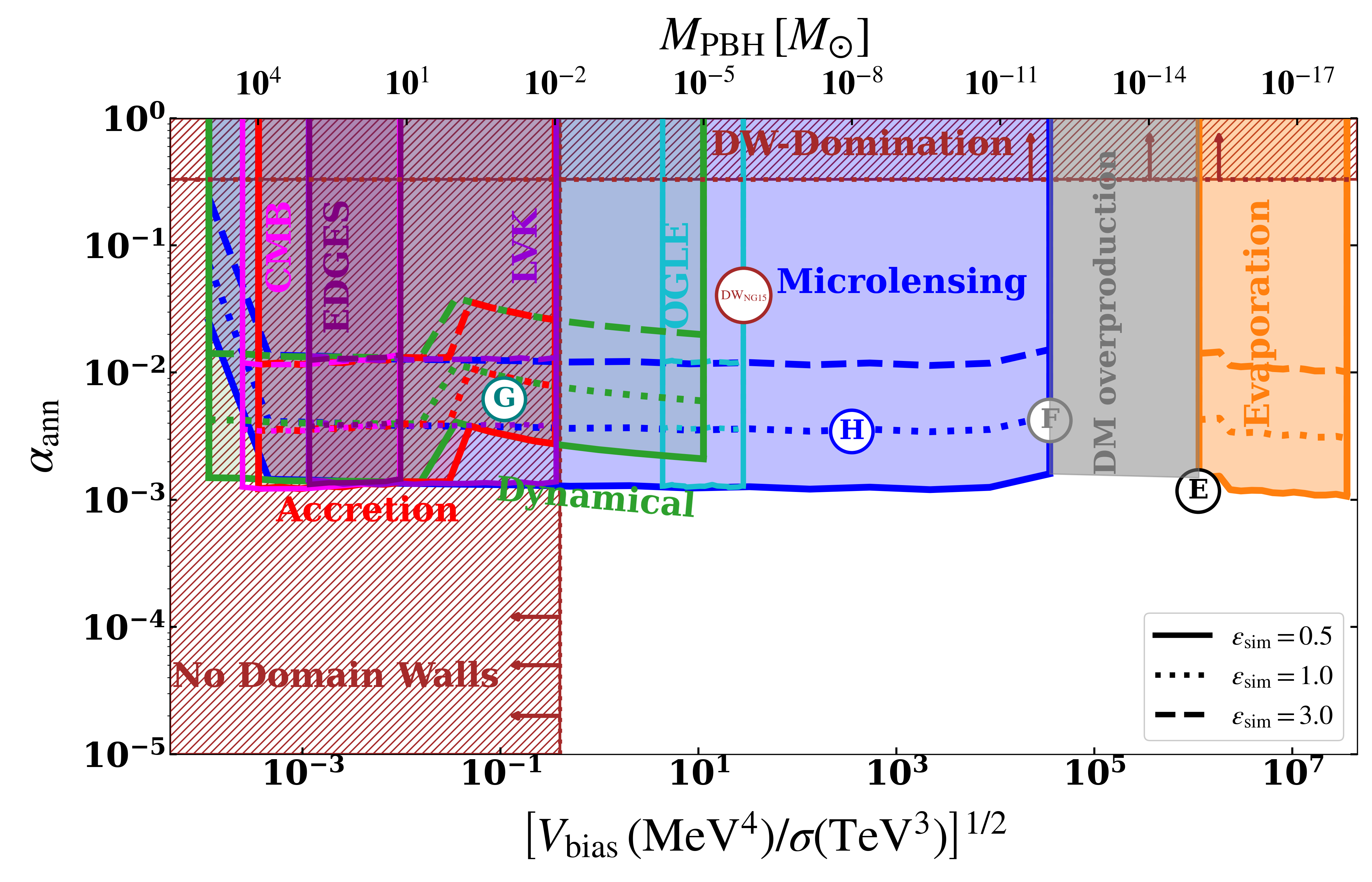}
\caption{\textit{Exclusion regions on FOPT parameters in the $\beta/H$-$T_\star$ plane (left) and on DW parameters in the $\alpha_{\rm ann}$-$(V_{\rm bias}/\sigma)^{1/2}$ plane (right); top axes give the corresponding PBH mass via \cref{eq:MPBHFOPT,eq:MPBHDW}. The solid, dotted, and dashed curves are contours of $f_{\rm PBH}=1$ for $\alpha\gg1$, $\alpha=1$, and $\alpha=10^{-1}$ (left), and for the lattice-calibrated horizon-crossing parameter $\epsilon_{\rm sim}=0.5$, $1.0$, and $3.0$ (right). See text for the colored/hatched regions and the circled benchmark points.}}
\label{fig:PT_exclusion}
\end{figure*}

\paragraph*{\textbf{Results and conditional constraints.-}}
We now translate the tensor-induced spectrum into constraints on the macroscopic FOPT and DW parameters, using the full landscape of existing PBH bounds spanning evaporation, lensing, accretion, and merger observables across the PBH mass range~\cite{Carr:2020gox,Green:2020jor,Saha:2021pqf,Laha:2019ssq,Ray:2021mxu,Carr:2026hot}.
The vertical bands in \Cref{fig:PT_exclusion} are mass-dependent PBH bounds, independent of $\beta/H$ or $\alpha_{\rm ann}$: Hawking \textbf{evaporation}~\cite{Carr:2009jm,Boudaud:2018hqb,DeRocco:2019fjq,Laha:2020ivk}, dark-matter \textbf{overproduction} ($f_{\rm PBH}>1$), \textbf{microlensing}/OGLE~\cite{Niikura:2019kqi,MACHO:2000qbb,EROS-2:2006ryy,Niikura:2017zjd}, \textbf{dynamical}-heating limits~\cite{Green:2020jor,Carr:2026hot}, \textbf{LVK} merger-rate bounds~\cite{DeLuca:2020qqa}, \textbf{EDGES} 21-cm heating~\cite{Mittal:2021egv,Dasgupta:2019cae}, and CMB $\mu$-distortion/\textbf{accretion}~\cite{Kohri:2014lza,Nakama:2016kfq,Khan:2025kag,Nakama:2017xvq,Ali-Haimoud:2016mbv,Poulin:2017bwe,Serpico:2020ehh}. In the left (FOPT) panel, the horizontal hatched bands exclude small $\beta/H$ where the GW background itself would violate the dark-radiation bound, currently from BBN+CMB~\cite{Planck:2018vyg,Fields:2019pfx} and prospectively from CMB-Bharat/CMB-HD~\cite{BICEP:2021xfz,SimonsObservatory:2018koc,Hazumi:2019lys,Abazajian:2019eic,SPHEREx:2014bgr,Sehgal:2019ewc,CMB-HD:2022bsz}, together with the low-reheating \textbf{BBN} band $T_\star\lesssim{\rm MeV}$~\cite{Rubakov:2017xzr}. In the right (DW) panel, the hatched ``DW domination'' band marks $t_{\rm ann}\gtrsim t_{\rm dom}$, where the bias energy would come to dominate the Universe~\cite{Ferreira:2022zzo}, and ``No Domain Walls'' marks $V_{\rm bias}$ large enough to preempt wall formation. Circled points A-D and ${\rm FOPT_{NG15}}$ (left), and E-H and ${\rm DW_{NG15}}$ (right), are the benchmarks tabulated in the Supplemental Material: A ($\alpha=10^3$)/B ($\alpha=10^{-1}$) and E/F sit on the $f_{\rm PBH}=1$ curve, realizing PBHs as the entire dark matter at asteroid mass within the Gaussian-tail benchmark; C, D and G, H give a sub-percent, future-testable fractional abundance; ${\rm FOPT_{NG15}}$ and ${\rm DW_{NG15}}$ are normalized to the NANOGrav 15-yr common-spectrum amplitude under a cosmological-source interpretation while remaining safely below overproduction.

All displayed benchmark points are required to satisfy the perturbativity checks $\max_k {\cal P}_{\chi,\rm ini}(k)<1$, $\max_k {\cal P}_{\Delta^{(2)}}(k)<1$, and $\max_k {\cal P}_{\zeta,\rm tot}(k)\lesssim {\cal O}(1)$. These conditions ensure, respectively, that the tensor metric perturbation remains perturbative, that the induced scalar density response does not enter a fully nonlinear regime, and that the scalar perturbations used in the abundance estimate remain within the domain of the second-order calculation.\footnote{The corresponding perturbativity masks and analytic scaling estimates are summarized in the Supplemental Material. Refer to Section 11 in SM.}

For FOPTs, strong and slow transitions are the most tightly constrained: for $\alpha\gtrsim\mathcal{O}(1)$, $\beta/H\lesssim6$ is excluded by PBH overproduction or non-observation across a wide span of $T_\star$ (left panel of \Cref{fig:PT_exclusion}), and for $\alpha<10^{-2}$ viability requires $\beta/H<1$, already in tension with BBN/CMB. Within the Gaussian-tail benchmark, this same region hosts all-dark-matter solutions: asteroid-mass PBHs saturate $\Omega_{\rm PBH}=\Omega_{\rm CDM}$ for $T_\star\in(4\times10^2,10^4)\,{\rm GeV}$ at $\beta/H\simeq6$, $\alpha\gtrsim\mathcal{O}(1)$, with GW peak $\Omega_{\rm GW}^{p}h^2\sim\mathcal{O}(10^{-8})$ at $f_p\in(10^{-5},10^{-2})\,{\rm Hz}$, squarely within reach of LISA and SKA and, for $\beta/H\lesssim5$, of next-generation CMB experiments~\cite{BICEP:2021xfz,SimonsObservatory:2018koc,Hazumi:2019lys,Abazajian:2019eic,SPHEREx:2014bgr}. FOPT parameters consistent with the NANOGrav/PTA common-spectrum signal~\cite{NANOGrav:2023gor,Antoniadis:2023rey,Reardon:2023gzh,Xu:2023wog} do not overproduce PBHs, while the recently claimed HSC-SUBARU lensing events~\cite{Sugiyama:2026kpv} (whose lensing origin has been questioned~\cite{Mroz:2026nez}) map onto $T_\star\approx(1$-$10)\,{\rm GeV}$ in our framework. Given that $k_{p}^{\rm TIS}$ differs from the $k_{p}^{\rm GW}$, the resulting $(\beta/H,\alpha,T_\star)$-$M_{\rm PBH}$ correlation differs cleanly from delayed-vacuum-transition scenarios~\cite{Gouttenoire:2023naa,Lewicki:2023ioy}, which have been argued to be gauge dependent~\cite{Franciolini:2025ztf}\footnote{A recent study argues the scenario can be rescued with an epoch of early matter domination~\cite{Ai:2026zrs}.}, offering a distinguishing observational signature.

The right panel of \Cref{fig:PT_exclusion} shows an analogous structure for DW annihilation; the lattice-calibrated horizon-crossing parameter $\epsilon_{\rm sim}$~\cite{Notari:2025kqq} plays a role analogous to $\alpha$ for FOPTs. PBH production is governed primarily by the wall energy fraction $\alpha_{\rm ann}$ and by $(V_{\rm bias}/\sigma)^{1/2}$, which fixes $T_{\rm ann}$ and hence $M_{\rm PBH}$; the allowed $\alpha_{\rm ann}$ range further depends on the lattice-calibrated crossing parameter $\epsilon_{\rm sim}$, larger values (later horizon crossing) requiring smaller $\alpha_{\rm ann}$ for fixed $f_{\rm PBH}$. Within the Gaussian-tail benchmark, PBHs can constitute all of the dark matter for $(V_{\rm bias}\,[{\rm MeV}^4]/\sigma\,[{\rm TeV}^3])^{1/2}\in[10^2,10^5]$, equivalently $\sigma^{1/3}\in[10^6,10^8]\,{\rm TeV}$ and $V_{\rm bias}^{1/4}\in[10^7,10^{10}]\,{\rm MeV}$, with GW peak $\Omega_{\rm GW}^{p}h^2\sim\mathcal{O}(10^{-9})$ for $\alpha_{\rm ann}\sim10^{-2}$ at $f_p\in(4\times10^{-4},10^{-1})\,{\rm Hz}$ and $T_{\rm ann}\in4.5\times[10^3,10^6]\,{\rm GeV}$, testable by $\mu$-ARES, LISA, DECIGO, and ET; DW parameters compatible with the PTA signal again do not overproduce PBHs. Relative to mechanisms invoking the DW first-order curvature or direct wall collapse~\cite{Ferreira:2023jbu,Ferreira:2022zzo,Ferreira:2024eru,Dunsky:2024zdo,Gouttenoire:2023gbn}, our tensor-induced channel predicts a different $M_{\rm PBH}$-$T_{\rm ann}$ correlation, providing a complementary way to discriminate between competing PBH-formation pathways with joint GW and PBH data.

\begin{figure*}[ht]
\includegraphics[width=8.5 cm,height=6.0 cm]{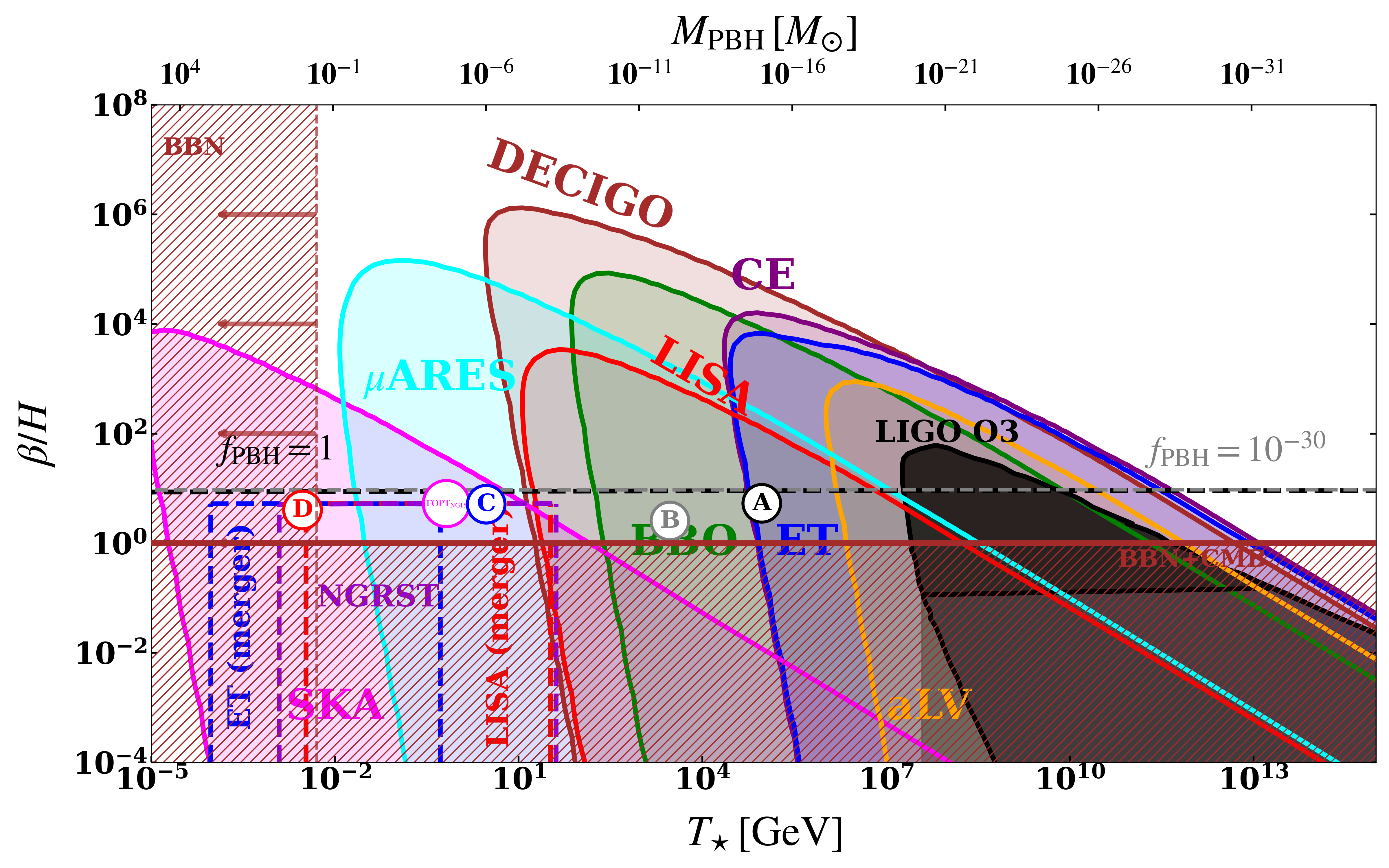}
\includegraphics[width=8.5 cm,height=6.0 cm]{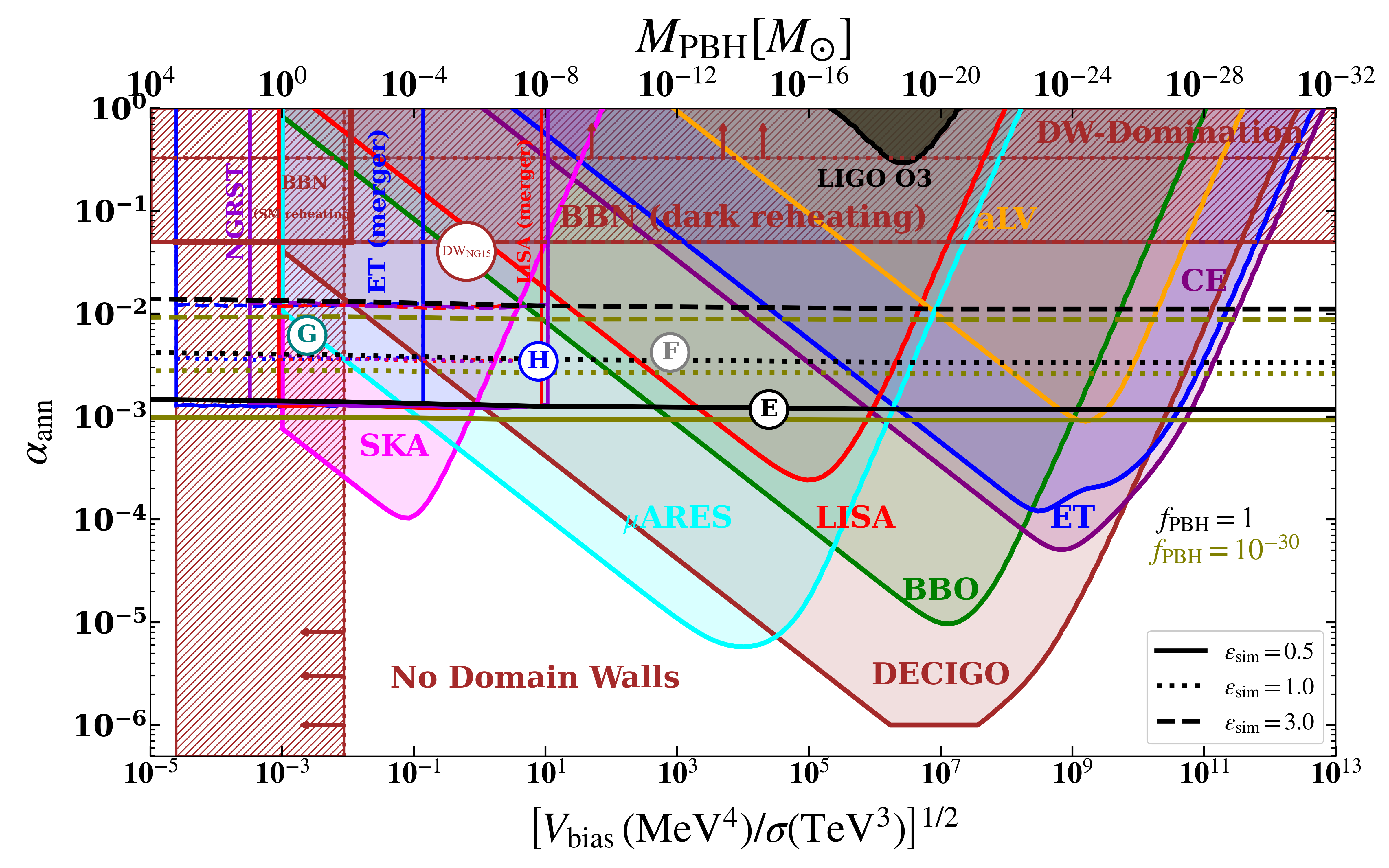}
\caption{\textit{Gravitational-wave detectability (${{\rm SNR}}>10$) of the FOPT (left) and DW (right) sources of \Cref{fig:PT_exclusion}, in the same respective planes. Black-dashed and olive dot-dashed curves reproduce the $f_{\rm PBH}=1$ and $f_{\rm PBH}=10^{-30}$ contours for the same line styles as in \Cref{fig:PT_exclusion} ($\alpha\gg1,1,10^{-1}$ left; $\epsilon_{\rm sim}=0.5,1.0,3.0$ right). See text for the detectors and hatched bands.}}
\label{fig:SNR_PT}
\end{figure*}
\Cref{fig:SNR_PT} shows where the FOPT bubble/sound-wave background of \cref{eq:OmegaFOPT} (left) and the DW-annihilation background of \cref{eq:OmegaDW} (right) are detectable by SKA~\cite{Janssen:2014dka,Weltman:2018zrl,Carilli:2004nx}, $\mu$-ARES~\cite{Sesana:2019vho}, LISA~\cite{amaroseoane2017laser,Baker:2019nia,LISACosmologyWorkingGroup:2022jok}, DECIGO~\cite{Yagi:2011yu,Kawamura:2020pcg}, BBO~\cite{Corbin:2005ny,Harry:2006fi}, ET~\cite{Punturo:2010zz,Hild:2010id}, and CE~\cite{Reitze:2019iox}, together with the current LVK-O3 sensitivity (aLV)~\cite{KAGRA:2021kbb}. Dashed vertical lines mark the projected reach of PBH-merger searches at ET/LISA and of NGRST microlensing~\cite{DeRocco:2023gde}, set by $M_{\rm PBH}$ alone; the hatched bands are excluded by the dark-radiation bound on the respective GW backgrounds and, in the right panel, by DW over-closure (``DW domination''). The PBH-viable regions for both sources overlap substantially with these detector bands, after also imposing the dark-radiation bound $\Delta N_{\rm eff}\lesssim0.3$ on $\int d\ln f\,\Omega_{\rm GW,0}(f)h^2$~\cite{Planck:2018vyg,Fields:2019pfx}. Since $M_{\rm PBH}$ tracks $T_\star$ or $T_{\rm ann}$ one-to-one, each detector band maps onto a characteristic PBH mass window (\Cref{business_table}): PTA/SKA probes $M_{\rm PBH}\sim10^{-6}$-$10^{3}\,M_\odot$, LISA and $\mu$-ARES the asteroid-to-solar mass range most relevant for dark matter, and BBO/DECIGO/ET/CE the sub-asteroid range down to the evaporation limit. This mapping makes a stochastic GW detection in a given band an immediate, sharply testable prediction for the associated PBH abundance, while stochastic-background non-detection excludes only spectra above the analysis-specific GW upper limit. Because this tensor-induced channel is generic to any transient radiation-era source, it constitutes an irreducible tensor-induced contribution under the stated source and statistical assumptions to the PBH abundance, on top of which source-specific channels (delayed nucleation, direct wall collapse) may add further contributions.
\paragraph{Viable Particle Theory Scenarios}
We discuss the FOPT and DW annihilation arising due to well-motivated particle theory models where parameters like $\alpha, \beta/H, T_{\star}$ for FOPT and $\alpha_{\rm ann}, \sigma, V_{\rm bias} $ arise from microscopic theory parameters involving mass and interaction strengthen of particles present in early Universe. We refer the reader to Supplementary materials for some discussion related to the particle microphysics.

\begin{table}
 \caption{\textit{Correlation between GW detectors and PBH.}}
\renewcommand{\arraystretch}{0.6}
\hspace{-0.75cm}
\begin{tabular}{|l|l|l|l|}
\hline
\textbf{Detector} & Freq. &  \textbf{$M_{\rm PBH}$} (FOPT) & \textbf{$M_{\rm PBH}$} (DW) \\
\hline
SKA & nHz & $(10^{-5} - 10^3) \, M_{\odot}$ & $(10^{-6} - 10^{0}) \, M_{\odot}$\\
$\mu$Ares & $\mu$Hz & $(10^{-15} - 10^1) \, M_{\odot}$ & $(10^{-18} - 10^{-4}) \, M_{\odot}$\\
LISA & mHz & $(10^{-18} - 10^6) \, M_{\odot}$ & $(10^{-18} - 10^{-12}) \, M_{\odot}$\\
BBO & dHz & $(10^{-21} - 10^{-8}) \, M_{\odot}$ & $(10^{-24} - 10^{-12}) \, M_{\odot}$\\
ET, CE & Hz & $(10^{-26} - 10^{-14}) \, M_{\odot}$ & $(10^{-26} - 10^{-18}) \, M_{\odot}$\\
LVK & Hz & $(10^{-26} - 10^{-16}) \, M_{\odot}$ & $(10^{-26} - 10^{-22}) \, M_{\odot}$\\
\hline
\end{tabular}
\label{business_table}
\end{table}

\paragraph{\textbf{Discussion \& Conclusion}}
\label{sec:conclusion}
We have identified a direct route from primordial tensor perturbations to PBHs. Gravitational waves produced by FOPT bubble collisions and acoustic motion, or by DW annihilation, source scalar density fluctuations at second order; if sufficiently large, these perturbations collapse after horizon re-entry. Extending the tensor-induced formalism of Ref.~\cite{Kumar:2025jfi} to causal, finite-time, sub-Hubble sources, we have derived the induced spectra, computed the corresponding PBH abundance, and translated PBH observations into constraints on $(\alpha,\beta/H,T_\star)$ and $(\alpha_{\rm ann},V_{\rm bias},\sigma)$. The analysis isolates the tensor-generated contribution, complementary to delayed-vacuum-transition and direct-defect-collapse channels~\cite{Gouttenoire:2023gbn,Gouttenoire:2023naa,Lewicki:2023ioy,Ferrer:2018uiu,Gelmini:2022nim,Gelmini:2023ngs}.

The phenomenological consequence is sharp: the same source fixes both a stochastic GW spectrum and a characteristic PBH mass scale. PBH limits can therefore test early-Universe phase transitions and topological defects independently of a resolved GW detection, while a future stochastic-background measurement would select a correlated PBH mass window and abundance. The detector-mass correspondence is summarized in \Cref{business_table}.

The macroscopic parameters entering this correlation are themselves calculable from particle physics. Strong FOPTs may arise, for example, in $U(1)_{B-L}$-extended type-I seesaw models, whereas metastable DW networks occur naturally in Dirac-seesaw and leptogenesis scenarios with spontaneously broken $\mathcal{Z}_2$ symmetry; representative realizations are presented in the Supplemental Material.\footnote{See Section 10 in Supplemental Material.} The accompanying semi-analytical relations connect these microscopic dynamics to $\mathcal{P}_{\delta^{(2)}}$, $M_{\rm PBH}$, and $f_{\rm PBH}$, turning PBH and GW observations into complementary probes of otherwise inaccessible early-Universe physics. We provide the symbolic expressions for allowed values of $\beta/H$ and $\alpha_{\rm ann}$ for FOPTs and annihilating DWs respectively in terms of other source specific parameters to relate them specific particle physics models \footnote{Symbolic formulae for FOPT and annihilating DWs are provided in SM section 8.}.

The key prediction is not a GW signal or a PBH population in isolation, but their correlation. A stochastic background in a given frequency band points to a definite PBH mass range; the absence of PBHs in that range constrains the source, while a coincident signal in both channels would identify the underlying cosmic scale. Tensor-induced PBH formation thus converts phase transitions and topological defects into jointly testable targets for GW observatories, PBH searches, and dark-matter experiments. If asteroid-mass PBHs constitute the dark matter in the Gaussian-benchmark viable regions identified here, the associated GW background is not an optional by-product but a second observable imprint of the same early-Universe event. This two-channel consistency relation is the decisive signature of the mechanism.

\paragraph{\textbf{Acknowledgement}}
\label{sec:Acknowledgement}
Authors thank Angelo Ricciardone, Gabriele Franciolini, Chris Byrnes, Archit Vidyarthi and Xavier Pritchard for helpful discussion and comments. A.G. acknowledges the support from the Royal Society, UK, Funding Reference: NIF\ R1\ 253963. 

\addtocontents{toc}{\protect\setcounter{tocdepth}{-1}}
\bibliography{ref}
\addtocontents{toc}{\protect\setcounter{tocdepth}{3}}

\onecolumngrid

\renewcommand{\theequation}{S\arabic{equation}}
\renewcommand{\thefigure}{S\arabic{figure}}
\renewcommand{\thetable}{S\arabic{table}}
\renewcommand{\thesection}{\arabic{section}}
\makeatletter
\renewcommand{\@seccntformat}[1]{\csname the#1\endcsname.\quad}
\makeatother

\setcounter{section}{0}
\setcounter{subsection}{0}
\setcounter{subsubsection}{0}
\setcounter{equation}{0}
\setcounter{figure}{0}
\setcounter{table}{0}

\begin{center}
\textbf{\Large Supplemental Material}
\end{center}
\tableofcontents

\renewcommand{\theequation}{S\arabic{equation}}
\renewcommand{\thefigure}{S\arabic{figure}}
\renewcommand{\thetable}{S\arabic{table}}
\renewcommand{\thesection}{\arabic{section}}
\makeatletter
\renewcommand{\@seccntformat}[1]{\csname the#1\endcsname.\quad}
\makeatother

\section{Gravitational-wave source spectra}\label{SM:GWsources}

We collect the source conventions used in the Letter. Natural units $c=\hbar=k_B=1$ are used throughout. Quantities labelled by $\star$ are evaluated at completion of a first-order phase transition (FOPT), while ``ann'' denotes domain-wall (DW) annihilation. We assume radiation domination, entropy conservation after production, and no prolonged vacuum- or wall-dominated stage. The present-day frequency corresponding to the Hubble scale at temperature $T$ is
\begin{equation}
f_{H,0}(T)=\frac{a(T)}{a_0}\frac{H(T)}{2\pi}
\simeq2.6\times10^{-8}\,{\rm Hz}
\left(\frac{T}{\rm GeV}\right)
\left(\frac{g_{*,\rho}(T)}{100}\right)^{1/2}
\left(\frac{g_{*,s}(T)}{100}\right)^{-1/3},
\label{SM:fH}
\end{equation}
where $g_{*,\rho}$ and $g_{*,s}$ are the effective energy and entropy degrees of freedom. All quoted spectra are understood as present-day spectra after the standard radiation-era redshift.

\subsection{First-order phase transitions}\label{SM:FOPT}
For source component $i\in\{\mathrm{col},\mathrm{sw}\}$, we write
\begin{equation}
\Omega_{{\rm GW},0}^{i}(f)=C_i(v_w)
\left(\frac{H_\star}{\beta}\right)^{p_i}
\left(\frac{\kappa_i\alpha}{1+\alpha}\right)^2
\Upsilon_i\,{\cal S}_i(f/f_{p,i}),
\label{SM:OmegaPTgeneral}
\end{equation}
where $\alpha\equiv\rho_{\rm vac}/\rho_r$, $\beta^{-1}$ is the transition duration, $H_\star$ is the Hubble rate, $v_w$ is the wall speed, $\kappa_i$ is the efficiency for transferring released vacuum energy into source $i$, $C_i$ fixes the peak normalization, and $\Upsilon_i$ accounts for finite source lifetime. Collision templates have $p_{\rm col}=2$, whereas a long-lived acoustic source has $p_{\rm sw}=1$ before lifetime suppression~\cite{Lewicki:2022pdb,LISACosmologyWorkingGroup:2022jok}. The numerical analysis uses the collision-dominated benchmark $p_i=2$, $v_w=1$, $\kappa_i=1$, and $\Upsilon_i=1$. Consequently, the constraints are conditional on this benchmark; Eq.~\eqref{SM:OmegaPTgeneral} displays how a different source prescription rescales the result.

For this benchmark,
\begin{equation}
\Omega_{\rm GW}^{\rm FOPT}(f)=
\left(\frac{\alpha}{1+\alpha}\right)^2
\left(\frac{\beta}{H_\star}\right)^{-2}
\frac{A(a+b)^cS_H(f,f_H)}{[b(f/f_p)^{-a/c}+a(f/f_p)^{b/c}]^c},
\label{SM:OmegaPT}
\end{equation}
with $a=b=2.4$, $c=4$, and $A=5.1\times10^{-2}$. Here $a$ and $b$ are the low- and high-frequency indices of the broken-power-law template, $c$ controls the turnover, and $A$ normalizes the peak. We use
\begin{equation}
f_p=0.7f_{H,0}(T_\star)\frac{\beta}{H_\star},
\label{SM:fpeakPT}
\end{equation}
and impose a causal infrared turnover through
\begin{equation}
S_H(f,f_H)=\left[1+
\left(\frac{\Omega_{\rm CT}(f)}{\Omega_{\rm CT}(f_H)}\right)^{-1/\delta}
\left(\frac{f}{f_H}\right)^{a/\delta}\right]^{-\delta},
\qquad \delta=1,
\label{SM:SH}
\end{equation}
where $\Omega_{\rm CT}$ is normalized at $f_H$ and enforces $\Omega_{\rm GW}\propto f^3$ sufficiently far in the infrared~\cite{Caprini:2009fx,Domenech:2020kqm}. For $\alpha\gg1$, the benchmark additionally requires successful percolation and completion without substantial vacuum domination; parameter points violating these assumptions are outside the calculation.

\subsection{\mbox{Annihilating} domain-wall networks}\label{SM:DW}
For a scaling DW network,
\begin{equation}
\Omega_{\rm GW}^{\rm DW}(f)=\frac{3\epsilon_{\rm GW}\alpha_{\rm ann}^2}{8\pi}
\left[\frac14\left(\frac{\Omega_{\rm CT}(f_p)}{\Omega_{\rm CT}(f)}\right)^{1/\delta}
+\frac34\left(\frac{f}{f_p}\right)^{1/\delta}\right]^{-\delta},
\label{SM:OmegaDW}
\end{equation}
where $\alpha_{\rm ann}\equiv\rho_{\rm DW}/\rho_r|_{\rm ann}$, $\epsilon_{\rm GW}\simeq0.7$ is the simulation efficiency, and $\delta=1$. The peak is $f_p=f_{H,0}(T_{\rm ann})$; the ultraviolet $f^{-1}$ tail is cut off at the inverse wall thickness~\cite{Hiramatsu:2013qaa,Ferreira:2022zzo,Ellis:2023oxs}. The integrated energy, not the peak amplitude, is used for the dark-radiation bound.

The bias pressure $V_{\rm bias}$ overcomes wall tension when $H_{\rm ann}\simeq V_{\rm bias}/(A\sigma)$, giving
\begin{equation}
T_{\rm ann}\simeq\frac{5\,{\rm MeV}}{\sqrt A}
\left(\frac{10.75}{g_{*,\rho}(T_{\rm ann})}\right)^{1/4}
\left(\frac{V_{\rm bias}^{1/4}}{10\,{\rm MeV}}\right)^2
\left(\frac{10^5\,{\rm GeV}}{\sigma^{1/3}}\right)^{3/2},
\label{SM:Tann}
\end{equation}
with $A=0.8$. We require $T_{\rm ann}>T_{\rm BBN}$ and $t_{\rm ann}<t_{\rm dom}$. These conditions remove the region in which walls dominate before annihilation. The macroscopic source variables may be taken as $(\alpha_{\rm ann},T_{\rm ann})$ or equivalently $(\alpha_{\rm ann},(V_{\rm bias}/\sigma)^{1/2})$.

\section{First-order scalar spectra associated with the sources}

This section compares source-generated first-order scalar perturbations with the tensor-tensor contribution. The comparison is performed at equal wavenumber and with the same power-spectrum convention; spectral slopes alone do not establish dominance.

\subsection{First-order phase transition}\label{appn:firstorder_PT}
We use the fit~\cite{Wang:2026zvz}
\begin{equation}
{\cal P}^{(1)}_\zeta(k)=A_\zeta\left(\frac{\alpha}{1+\alpha}\right)^2
\frac{(k/k_{\max})^3(\beta/H_n)^{-3}}
{[a_\zeta+(k/k_{\max})^2(\beta/H_n)^{-2}]^3(b_\zeta+\beta/H_n)^2}
\Theta(k_{\max}-k),
\label{eq:pr}
\end{equation}
where $A_\zeta=0.0038$, $a_\zeta=0.043$, $b_\zeta=1.77$, $H_n$ is the Hubble rate at nucleation, and $\Theta$ is the Heaviside function. The fit has a causal $k^3$ infrared branch and an explicit ultraviolet termination. Its domain of validity and thermal history must match the FOPT benchmark before a numerical comparison is made~\cite{Elor:2023xbz,Greene:2026gnw}. The comoving scale associated with reheating is
\begin{equation}
k_{\max}=1.61\times10^7\,{\rm Mpc}^{-1}
\left(\frac{g_{*,s}(T_{\rm reh})}{68.74}\right)^{-1/3}
\left(\frac{g_{*,\rho}(T_{\rm reh})}{69.76}\right)^{1/2}
\left(\frac{T_{\rm reh}}{\rm GeV}\right).
\label{eq:kmaxFOPT}
\end{equation}
We distinguish $T_{\rm reh}$ from $T_\star$ unless reheating is instantaneous.

\subsection{Domain walls}\label{appn:firstorder_DW}
The quoted wall-induced potential spectrum~\cite{Ferreira:2024eru,Lu:2024dzj},
\begin{equation}
{\cal P}_\phi(k)=\left(\frac{k_B}{k}\right)^8,
\qquad k_B^4\equiv\frac{B^2}{2\pi^2},
\label{eq:mcp}
\end{equation}
is an asymptotic branch, not a physical infrared spectrum extending to $k=0$. Causality and the finite network correlation length impose a cutoff near the comoving Hubble scale at annihilation. We therefore use
\begin{equation}
{\cal P}^{(1)}_\phi(k)={\cal A}_{\phi}
\left(\frac{k_B}{k}\right)^8
\Theta(k-k_{\rm IR})\Theta(k_{\rm UV}-k),
\label{eq:DWcut}
\end{equation}
where $k_{\rm IR}\sim a_{\rm ann}H_{\rm ann}$ and $k_{\rm UV}$ is set by the wall thickness or simulation resolution. The apparent $k^{-8}$ divergence is therefore never extrapolated into the acausal infrared. Wall domination occurs approximately at
\begin{equation}
T_{\rm dom}\simeq282.8\,{\rm GeV}
\left(\frac{{\cal A}}{1.0}\right)^{1/2}
\left(\frac{107}{g_*}\right)^{1/4}
\left(\frac{\sigma^{1/3}}{10^8\,{\rm GeV}}\right)^{3/2},
\label{eq:Tdom}
\end{equation}
and we retain only $T_{\rm ann}>T_{\rm dom}$. The tensor-induced contribution is compared with Eq.~\eqref{eq:DWcut} only within the overlapping physical interval $k_{\rm IR}<k<k_{\rm UV}$.

\section{Tensor-induced comoving density perturbation}\label{sec:TIDP}
We use a spatially flat radiation-dominated FLRW background and retain the first-order transverse-traceless perturbation $\chi_{ij}$,
\begin{equation}
ds^2=a^2(\eta)\left[-d\eta^2+\left(\delta_{ij}+\chi_{ij}+\frac12\delta g_{ij}^{(2)}\right)dx^idx^j\right],
\qquad \partial_i\chi_{ij}=\chi_{ii}=0.
\label{eq:metricSM}
\end{equation}
The collapse variable is the second-order comoving density contrast $\Delta^{(2)}$. Since the first-order tensor is gauge invariant and $\Delta^{(2)}$ is defined on comoving hypersurfaces, the tensor-tensor contribution is gauge invariant at the order retained~\cite{Bari:2022grh}. Primes denote $d/d\eta$, and ${\cal H}=a'/a$.

The tensor obeys
\begin{equation}
\chi_{ij}''+2{\cal H}\chi_{ij}'-\nabla^2\chi_{ij}=8\pi Ga^2\Pi_{ij}^{\rm TT}.
\label{eq:tensor_evolution}
\end{equation}
After source switch-off, $\Pi_{ij}^{\rm TT}=0$. We expand
\begin{equation}
\chi_{ij}(\mathbf x,\eta)=\sum_{\lambda=+,\times}\int\frac{d^3k}{(2\pi)^{3/2}}
 e^{i\mathbf k\cdot\mathbf x}\epsilon_{ij}^{\lambda}(\hat{\mathbf k})
 \chi_\lambda(\mathbf k,\eta),
\label{eq:tensor_modes}
\end{equation}
with $\epsilon_{ij}^{\lambda}\epsilon^{\lambda'ij}=2\delta_{\lambda\lambda'}$. In radiation domination, the free transfer function is $T_h(x)=\sin x/x=\sqrt{\pi/(2x)}J_{1/2}(x)$.

Combining the radiation conservation equations and Einstein constraints gives a sourced third-order equation for the scalar velocity potential~\cite{Bari:2022grh}. Its Green-function solution may be written as
\begin{equation}
\Delta^{(2)}(\mathbf k,x)=-\frac{\pi}{12}\sum_{\lambda\lambda'}
\int\frac{d^3q}{(2\pi)^{3/2}}
\epsilon_{ij}^{\lambda}(\hat{\mathbf q})
\epsilon^{\lambda'ij}(\widehat{\mathbf k-\mathbf q})
(uv)^{-1/2}{\cal I}(u,v,x;x_{\rm ini})
\chi_{\lambda,\rm ini}(\mathbf q)
\chi_{\lambda',\rm ini}(\mathbf k-\mathbf q),
\label{eq:delta}
\end{equation}
where $v=q/k$, $u=|\mathbf k-\mathbf q|/k$, $x=k\eta$, and $x_{\rm ini}=k\eta_{\rm ini}$. Equation~\eqref{eq:delta} corrects the momentum argument of the first tensor mode: the convolution contains $\chi(\mathbf q)$ and $\chi(\mathbf k-\mathbf q)$.

We define
\begin{equation}
\langle\Delta^{(2)}(\mathbf k)\Delta^{(2)}(\mathbf k')\rangle=
\delta^{(3)}(\mathbf k+\mathbf k')\frac{2\pi^2}{k^3}{\cal P}_{\Delta^{(2)}}(k),
\label{eq:PSD}
\end{equation}
and obtain
\begin{equation}
{\cal P}_{\Delta^{(2)}}(k,x)=\frac12\int_0^\infty dv\int_{|1-v|}^{1+v}du\,
\frac{F(u,v)}{(uv)^3}\overline{{\cal I}^2(u,v,x;x_{\rm ini})}
{\cal P}_{\chi,\rm ini}(ku){\cal P}_{\chi,\rm ini}(kv),
\label{eq:PSDf}
\end{equation}
where
\begin{equation}
F(u,v)=\frac{v^8+(u^2-1)^4+4v^6(7u^2-1)+4v^2(u^2-1)^2(7u^2-1)+v^4(70u^4-60u^2+6)}{u^4v^4}.
\label{eq:polarizationF}
\end{equation}
The time kernel is built from
\begin{align}
{\cal I}_a&=\int_{x_{\rm ini}}^x d\bar x\,\bar x^{-1/2}J_{3/2}(\bar x/\sqrt3)J_{3/2}(u\bar x)J_{3/2}(v\bar x),\\
{\cal I}_b&=\int_{x_{\rm ini}}^x d\bar x\,\bar x^{-1/2}Y_{3/2}(\bar x/\sqrt3)J_{3/2}(u\bar x)J_{3/2}(v\bar x),
\label{eq:kernelab}
\end{align}
with $\overline{{\cal I}^2}=\pi^3({\cal I}_a^2+{\cal I}_b^2)/144$ after late-time oscillation averaging. Importantly, $x_{\rm ini}$ is a lower limit of the time integral, not of the $u,v$ convolution. The $x\to\infty$ approximation is accepted only after numerical convergence with finite $x$ is verified. The integrand in \cref{eq:PSDf} is manifestly symmetric in $u \leftrightarrow v$ due to contraction of polarization tensors in \cref{eq:polarizationF} in similar manner the scalar induced gravitational waves case.

\section{Evaluation and validation of the response kernels}\label{sec:kernels}

\subsubsection{Super-Hubble GWs:} For super-Hubble sources of GWs, including inflationary GWs, the lower limit of \cref{eq:kernelab} is taken to be zero. In order to obtain an analytical form of the kernel in such cases, we use known analytical results for the integral of the product of three Bessel functions, expressed in terms of the Legendre functions on the cut and the associated Legendre functions of the second kind. For $|a-b|<c<a+b$, one finds

\begin{align}
	\int_0^{\infty} d\tilde x \tilde x^{1-\beta}
	\left\{	
	\begin{aligned}
		J_\beta(c\tilde x)\\
		Y_\beta(c\tilde x)
	\end{aligned}
	\right\}
	J_{\nu}(a\tilde x)J_{\nu}(b\tilde x)=\frac{1}{\pi}\sqrt{\frac{2}{\pi}}\frac{(ab)^{\beta-1}}{c^\beta}\left(\sin\varphi\right)^{\beta-1/2}\left\{	
	\begin{aligned}
		\frac{\pi}{2}\mathsf{P}^{-\beta+1/2}_{\nu-1/2}(\cos\varphi)\\
		-\mathsf{Q}^{-\beta+1/2}_{\nu-1/2}(\cos\varphi)
	\end{aligned}
	\right\} \label{eq:besslint}
\end{align}
where
\begin{align}
	16\Delta^2\equiv\left(c^2-(a-b)^2\right)\left((a+b)^2-c^2\right)\quad,\quad
	\cos\varphi=\frac{a^2+b^2-c^2}{2ab}\quad,\quad\sin\varphi=\frac{2\Delta}{ab}\,.
\end{align}
and for $c>a+b$
\begin{align}
	\int_0^{\infty} d\tilde x \tilde x^{1-\beta}
	&\left\{	
	\begin{aligned}
		J_\beta(c\tilde x)\\
		Y_\beta(c\tilde x)
	\end{aligned}
	\right\}
	J_{\nu}(a\tilde x)J_{\nu}(b\tilde x)\nonumber\\&
	=\frac{1}{\pi}\sqrt{\frac{2}{\pi}}\frac{(ab)^{\beta-1}}{c^\beta}\left(\sinh\phi\right)^{\beta-1/2}\Gamma[\nu-\beta+1]{\cal Q}^{-\beta+1/2}_{\nu-1/2}(\cosh\phi)\left\{	
	\begin{aligned}
		-\sin\left[(\nu-\beta)\pi\right]\\
		\cos\left[(\nu-\beta)\pi\right]
	\end{aligned}
	\right\} \label{eq:besslint1}
\end{align} 
\begin{align}
	16\tilde\Delta^2\equiv\left(c^2-(a-b)^2\right)\left(c^2-(a+b)^2\right) \quad {,}\quad 
	\cosh\phi=\frac{c^2-(a^2+b^2)}{2ab}\quad,\quad\sinh\phi=\frac{2\tilde\Delta}{ab}\,.
\end{align}
In the above expressions $\mathsf{P}_{\mu}^{\nu}(x)$ and $\mathsf{Q}_{\mu}^{\nu}(x)$ are the Legendre functions on the cut (or Ferrers functions), while $\mathcal{Q}_{\mu}^{\nu}(x)$ is the associated Legendre function of the second kind, whose definitions can be found in the NIST database \cite{NIST:DLMF}. The Legendre functions on the cut are defined for $|x| < 1$ as
\begin{align}
\mathsf{P}^{\mu}_{\nu}\left(x\right)=\left(\frac{1+x}{1-x}\right)^{\mu/2}%
\mathbf{F}\left(\nu+1,-\nu;1-\mu;\tfrac{1}{2}-\tfrac{1}{2}x\right)\,,
\end{align}
\begin{align}
\mathsf{Q}^{\mu}_{\nu}\left(x\right)=\frac{\pi}{2\sin\left(\mu\pi\right)}&\Bigg\{%
\cos\left(\mu\pi\right)\left(\frac{1+x}{1-x}\right)^{\mu/2}\mathbf{F}\left(%
\nu+1,-\nu;1-\mu;\tfrac{1}{2}-\tfrac{1}{2}x\right)\\&-\frac{\Gamma\left(\nu+\mu+1%
\right)}{\Gamma\left(\nu-\mu+1\right)}\left(\frac{1-x}{1+x}\right)^{\mu/2}%
\mathbf{F}\left(\nu+1,-\nu;1+\mu;\tfrac{1}{2}-\tfrac{1}{2}x\right)\Bigg\}\,,
\end{align}
where
\begin{align}
\mathbf{F}\left(a,b;c;x\right)=\frac{1}{\Gamma\left(c\right)}F\left(a,b;c;x\right)
\end{align}
and $F\left(a,b;c;x\right)$ is the Gauss's hypergeometric function.

The associated Legendre polynomials are defined for $|x|>1$ by 
\begin{align}
P^{\mu}_{\nu}\left(x\right)=\left(\frac{x+1}{x-1}\right)^{\mu/2}\mathbf{F}%
\left(\nu+1,-\nu;1-\mu;\tfrac{1}{2}-\tfrac{1}{2}x\right)\,,
\end{align}
and
\begin{align}
Q^{\mu}_{\nu}\left(x\right)=e^{\mu\pi i}\frac{\pi^{1/2}\Gamma\left(\nu+\mu+1%
\right)\left(x^{2}-1\right)^{\mu/2}}{2^{\nu+1}x^{\nu+\mu+1}}\mathbf{F}\left(%
\tfrac{1}{2}\nu+\tfrac{1}{2}\mu+1,\tfrac{1}{2}\nu+\tfrac{1}{2}\mu+\tfrac{1}{2}%
;\nu+\tfrac{3}{2};\frac{1}{x^{2}}\right)\,.
\end{align}
It is more convenient to work with a real-valued version of the associated Legendre function of the second kind, given by
\begin{align}
{\cal Q}_{\nu}^\mu(x)\equiv e^{-\mu\pi i}\frac{Q_{\nu}^\mu(x)}{\Gamma[\mu+\nu+1]}\,.
\end{align}

Identifying $\beta = \nu = 3/2$, $a=v$, $b=u$, and $c= 1 / \sqrt{3}$ from \cref{eq:kernelab,eq:besslint,eq:besslint1}, it is straightforward to obtain the final form of the kernels as follows:
\begin{equation}
    \mathcal{I}_{a}(u,v) = \frac{3^{3/4}}{2} \,\sqrt{\frac{2\,u v}{\pi}}\,\left(1 - y^{2}\right)\,\Theta \left(u + v - \frac{1}{\sqrt{3}}\right)\,, \label{eq:anakernela}
\end{equation}
\begin{equation}
        \mathcal{I}_{b}(u,v) = \frac{3^{3/4}}{2\pi}\, \sqrt{\frac{2\,u v}{\pi}} \, \left[ y +\frac{\left(1 - y^{2}\right)}{2}\, \log\left(\frac{1+y}{1-y}\right)\right] \,,\label{eq:anakernelb}
\end{equation}
where $y = \left(u^2 + v^2 -1/3\right) / \left(2 u v\right)$.

\subsubsection{Sub-Hubble GWs:}  \label{sec:subgws}
Now we turn our attention to the computation of the kernels given in \cref{eq:kernelab} for sub-Hubble GW sources such as PTs, domain walls, and cosmic strings. First, we need to determine the value of $x_{\rm ini}$ at which these GW sources become sub-Hubble. In the case of FOPT, the determination of $x_{\rm ini}$ is directly related to the duration of the phase transition, $\beta/H$. For DWs, the situation is less clear: the value of $x_{\rm ini}$ should be obtained from numerical simulations. In \cite{Notari:2025kqq}, numerical simulations were used to determine $x_{\rm ini} = k_p / (2\pi a H)$. In their analysis, the viable values of $x_{\rm ini}$ lie in the range $[1,4]$. In order to make our study general, we introduce a new variable $x_{\rm ini} \equiv \epsilon_{\rm sim}$. Using their respective values of $x_{\rm ini}$ for FOPT and DWs, we express the product of three Bessel functions in terms of trigonometric functions as:

\begin{equation}
\begin{split}
    &   x^{-1/2} J_{3/2}\left(a x\right) J_{3/2}\left(b x\right) J_{3/2}\left(c x \right) = \frac{1}{\pi^{3/2}\,\sqrt{2 a b c} x^2} \\& \, \Bigg[\cos\left[\left(a-b-c\right)x\right]\,\left\{ -1 - \frac{1}{a b x^{2}} - \frac{1}{a c x^{2}} + \frac{1}{b c x^{2}}\right\} + \cos\left[\left(a+b-c\right)x\right]\,\left\{ -1 + \frac{1}{a b x^{2}} - \frac{1}{a c x^{2}} - \frac{1}{b c x^{2}}\right\} \\& + \cos\left[\left(a-b+c\right)x\right]\,\left\{ -1 - \frac{1}{a b x^{2}} + \frac{1}{a c x^{2}} - \frac{1}{b c x^{2}}\right\} + \cos\left[\left(a+b+c\right)x\right]\,\left\{ -1 + \frac{1}{a b x^{2}} + \frac{1}{a c x^{2}} + \frac{1}{b c x^{2}}\right\} 
    \\& + \frac{\sin\left[\left(a-b-c\right)x\right]}{x}\,\left\{-\frac{1}{a b c \, x^{2}} + \frac{1}{a} - \frac{1}{b} - \frac{1}{c} \right\} + \frac{\sin\left[\left(a+b-c\right)x\right]}{x}\,\left\{\frac{1}{a b c \, x^{2}} + \frac{1}{a} + \frac{1}{b} - \frac{1}{c} \right\} \\& + \frac{\sin\left[\left(a-b+c\right)x\right]}{x}\,\left\{\frac{1}{a b c \, x^{2}} + \frac{1}{a} - \frac{1}{b} + \frac{1}{c} \right\} + \frac{\sin\left[\left(a+b+c\right)x\right]}{x}\,\left\{-\frac{1}{a b c \, x^{2}} + \frac{1}{a} + \frac{1}{b} + \frac{1}{c} \right\}\Bigg] 
\end{split}
\end{equation}
\begin{equation}
\begin{split}
    &   x^{-1/2} Y_{3/2}\left(a x\right) J_{3/2}\left(b x\right) J_{3/2}\left(c x \right) = \frac{1}{\pi^{3/2}\,\sqrt{2 a b c} x^2} \\& \, \Bigg[\frac{\cos\left[\left(a-b-c\right)x\right]}{x}\,\left\{ \frac{1}{a b c \, x^{2}} - \frac{1}{a} + \frac{1}{b} + \frac{1}{c}\right\} + \frac{\cos\left[\left(a+b-c\right)x\right]}{x}\,\left\{-\frac{1}{a b c \, x^{2}} - \frac{1}{a} - \frac{1}{b} + \frac{1}{c}\right\} \\& + \frac{\cos\left[\left(a-b+c\right)x\right]}{x}\,\left\{-\frac{1}{a b c \, x^{2}} - \frac{1}{a} + \frac{1}{b} - \frac{1}{c}\right\} + \frac{\cos\left[\left(a+b+c\right)x\right]}{x}\,\left\{ \frac{1}{a b c \, x^{2}} - \frac{1}{a} - \frac{1}{b} - \frac{1}{c}\right\} 
    \\& + \sin\left[\left(a-b-c\right)x\right]\,\left\{-1 -\frac{1}{a b \, x^{2}} - \frac{1}{ac\,x^2} + \frac{1}{b c \, x^2} \right\} + \sin\left[\left(a+b-c\right)x\right]\,\left\{-1 +\frac{1}{a b \, x^{2}} - \frac{1}{ac\,x^2} - \frac{1}{b c \, x^2} \right\} \\& + \sin\left[\left(a-b+c\right)x\right]\,\left\{-1 -\frac{1}{a b \, x^{2}} + \frac{1}{ac\,x^2} - \frac{1}{b c \, x^2}\right\} + \sin\left[\left(a+b+c\right)x\right]\,\left\{-1 +\frac{1}{a b \, x^{2}} + \frac{1}{ac\,x^2} + \frac{1}{b c \, x^2} \right\}\Bigg] 
\end{split}
\end{equation}
In order to obtain the closed form of the kernel integrals in \cref{eq:kernelab}, we make use of the following:
\begin{eqnarray}
    \int_{x_{\rm ini}}^{\infty}\, \frac{\cos\left(a x\right)}{x^2} &=& - \frac{a \, \pi}{2} + \frac{\cos\left(a\, x_{\rm ini}\right)}{x_{\rm ini}} + a \, \text{Si} \left(a \, x_{\rm ini}\right) \\ 
     \int_{x_{\rm ini}}^{\infty}\, \frac{\cos\left(a x\right)}{x^3} &=& \frac{\cos\left(a\,x_{\rm ini}\right)}{2\,x^{2}_{\rm ini}} + \frac{1}{2} \, a^2 \, \text{Ci} \left(a\,x_{\rm ini}\right) - \frac{a\, \sin \left(a\,x_{\rm ini}\right)}{2\,x_{\rm ini}} \\
     \int_{x_{\rm ini}}^{\infty}\, \frac{\cos\left(a x\right)}{x^4} &=& \frac{a^3 \, \pi}{12} + \frac{\cos\left(a\,x_{\rm ini}\right)}{3\,x^{3}_{\rm ini}} - \frac{a^2 \cos\left(a\,x_{\rm ini}\right)}{6\,x_{\rm ini}} - \frac{a \sin \left(a x_{\rm ini}\right)}{6\,x_{\rm ini}^2} - \frac{1}{6}\,a^3\, \text{Si}\left(a x_{\rm ini}\right) \\
     \int_{x_{\rm ini}}^{\infty}\, \frac{\cos\left(a x\right)}{x^5} &=& \frac{\cos\left(a\,x_{\rm ini}\right)}{4\,x^{4}_{\rm ini}} - \frac{a^2 \cos\left(a\,x_{\rm ini}\right)}{24\,x_{\rm ini}^2} - \frac{a \sin \left(a x_{\rm ini}\right)}{12\,x_{\rm ini}^3} + \frac{a^3 \sin \left(a x_{\rm ini}\right)}{24\,x_{\rm ini}}- \frac{1}{24}\,a^4\, \text{Ci}\left(a x_{\rm ini}\right) \nonumber \\
    \int_{x_{\rm ini}}^{\infty}\, \frac{\sin\left(a x\right)}{x^2} &=&  \frac{\sin\left(a\, x_{\rm ini}\right)}{x_{\rm ini}} - a \, \text{Ci} \left(a \, x_{\rm ini}\right) \\ 
     \int_{x_{\rm ini}}^{\infty}\, \frac{\sin\left(a x\right)}{x^3} &=&  - \frac{a^2 \, \pi}{4} + \frac{a \cos\left(a x_{\rm ini}\right)}{2 x_{\rm ini}} + \frac{\sin\left(a x_{\rm ini}\right)}{2\,x_{\rm ini}^2} + \frac{1}{2} \, a^2 \, \text{Si} \left(a x_{\rm ini}\right) \\
     \int_{x_{\rm ini}}^{\infty}\, \frac{\sin\left(a x\right)}{x^4} &=&  \frac{a \cos\left(a x_{\rm ini}\right)}{6 \, x_{\rm ini}^2} + \frac{\sin \left(a x_{\rm ini}\right)}{3 x_{\rm ini}^3} - \frac{a^2\, \sin \left(a x_{\rm ini}\right)}{6 x_{\rm ini}} + \frac{a^3}{6} \, \text{Ci}\left(a x_{\rm ini}\right)\\
     \int_{x_{\rm ini}}^{\infty}\, \frac{\sin\left(a x\right)}{x^5} &=&  \frac{a^4 \, \pi}{48} + \frac{a\, \cos\left(a x_{\rm ini}\right)}{12 \, x_{\rm ini}^3} - \frac{a^3\, \cos\left(a x_{\rm ini}\right)}{24\,x_{\rm ini}} + \frac{\sin\left(a x_{\rm ini}\right)}{4 x_{\rm ini}^4} - \frac{a^2\, \sin\left(a x_{\rm ini}\right)}{24\,x_{\rm ini}^2} - \frac{1}{24}\,a^4\,\text{Si} \left(a x_{\rm ini} \right) \nonumber \\
\end{eqnarray}
where $\text{Si}\left(x_{\rm ini}\right)$ and $\text{Ci}\left(x_{\rm ini}\right)$ are the sine and cosine integrals, defined as:
\begin{eqnarray}
    \text{Si}\left(x_{\rm ini}\right) &=& \int_{0}^{x_{\rm ini}} \frac{\sin(x)}{x}\, dx \\ 
    \text{Ci}\left(x_{\rm ini}\right) &=& -\int_{x_{\rm ini}}^{\infty} \frac{\cos(x)}{x}\, dx
\end{eqnarray}
It is possible to recover the \cref{eq:anakernela,eq:anakernelb} by taking the $x_{\rm ini} \rightarrow 0$. Note that the integrand in \cref{eq:kernelab} are finite at $\bar{x}$. To see this, consider the following:
\begin{equation}
\mathcal{I}_a(x_{\mathrm{ini}})=\mathcal{I}_\infty(u,v)-
\int_{0}^{x_{\mathrm{ini}}} d\bar x\;
\bar x^{-1/2} J_{3/2}\!\left(\frac{\bar x}{\sqrt3}\right)
J_{3/2}(u\bar x) J_{3/2}(v\bar x).
\end{equation}
For small arguments,
\[
J_{3/2}(z)=\sqrt{\frac{2}{\pi}}z^{3/2}
\left(\frac{1}{3}-\frac{z^{2}}{30}+O(z^{4})\right).
\]
Set \(a=1/\sqrt3\), \(b=u\), \(c=v\).  Then as \(\bar x\to0\),
\[
J_{3/2}(a\bar x)J_{3/2}(b\bar x)J_{3/2}(c\bar x)
= \Bigl(\frac{2}{\pi}\Bigr)^{3/2}(abc)^{3/2}\bar x^{9/2}
\left[\frac{1}{27}+O(\bar x^{2})\right].
\]
Multiplying by the remaining factor \(\bar x^{-1/2}\) gives the full integrand
\[
f(\bar x)\equiv\bar x^{-1/2} J_{3/2}(a\bar x)J_{3/2}(b\bar x)J_{3/2}(c\bar x)
\sim C\,\bar x^{4}+O(\bar x^{6}),
\]
with \(C=\dfrac{2^{3/2}}{27\,\pi^{3/2}}(abc)^{3/2}\).  
Thus the integral over \([0,x_{\mathrm{ini}}]\) satisfies
\[
\int_{0}^{x_{\mathrm{ini}}} f(\bar x)\,d\bar x
= \frac{C}{5}x_{\mathrm{ini}}^{5}+O(x_{\mathrm{ini}}^{7}).
\]
Consequently,
\[
\lim_{x_{\mathrm{ini}}\to0}
\int_{0}^{x_{\mathrm{ini}}} f(\bar x)\,d\bar x = 0,
\]
and therefore
\[
\lim_{x_{\mathrm{ini}}\to0}\mathcal{I}_a(x_{\mathrm{ini}})
= \mathcal{I}_\infty(u,v)
= \frac{3^{3/4}}{2}\sqrt{\frac{2uv}{\pi}}\,(1-y^{2})\,
\Theta\!\left(u+v-\frac1{\sqrt3}\right).
\]
\paragraph{Validation.}
The numerical implementation is validated through the following checks:
\begin{enumerate}  
\item the expected symmetry under interchange of the two tensor momenta is recovered after accounting for the polarization contraction in \cref{eq:polarizationF};  
\item the super‑Hubble limit \eqref{eq:anakernela}-\eqref{eq:anakernelb} is exactly reproduced as \(x_{\rm ini}\to0\);
\item the result is stable under increasing the upper integration limit \(x_{\max}\) and tightening the quadrature tolerances;  
\item direct integration of the Bessel functions agrees with the half‑integer trigonometric representation;  
\item the outer \(u,v\) convolution is convergent after varying the ultraviolet truncation imposed by the physical source spectrum.  
\end{enumerate}

\section{Tensor-to-scalar transfer and numerical spectra}\label{sec:results}

The energy density of freely propagating tensor modes after source switch-off is related to the tensor power spectrum by~\cite{Greene:2026one}
\begin{equation}
    \Omega_{\rm GW}(\tau,k)=\frac{1}{3}\left(\frac{k}{aH}\right)^2 {\cal P}_{\chi}(\tau,k),
    \label{eq:Omega_Pchi}
\end{equation}
where $k=2\pi f$ in units with $c=1$. The spectrum ${\cal P}_{\chi}(\tau,k)$ is the time-evolved tensor spectrum. After the active source has switched off we write
\begin{equation}
    {\cal P}_{\chi}(\tau,k)={\cal T}^{2}(\tau,k){\cal P}_{\chi_{\rm ini}}(k),
\end{equation}
with ${\cal T}$ obtained from the sourceless tensor equation, Eq.~\eqref{eq:tensor_evolution}. For modes inside the horizon during radiation domination, this gives the matching estimate
\begin{equation}
    \Omega_{\rm GW}(x_{\rm ini},k)\simeq \frac{1}{3}{\cal P}_{\chi_{\rm ini}}(k).
    \label{apneq:GW_PT}
\end{equation}
Equation~\eqref{apneq:GW_PT} is used as a free-propagation matching approximation. If the unequal-time correlator of a long-lived incoherent source changes the matched tensor amplitude by a factor $A_{\rm match}$, then ${\cal P}_{\Delta^{(2)}}$ scales approximately as $A_{\rm match}^2$, while the PBH abundance changes exponentially through the variance. We therefore treat this normalization as a systematic uncertainty rather than as a universal source-independent constant.

The numerical pipeline used in the Letter is: construct $\Omega_{\rm GW}$ for the chosen source prescription; match to ${\cal P}_{\chi,\rm ini}$ with Eq.~\eqref{apneq:GW_PT}; evaluate the finite-$x_{\rm ini}$ kernel; perform the $u,v$ convolution in Eq.~\eqref{eq:PSDf}; and locate $k_p^{\rm TIS}$ directly from the induced spectrum. 

\begin{figure}[t]
    \centering
    \includegraphics[width=0.47\linewidth]{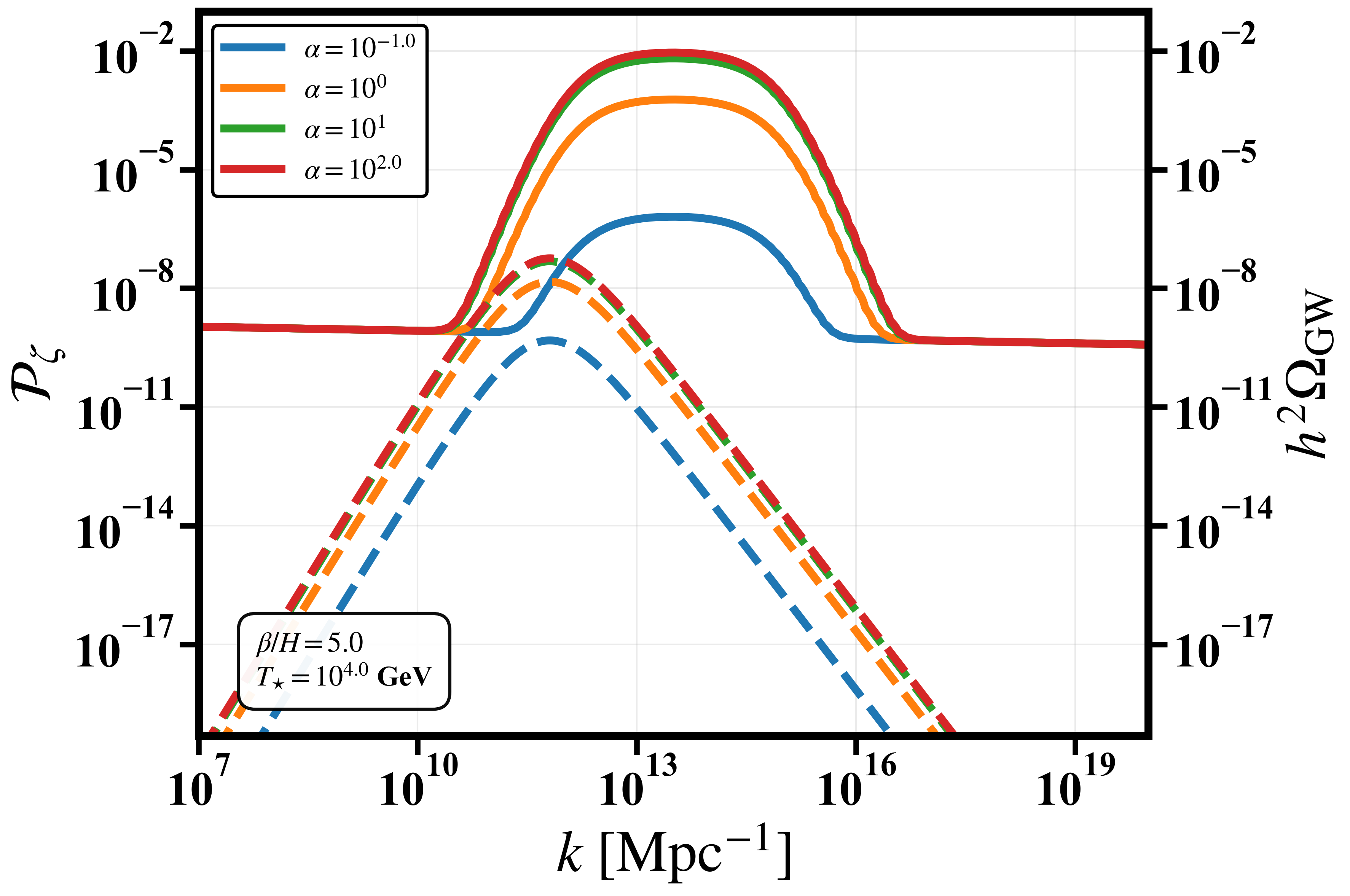}
    \includegraphics[width=0.47\linewidth]{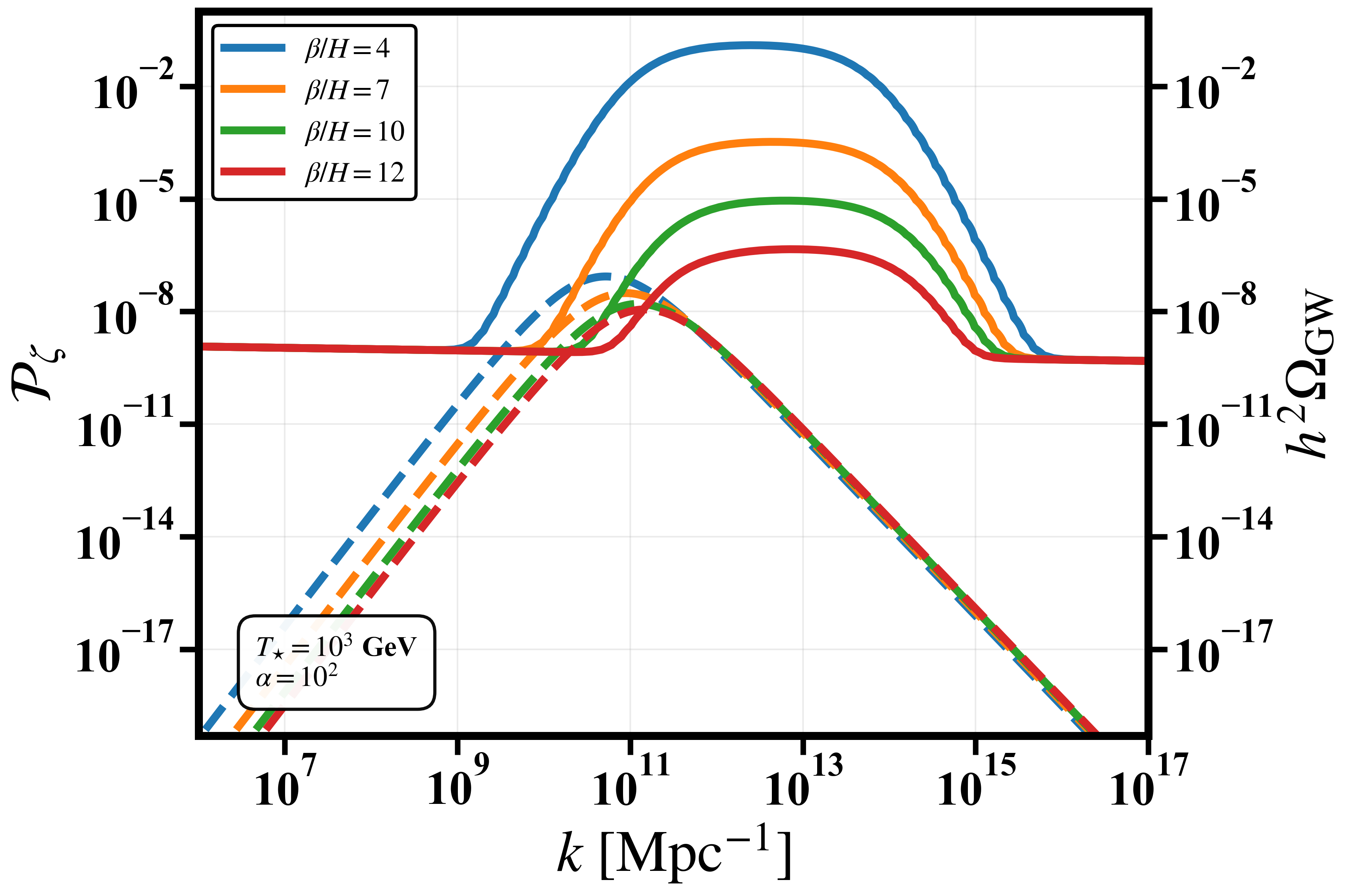}
    \includegraphics[width=0.47\linewidth]{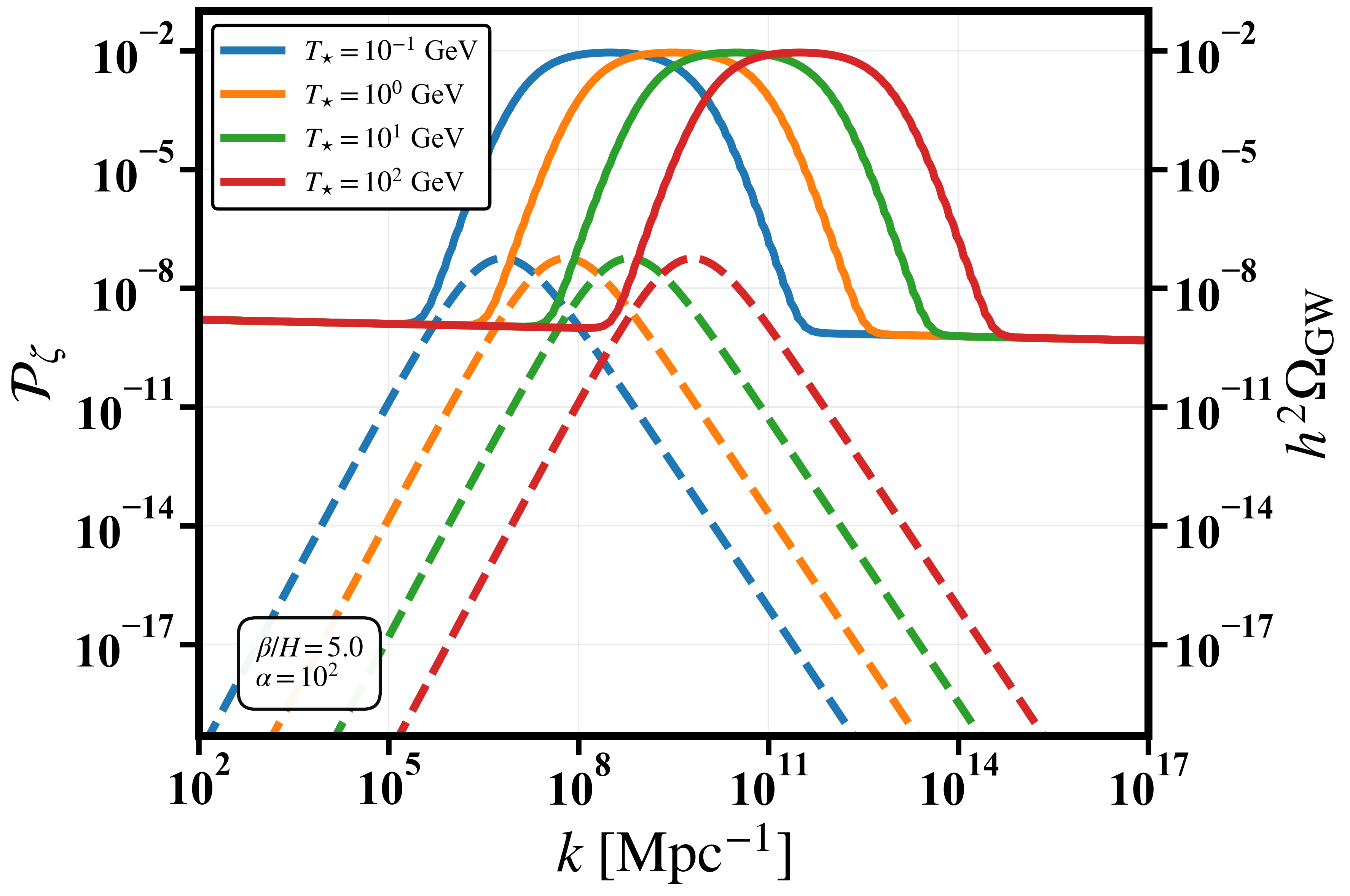}
    \caption{\it Gravitational-wave energy density spectrum $\Omega_{\rm GW}$ from FOPT (dashed) and the corresponding scalar power spectrum including the tensor-induced contribution, as functions of wavenumber $k$ (with $k=2\pi f$). The panels show the dependence on $\alpha$, $\beta/H_\star$, and $T_\star$ for the benchmark spectral shape $a=b=2.4$, $c=4$.}
    \label{fig:PS_PT}
\end{figure}

\begin{figure}[t]
    \centering
    \includegraphics[width=0.47\linewidth]{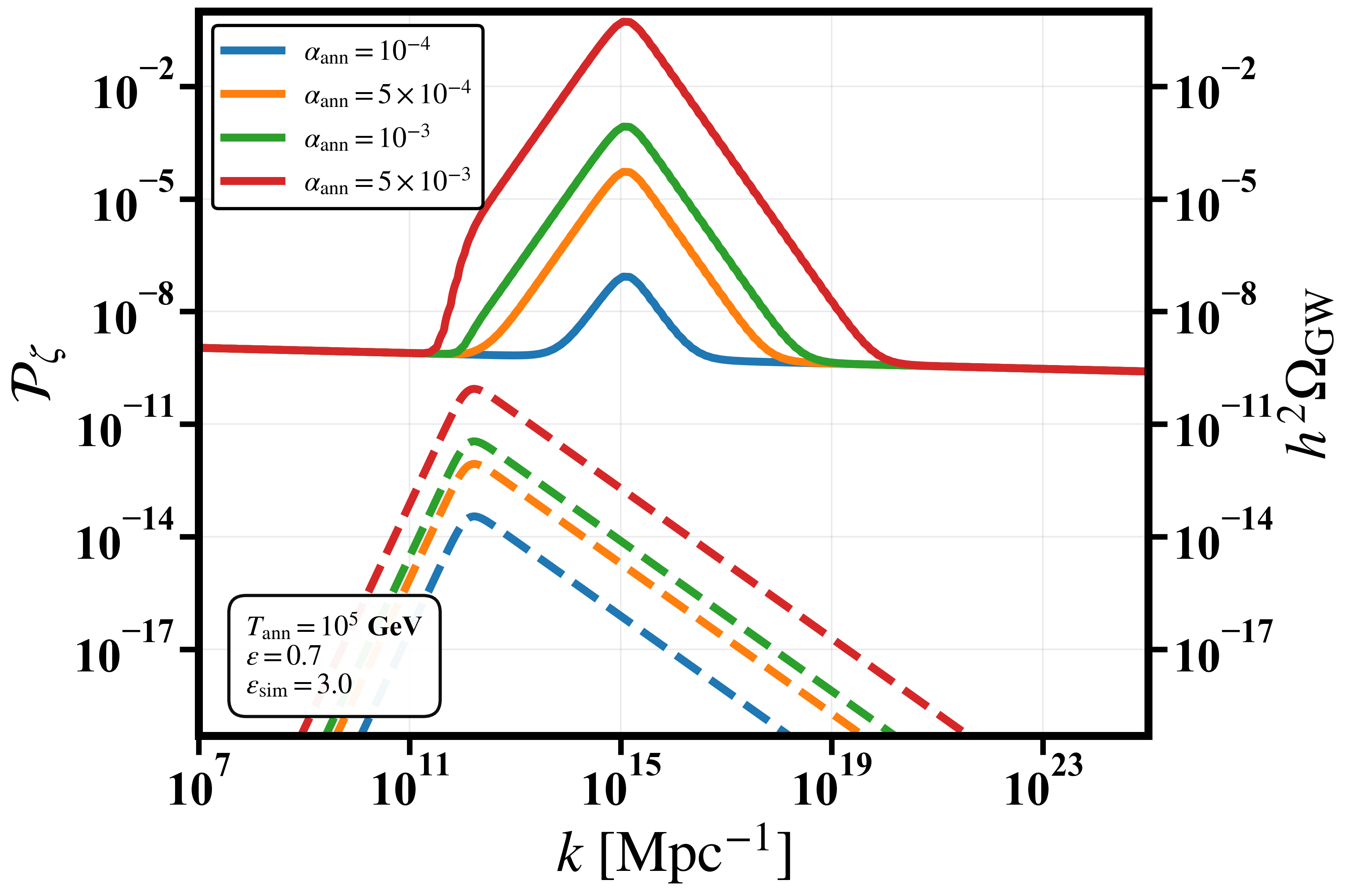}
    \includegraphics[width=0.47\linewidth]{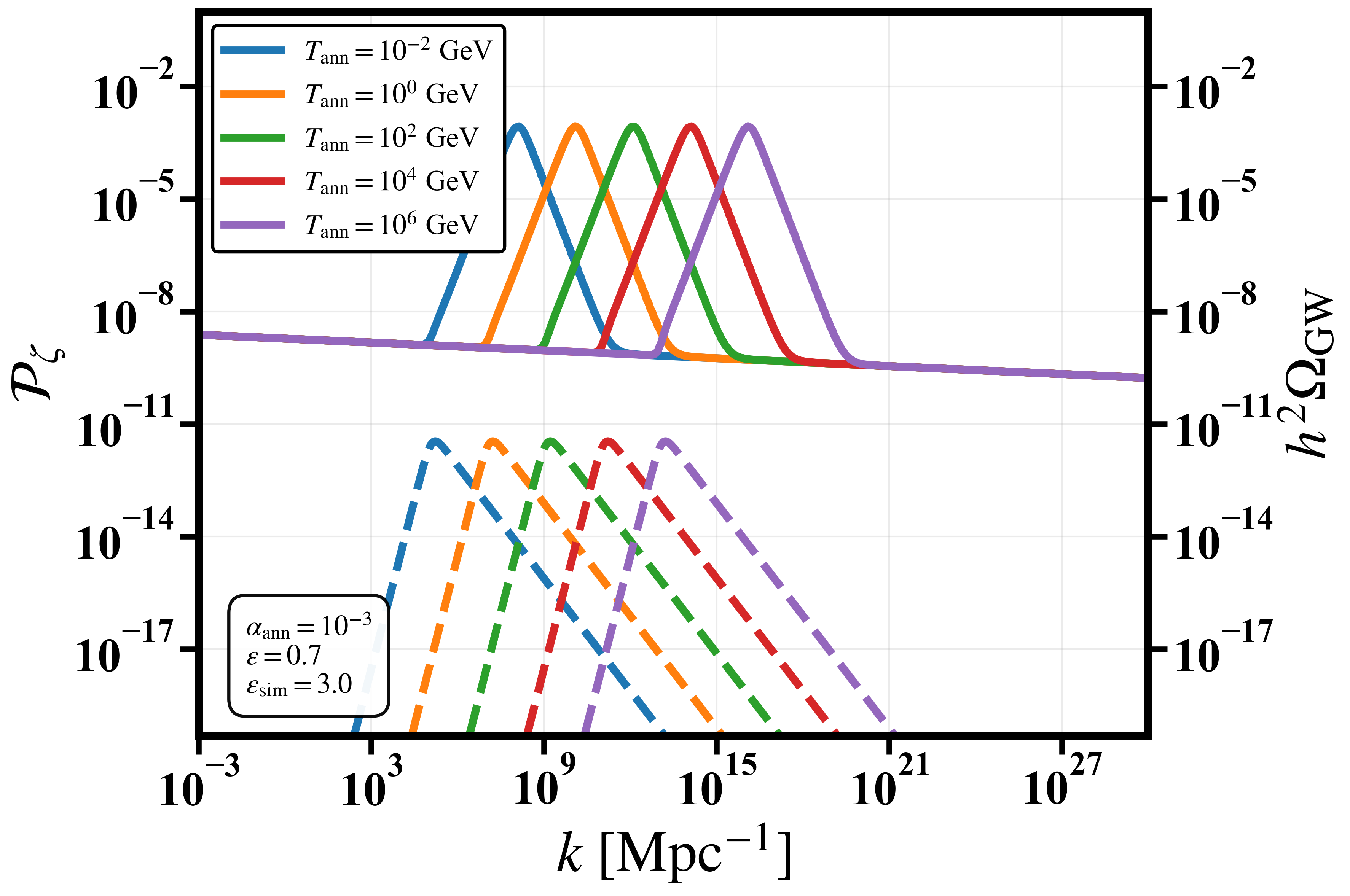}
    \caption{\it Gravitational-wave energy density spectrum $\Omega_{\rm GW}$ from DW annihilation (dashed) and the corresponding scalar power spectrum including the tensor-induced contribution. The panels show the dependence on $\alpha_{\rm ann}$ and $T_{\rm ann}$, with $T_{\rm ann}$ fixed by the bias-to-tension ratio through Eq.~\eqref{SM:Tann}.}
    \label{fig:PS_DW}
\end{figure}

Figure~\ref{fig:PS_PT} shows the expected parameter dependence for FOPTs. Increasing $\alpha$ raises the GW amplitude and therefore the induced scalar power until the spectrum saturates for $\alpha\gg1$, as implied by Eq.~\eqref{SM:OmegaPT}. Increasing $T_\star$ shifts both the GW peak and the tensor-induced scalar peak to larger wavenumber. Decreasing $\beta/H_\star$ enhances the amplitude and shifts the peak to lower frequency for fixed $T_\star$. The induced scalar peak is found numerically at $k_p^{\rm TIS}\simeq3.88\times10^2 k_p^{\rm GW}$ for the benchmark choices used in Fig.~\ref{fig:PS_PT}; this number is not universal and changes with the matching time and source scales. The local branch scaling is
\begin{align}
 {\cal P}_{\delta^{(2)}}^{\rm FOPT}(k) \propto
\left(\frac{\kappa_i\alpha}{1+\alpha}\right)^4
\left(\frac{H_\star}{\beta}\right)^{2p_i}
\begin{cases}
\displaystyle \left(\frac{k}{k_{p}}\right)^{2a}, & k\ll k_{p}, \\[2ex]
\displaystyle \left(\frac{k}{k_{p}}\right)^{a-b}, & k_{p}\leq k\leq k_{p}^{\rm TIS}, \\[2ex]
\displaystyle \left(\frac{k}{k_{p}}\right)^{-2b}, & k\gg k_{p}^{\rm TIS}.
\end{cases}
\label{apneq:ind_PT_ana}
\end{align}
Figure~\ref{fig:PS_DW} shows the analogous behavior for DW annihilation. Increasing $\alpha_{\rm ann}$ increases the GW amplitude and hence the tensor-induced density spectrum. Increasing $T_{\rm ann}$ moves the spectral features to larger wavenumber. For the displayed DW benchmark the induced peak occurs at $k_p^{\rm TIS}\simeq10^3 k_p^{\rm GW}$, again as a benchmark output rather than a universal resonance factor. The corresponding local branch scaling is
\begin{align}
 {\cal P}_{\delta^{(2)}}^{\rm DW}(k) \propto
\frac{9\epsilon_{\rm GW}^2\alpha_{\rm ann}^4}{64\pi^2}
\begin{cases}
\displaystyle \left(\frac{k}{k_{p}}\right)^6, & k\ll k_{p}, \\[2ex]
\displaystyle \left(\frac{k}{k_{p}}\right)^2, & k_{p}\leq k\leq k_{p}^{\rm TIS}, \\[2ex]
\displaystyle \left(\frac{k}{k_{p}}\right)^{-2}, & k\gg k_{p}^{\rm TIS}.
\end{cases}
\label{eq:ind_DW_ana}
\end{align}

\subsection{Spectral-distortion consistency}
CMB spectral distortions constrain an integral of scalar acoustic power over thermalization windows, not a pointwise density-spectrum amplitude. We convolve the scalar spectrum with the $\mu$- and $y$-distortion windows~\cite{Chluba:2012we}. COBE/FIRAS gives $|\mu|<9.0\times10^{-5}$ and $|y|<1.5\times10^{-5}$~\cite{Fixsen:1996nj}. Planck and Lyman-$\alpha$ limits are shown separately and are not called spectral-distortion bounds.
\begin{figure}[t]
    \centering
    \includegraphics[width=0.47\linewidth]{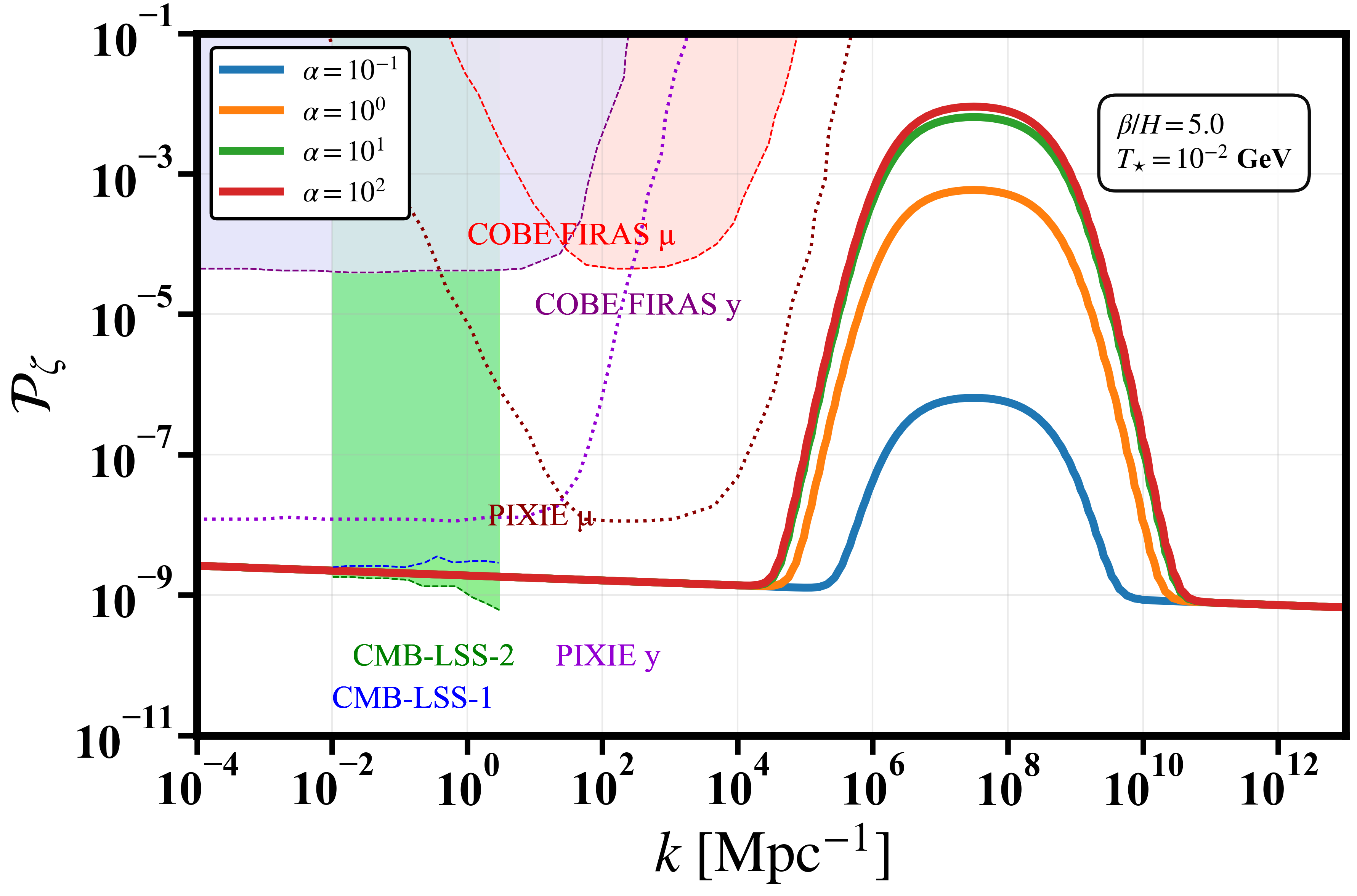}
    \includegraphics[width=0.47\linewidth]{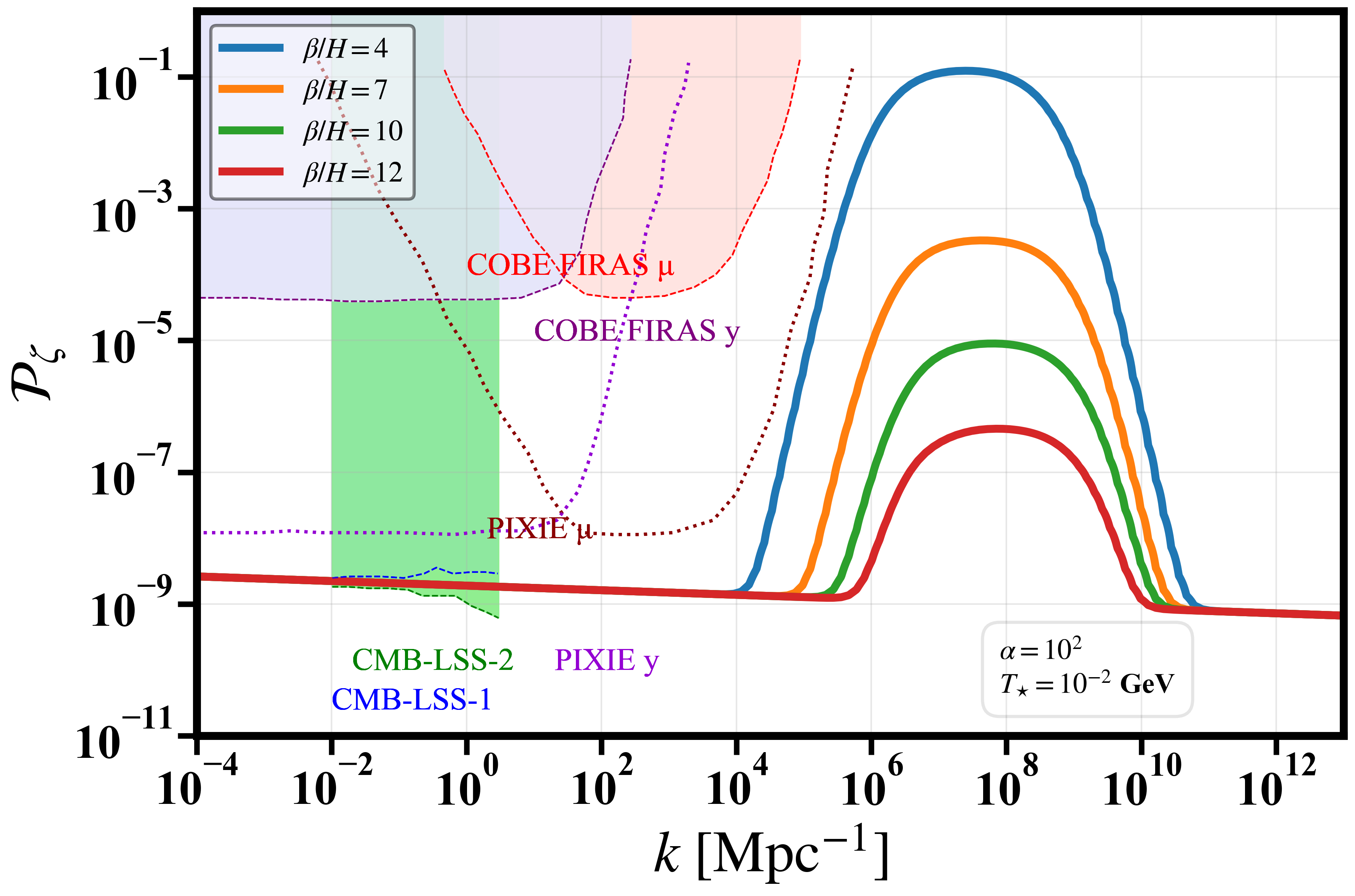}
    \includegraphics[width=0.50\linewidth]{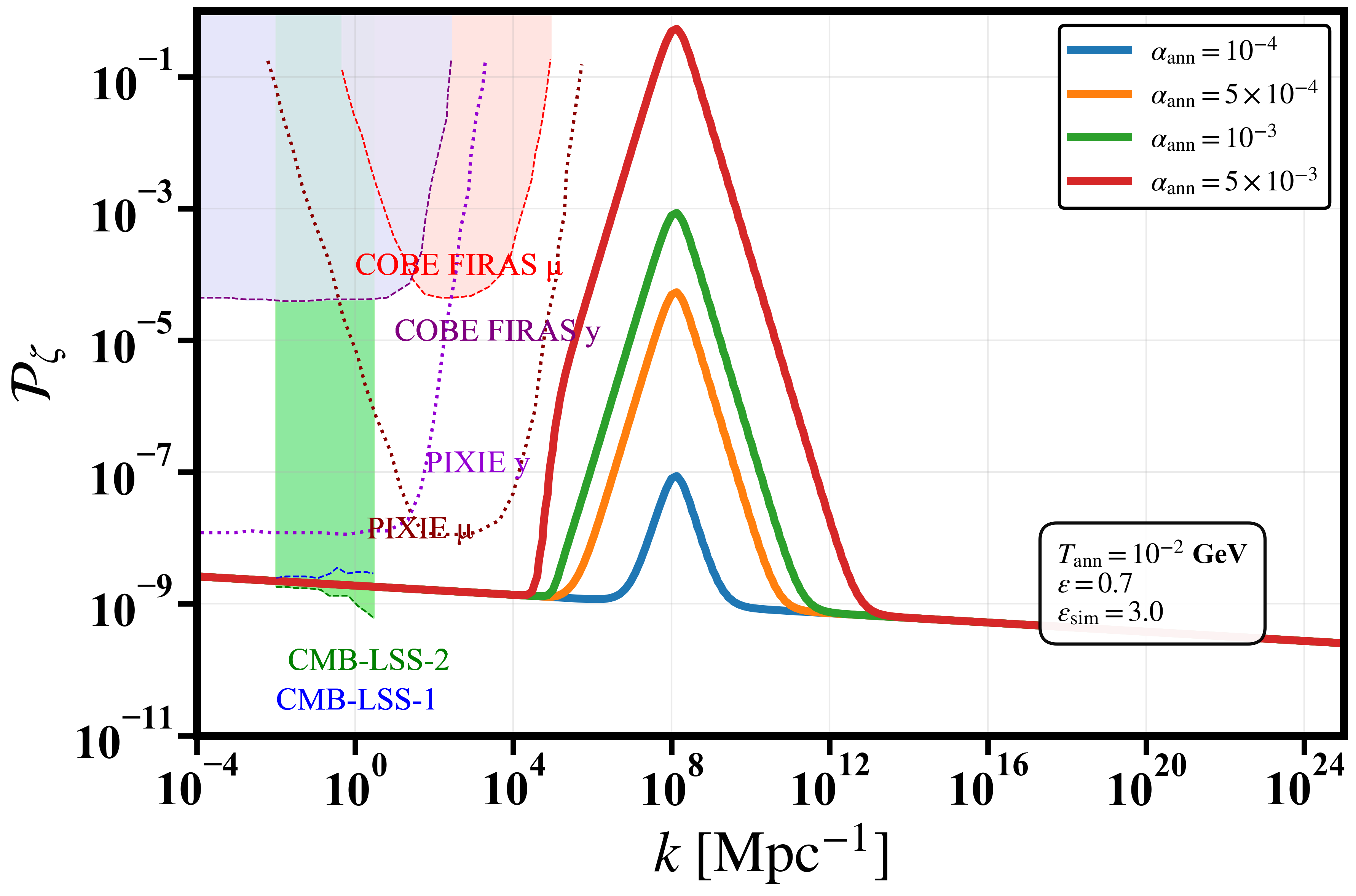}
    \caption{\it Induced scalar power spectrum for representative choices of the phase-transition parameters $T_{\star}$, $\beta/H$, and $\alpha$, and the domain-wall parameters $\alpha_{\rm ann}$ and $\epsilon$. The shaded areas represent constraints from current (solid line) and future (dashed line) experiments.}
    \label{fig:mu_constraints}
\end{figure}
\cref{fig:mu_constraints} displays the induced density spectrum for some illustrative choices of FOPT and DW parameters. The illustrative parameter choices shown in the figure are selected to lie outside the displayed exclusion regions. The figure is intended as a consistency check and as a visualization of how the induced spectrum moves when the macroscopic source parameters are varied. For comparison, and to check consistency with current observational constraints, we also include current and future bounds from various observational and experimental sources, such as Planck \cite{Planck:2018jri}, Lyman-$\alpha$ forest data \cite{Bird:2010mp}, PIXIE \cite{A_Kogut_2011}, COBE/FIRAS \cite{Fixsen:1996nj}, and Super-PIXIE \cite{Chluba:2019kpb}.
At scales $10^{-4} \lesssim k/\text{Mpc}^{-1} \lesssim 1$, the power spectrum is constrained by the angular resolution of current CMB measurements. Inhomogeneities at these scales produce deviations from the blackbody spectrum of the CMB, known as CMB spectral distortions \cite{Chluba:2012we}. These distortions are typically classified as $\mu$-distortions, associated with a non-zero chemical potential occurring at early times, and Compton $y$-distortions, which arise at redshifts $z \lesssim 5 \times 10^4$. We associate a $\mu$-distortion with a Bose-Einstein distribution with $\mu \neq 0$. The most stringent present constraints on spectral distortions are obtained from the COBE/FIRAS experiment, which yields $\lvert \mu \rvert \lesssim 9.0 \times 10^{-5}$ and $\lvert y \rvert \lesssim 1.5 \times 10^{-5}$ at 95\% confidence level \cite{Fixsen:1996nj}, plotted as shaded regions in \cref{fig:mu_constraints}.
The solid lines represent current experiments, while the dashed lines represent future experiments. Future detectors such as PIXIE are expected to probe distortions down to sensitivities of $\mu \lesssim 2 \times 10^{-8}$ and $y \lesssim 4 \times 10^{-9}$ \cite{A_Kogut_2011}. Throughout the rest of the analysis, we always choose both the FOPT and DW parameters such that these bounds are respected.

\section{Origin of the spectral enhancement}\label{enhance}
The acoustic kernel is enhanced near $c_s(u+v)=1$, i.e. $u+v=\sqrt3$, while the source spectra weight configurations in which $ku$ and $kv$ lie near the characteristic source features. The resonance condition is therefore an order-unity kinematic statement in the convolution variables; the much larger benchmark ratios $k_p^{\rm TIS}/k_p^{\rm GW}\simeq3.88\times10^2$ for FOPTs and $\simeq10^3$ for DWs also encode the finite matching time and the hierarchy between the Hubble scale and the microscopic source scale.

For a local tensor branch ${\cal P}_\chi\propto k^n$, the induced density spectrum scales as ${\cal P}_{\Delta^{(2)}}\propto k^{2n}$ only when the dimensionless convolution integral is dominated by the same branch and is regulated by the physical source cutoffs. A useful schematic quantity is
\begin{equation}
{\cal Q}_{n_1n_2}(x_{\rm ini})=\frac12\int_0^\infty dv\int_{|1-v|}^{1+v}du\,F(u,v)\overline{{\cal I}^2}(uv)^{-3}u^{n_1}v^{n_2},
\label{eq:Qgeneral}
\end{equation}
with the physical ultraviolet and infrared cutoffs of the source spectrum understood. Near resonance the formal logarithmic feature of the kernel is regulated by finite duration, finite observation time, and finite spectral width. Tensor-induced dominance is established by the amplitude ratio ${\cal R}(k)={\cal P}_{\Delta^{(2)}}/{\cal P}_{\Delta^{(1)}}$ across the scales contributing to $\sigma_R$, rather than by comparing slopes alone.

The scaling laws quoted in the Letter follow from applying this local-branch argument to the source spectra. For the FOPT infrared branch,
\begin{equation}
    {\cal P}^{\rm FOPT}_{\delta^{(2)},\rm IR}(k)
    \simeq \left(\frac{\alpha}{1+\alpha}\right)^4\left(\frac{\beta}{H}\right)^{-4}
    \left(\frac{k}{k_p}\right)^{2a}{\cal Q}^{\rm FOPT}_{\rm IR},
    \qquad
    {\cal Q}^{\rm FOPT}_{\rm IR}\equiv \frac12\int_0^\infty dv\int_{|1-v|}^{1+v}du\,F(u,v)\overline{{\cal I}^2}(uv)^{-3}u^a v^a .
    \label{eq:QIRFOPT}
\end{equation}
For the DW infrared branch,
\begin{equation}
    {\cal P}^{\rm DW}_{\delta^{(2)},\rm IR}(k)
    \simeq \frac{9\epsilon_{\rm GW}^2\alpha_{\rm ann}^4}{64\pi^2}
    \left(\frac{k}{k_p}\right)^6 {\cal Q}^{\rm DW}_{\rm IR},
    \qquad
    {\cal Q}^{\rm DW}_{\rm IR}\equiv \frac12\int_0^\infty dv\int_{|1-v|}^{1+v}du\,F(u,v)\overline{{\cal I}^2}.
    \label{eq:QIRDW}
\end{equation}
Here ``IR'' denotes the source branch below the GW peak, while the ultimate far-infrared behavior is fixed by causality. The constants ${\cal Q}_{\rm IR}$ and ${\cal Q}_{\rm int}$ absorb the finite integration range, source cutoffs, and $u,v$ weights.
For the intermediate branches one obtains
\begin{equation}
    {\cal P}^{\rm FOPT}_{\delta^{(2)},\rm int}(k)
    \simeq \left(\frac{\alpha}{1+\alpha}\right)^4\left(\frac{\beta}{H}\right)^{-4}
    \left(\frac{k}{k_p}\right)^{a-b}{\cal Q}^{\rm FOPT}_{\rm int},
    \label{eq:QintFOPT}
\end{equation}
\begin{equation}
    {\cal P}^{\rm DW}_{\delta^{(2)},\rm int}(k)
    \simeq \frac{9\epsilon_{\rm GW}^2\alpha_{\rm ann}^4}{64\pi^2}
    \left(\frac{k}{k_p}\right)^2 {\cal Q}^{\rm DW}_{\rm int}.
    \label{eq:Qint_DW}
\end{equation}
The UV branches are obtained analogously and are suppressed by the power-law falloff of the source spectra.

\begin{figure}[t]
    \centering
    \includegraphics[width=0.48\linewidth]{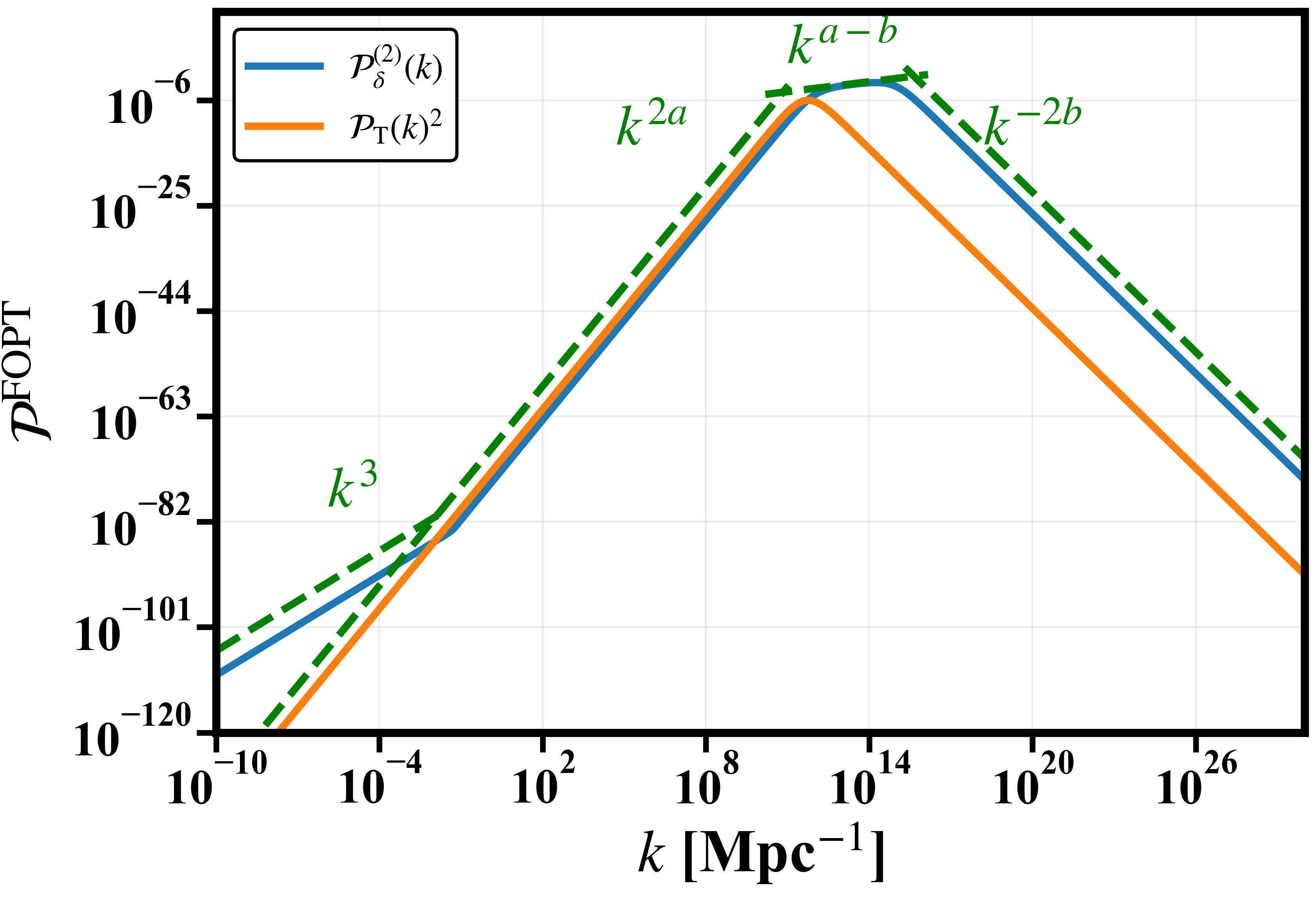}
    \includegraphics[width=0.48\linewidth]{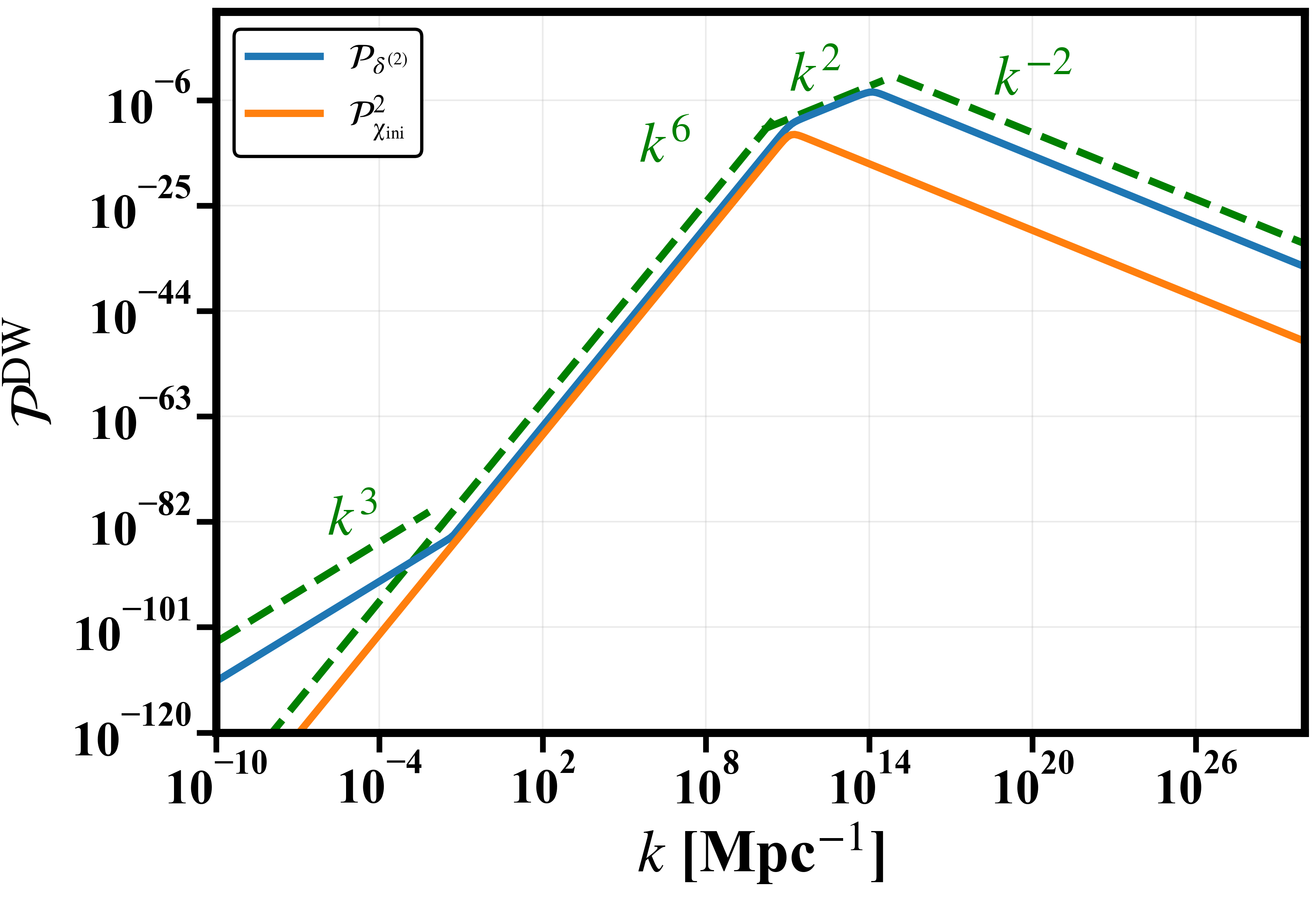}
    \caption{\it Comparison of the numerical tensor-induced density spectra with the local branch approximations for the FOPT (left) and DW (right) benchmark points. The solid orange curves show ${\cal P}_{\chi,\rm ini}^2$, while the green dashed curves show the analytic branch scalings.}
    \label{fig:scale}
\end{figure}

Figure~\ref{fig:scale} compares the numerical convolution with the branch scalings above. The induced spectrum follows the squared tensor spectrum only locally, when the same source branch dominates the integral and the dimensionless integral in Eq.~\eqref{eq:Qgeneral} is convergent with the imposed physical cutoffs. At wavenumbers well below all source correlation scales, the full induced spectrum returns to the causal $k^3$ behavior. The comparison with the first-order scalar spectra in Secs.~\ref{appn:firstorder_PT} and~\ref{appn:firstorder_DW} is made at equal wavenumber and with the same power-spectrum convention; tensor-induced dominance is therefore an amplitude statement over the scales contributing to $\sigma_R$, not a conclusion drawn from slopes alone.

\section{PBH formation from tensor-induced density fluctuations}\label{sec:PBH}

We define the present-day differential fraction
\begin{equation}
f_{\rm PBH}(M)=\frac{1}{\Omega_{\rm CDM}}\frac{d\Omega_{\rm PBH}}{d\ln M}.
\label{eq:fpbh-1}
\end{equation}
The collapse variable is the smoothed comoving density contrast at horizon entry. With a real-space top-hat,
\begin{equation}
W(x)=3\frac{\sin x-x\cos x}{x^3},
\label{eq:W}
\end{equation}
we calculate the variance directly from the induced density spectrum,
\begin{equation}
\sigma_R^2(\eta_R)=\int d\ln q\,W^2(qR){\cal P}_{\Delta^{(2)}}(q,\eta_R),
\qquad aH|_{\eta_R}=R^{-1}.
\label{apneq:variance}
\end{equation}
This avoids applying a linear $\Delta$-$\zeta$ conversion to an intrinsically second-order field. We adopt $\Delta_c=0.51$, ${\cal K}=4.36$, and $\gamma_c=0.38$ in
\begin{equation}
M={\cal K}M_H(\Delta-\Delta_c)^{\gamma_c}.
\label{apneq:MPBH}
\end{equation}
The threshold is profile and window dependent; varying $\Delta_c$ and the smoothing prescription provides a leading systematic uncertainty~\cite{Musco:2020jjb,Niemeyer:1997mt,Musco:2018rwt}. Unless stated otherwise, all quoted benchmark abundances use $\Delta_c=0.51$ and the real-space top-hat; changing these choices should be interpreted as a systematic displacement of the abundance contours rather than a change in the tensor-induced mechanism itself.

Even for Gaussian tensors, $\Delta^{(2)}$ is a quadratic random field and is therefore non-Gaussian. In this first analysis we deliberately retain the Gaussian one-point benchmark
\begin{equation}
P_G(\Delta_R)=\frac{1}{\sqrt{2\pi}\sigma_R}
\exp\left[-\frac{\Delta_R^2}{2\sigma_R^2}\right].
\label{eq:GaussianPDF}
\end{equation}
This permits a direct comparison with conventional PBH calculations, but PBHs sample the far tail, so the tensor-induced bispectrum and higher cumulants may change the abundance exponentially. All abundance contours and benchmark values in this work are therefore conditional Gaussian estimates. A full quadratic-field probability distribution and compaction-function calculation is deferred to future work.

Critical collapse generates an extended mass function even for a narrow horizon-mass spectrum:
\begin{equation}
\frac{d\beta}{d\ln M}(M|M_H)=\frac{M}{M_H}
P_G[\Delta(M)]\left|\frac{d\Delta}{d\ln M}\right|,
\quad \Delta(M)=\Delta_c+\left(\frac{M}{{\cal K}M_H}\right)^{1/\gamma_c}.
\label{apneq:massfunction}
\end{equation}
Redshifting as matter during radiation domination gives
\begin{equation}
f_{\rm PBH}(M)=\frac{1}{\Omega_{\rm CDM}}
\int d\ln M_H\left(\frac{M_{\rm eq}}{M_H}\right)^{1/2}
\frac{d\beta}{d\ln M}(M|M_H),
\label{eq:fPBH}
\end{equation}
up to the standard mild $g_*$ factor. We use $M_{\rm eq}\simeq3\times10^{17}M_\odot$. For observables linear in abundance, monochromatic limits are recast via
\begin{equation}
\int d\ln M\frac{f_{\rm PBH}(M)}{f_{\max}^{\rm mono}(M)}\le1.
\label{apneq:extendedconstraint}
\end{equation}
Accretion, merger-rate, and dynamical limits are not combined through Eq.~\eqref{apneq:extendedconstraint}; their model dependence is displayed separately.

The horizon mass is tied to the actual induced peak,
\begin{equation}
M_H(k)\simeq17M_\odot\left(\frac{k}{10^6\,{\rm Mpc}^{-1}}\right)^{-2}
\left(\frac{g_*}{10.75}\right)^{-1/6},
\qquad k=k_p^{\rm TIS}.
\label{eq:MHk}
\end{equation}

The benchmark tables are retained as representative points evaluated with the numerical $k_p^{\rm TIS}$ prescription used for the plotted contours. They should not be extrapolated outside the displayed scan domain. PTA-labelled points are conditional cosmological-source benchmarks, not unique fits, and spin values are illustrative because they depend on the peak profile and on the non-Gaussian statistics of the quadratic field.

\begin{table}[t] 
\centering
\caption{\it Benchmark points for FOPT, as shown in Fig. \ref{fig:PT_BP}.}
\label{tab:PT}
\begin{tabular}{|c|c|c|c|c|c|c|}
\hline
\hline
BPs & $\alpha$ & $\beta / H$ & $ T_{\star}$ [GeV] & $M_{\text{PBH}} [M_{\odot}]$ & $f_{\rm PBH}$ & $a_{\star}$ \\
\hline
A & $10^{3}$ & 5.33 & $9.33 \times 10^{4}$ & $10^{-15}$ & 1 & $9.9 \times 10^{-4}$\\
\hline
B & $10^{-1}$ & 2.54 & $2.95 \times 10^{3}$ & $10^{-12}$ & 1 & $1.1 \times 10^{-3}$\\
\hline
C & $10^{1}$ & 5.13 & 2.95 & $10^{-6}$ & $10^{-4}$ &$1.02 \times 10^{-3}$\\
\hline
D & $1$ & 4.03 & $2.95 \times 10^{-3}$ & 1 & $3 \times 10^{-3}$ & $1.45 \times 10^{-3}$\\
\hline
$\rm FOPT_{NG15}$ & $10^{2}$ & 5.27 & $0.66$ & $2.08 \times 10^{-5}$ & $8 \times 10^{-4}$ & $1.01 \times 10^{-3}$\\
\hline
\hline
\end{tabular}
\end{table}

\begin{figure}[t]
    \centering
\includegraphics[width=0.8\linewidth]{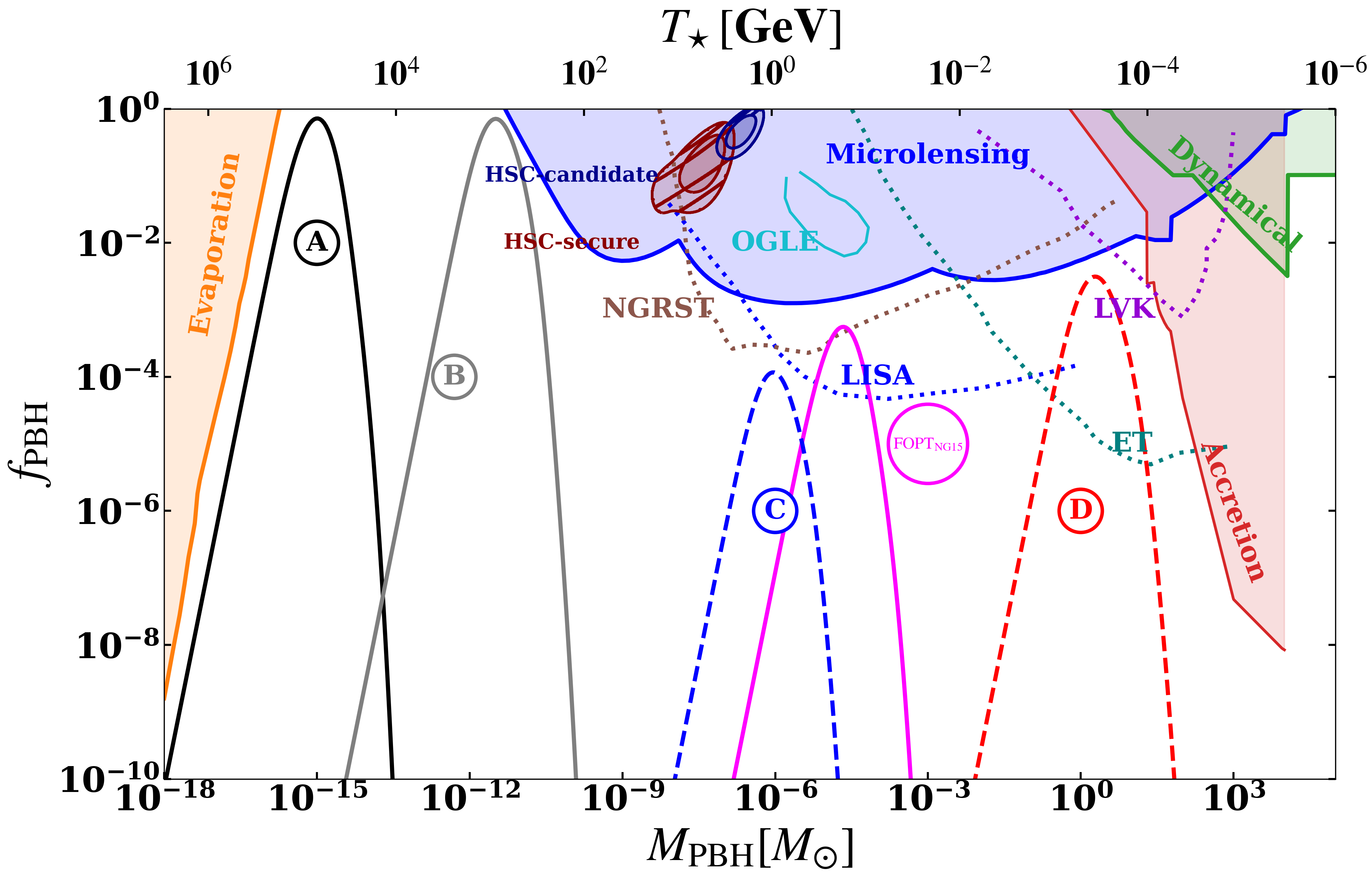}
    \caption{\it Summary of the PBH constraints. The colored shaded areas are excluded by BBN, CMB, cosmic rays, microlensing and GW observations, as discussed in the text. The future sensitivities of NGRST, LISA and ET are shown by the dotted curves. The FOPT parameters corresponding to each PBH mass and spin are shown in Table~\ref{tab:PT}, as per benchmark points A, B, C, and D. The green shaded region on the right edge with $T_{\star} \lesssim 10$ MeV is also excluded by BBN. The PBH mass on the top axis is obtained from the numerical peak $k_p^{\rm TIS}$; for FOPTs it depends on $T_{\star}$, $\beta/H_\star$, and $C_{\rm TIS}^{\rm FOPT}$ through $M_{\rm PBH}^{\rm FOPT} =  \left({C_{\rm TIS}^{\rm FOPT}}\right)^{-2}T_\star^{-2}(\beta/H_\star)^{-2}$. Benchmark points A and B can account for the entire DM abundance, while C and D will be testable in future experiments.}
    \label{fig:PT_BP}
\end{figure}

\begin{table}[t]
\centering
\caption{\it Benchmark points for Domain Walls, as shown in Fig. \ref{fig:DW_BP}.}
\label{tab:DW}
\begin{tabular}{|c|c|c|c|c|c|c|c|}
\hline
\hline
BPs & $\alpha_{\text{ann}}$ & $\epsilon_{\text{sim}}$ & $ T_{\text{ann}} [GeV]$ & $\left(\frac{V_{\rm bias} \left(\rm MeV^4\right)}{\sigma \left(\rm TeV^3\right)}\right)^{1/2} $& $M_{\text{PBH}} [M_{\odot}]$ & $f_{\rm PBH}$ & $a_{\star}$ \\
\hline
E & $1.17 \times 10^{-3}$ & 0.5 & $2.47 \times 10^{4}$ & $1.12 \times 10^{6}$& $10^{-15}$ & 1 & $9.9 \times 10^{-4}$\\
\hline
F & $4.2 \times 10^{-3}$ & 1.5 & $7.8 \times 10^{2}$& $3.54 \times 10^{4}$ & $10^{-12}$ & 1 & $1.1 \times 10^{-3}$\\
\hline
G & $6.16 \times 10^{-3}$ & 2.0 & $2.4 \times 10^{-3} $ & $1.09 \times 10^{-1}$ & $10^{-1}$ & $10^{-4}$ & $1.22 \times 10^{-3}$\\
\hline
H & $3.44 \times 10^{-3}$ & 1.0 &7.8 &$3.54 \times 10^{2}$ & $10^{-8}$ & $5 \times 10^{-3}$ &$1.07 \times 10^{-3}$\\
\hline
$\rm DW_{\rm NG15}$ & $4.04 \times 10^{-2}$ & 4.5 &0.63 &$28.63$ & $1.53 \times 10^{-6}$ & $1 \times 10^{-5}$ &$1.0 \times 10^{-3}$\\
\hline
\hline
\end{tabular}
\end{table}

\begin{figure}[t]
    \centering
    \includegraphics[width=0.8\linewidth]{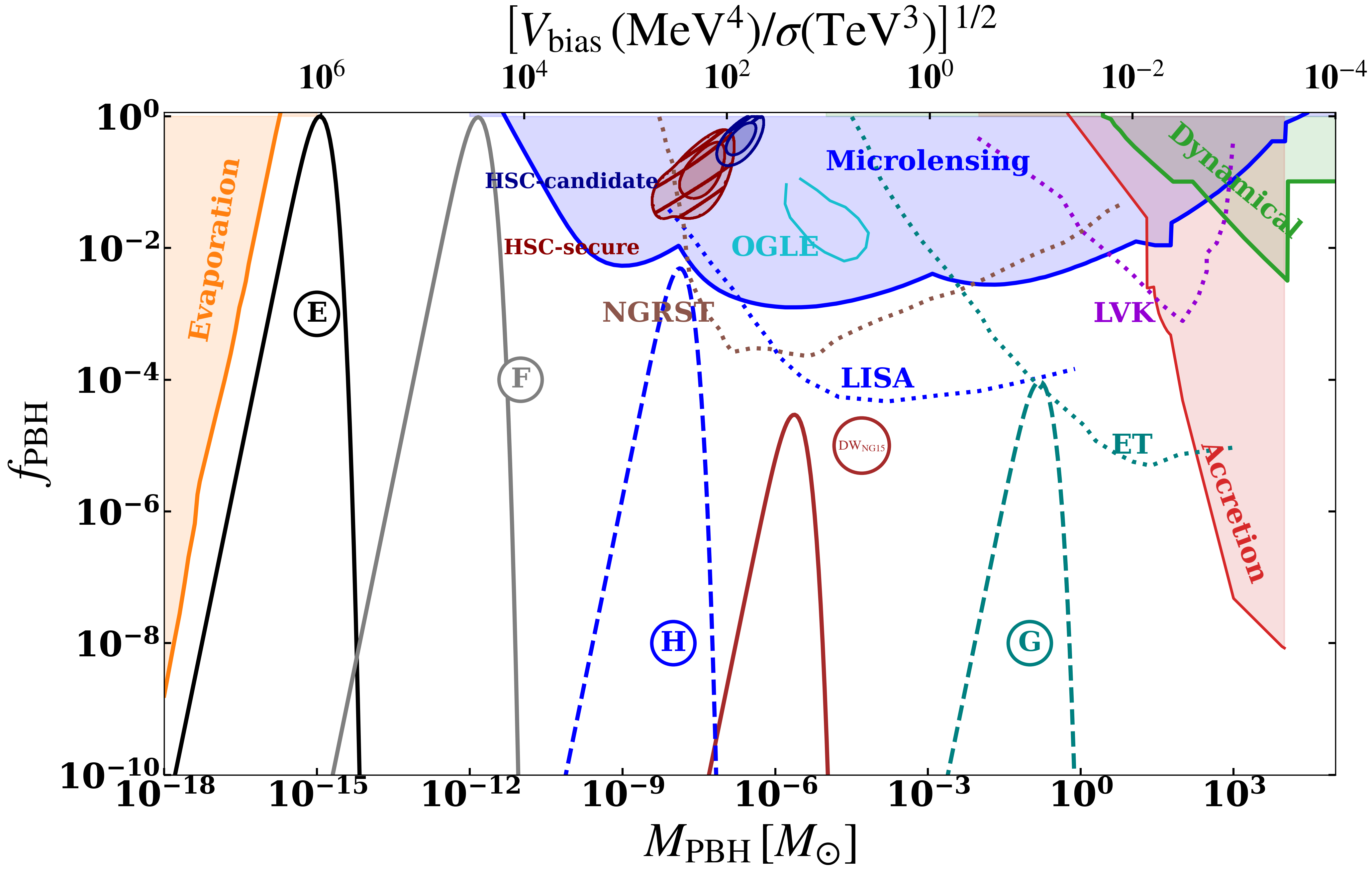}
    \caption{\it  Summary of the PBH constraints. The colored shaded areas are excluded by BBN, CMB, cosmic rays, microlensing and GW observations, as discussed in the text. The future sensitivities of NGRST, LISA and ET are shown by the dotted curves. The domain wall parameters corresponding to each PBH mass and spin are shown in Table~\ref{tab:DW}, as per benchmark points E, F, G, and H. The green shaded region on the right edge with $T_{\rm ann} \lesssim 10$ MeV is also excluded by BBN. The PBH mass on the top axis is obtained from $k_p^{\rm TIS}=C_{\rm TIS}^{\rm DW}k_H(T_{\rm ann})$ and therefore scales as $M_p^{\rm DW}\propto (C_{\rm TIS}^{\rm DW})^{-2}T_{\rm ann}^{-2}$, with $T_{\rm ann}$ fixed by $(V_{\rm bias}/\sigma)^{1/2}$. Benchmark points E and F can account for the entire DM abundance, while G and H will be testable in future experiments.}
    \label{fig:DW_BP}
\end{figure}
\section{Semi-analytical parameter scalings and numerical interpolation}\label{sec:fits}

The numerical abundance is exponentially sensitive to $\sigma_R$ and determined from the numerical integration of \cref{apneq:variance}. The peak scalings for variance are
\begin{equation}
\sigma_{R,p}^2\propto
\left(\frac{\kappa_i\alpha}{1+\alpha}\right)^4
\left(\frac{H_\star}{\beta}\right)^{2p_i}
\quad{\rm(FOPT)},
\qquad
\sigma_{R,p}^2\propto\epsilon_{\rm GW}^2\alpha_{\rm ann}^4
\quad{\rm(DW)}.
\label{eq:sigmascaling}
\end{equation}

Now, we obtain the analytical formulae relating the GW parameters to the PBH abundance.
Firstly, we present the analytical formulae relating the FOPT parameters to the fraction $f_{\rm PBH}$ of DM in the form of PBHs today.

We use the \cref{eq:MHk} to relate $M_{\rm PBH}$ to $T_{\star}$ and $\beta / H$ via the following relation:
\begin{eqnarray}
     M_{\rm PBH}^{\rm FOPT} =  \left({C_{\rm TIS}^{\rm FOPT}}\right)^{-2} \, T_\star^{-2} \,\left(\beta / H\right)^{-2}\simeq 7.66 \times 10^{-5}\, M_{\rm eq} \left(\frac{20}{g_{\star}\left(T_{\rm eq}\right)}\right)^{1/2}\, \left(\frac{T_{\rm eq}}{T_{\star}}\right)^{2} \, \left(\frac{\beta}{H}\right)^{-2}\label{apneq:MPBH_fopt}\,.
\end{eqnarray}
Similarly the $f_{\rm PBH}$ can be written as follows:
\begin{equation} 
   f_{\rm PBH} = \mathcal{A}_{\rm FOPT} \left(\alpha, \beta / H\right) \, \left(\frac{T_{\star}}{\rm 1 \, GeV} \right)\,, \label{eq:fpbh_fopt}
\end{equation}
where $\mathcal{A}_{\rm FOPT}$ is the proportionality constant responsible for normalization with respect to $T_{\star}$. It depends only on $\alpha$ and $\beta / H$. $M_{\rm PBH}$ is determined from the location of the peak of the second-order scalar perturbations induced by the FOPT tensors (see \cref{apneq:ind_PT_ana}).

\begin{figure}[]
    \centering
    \includegraphics[scale = 0.41]{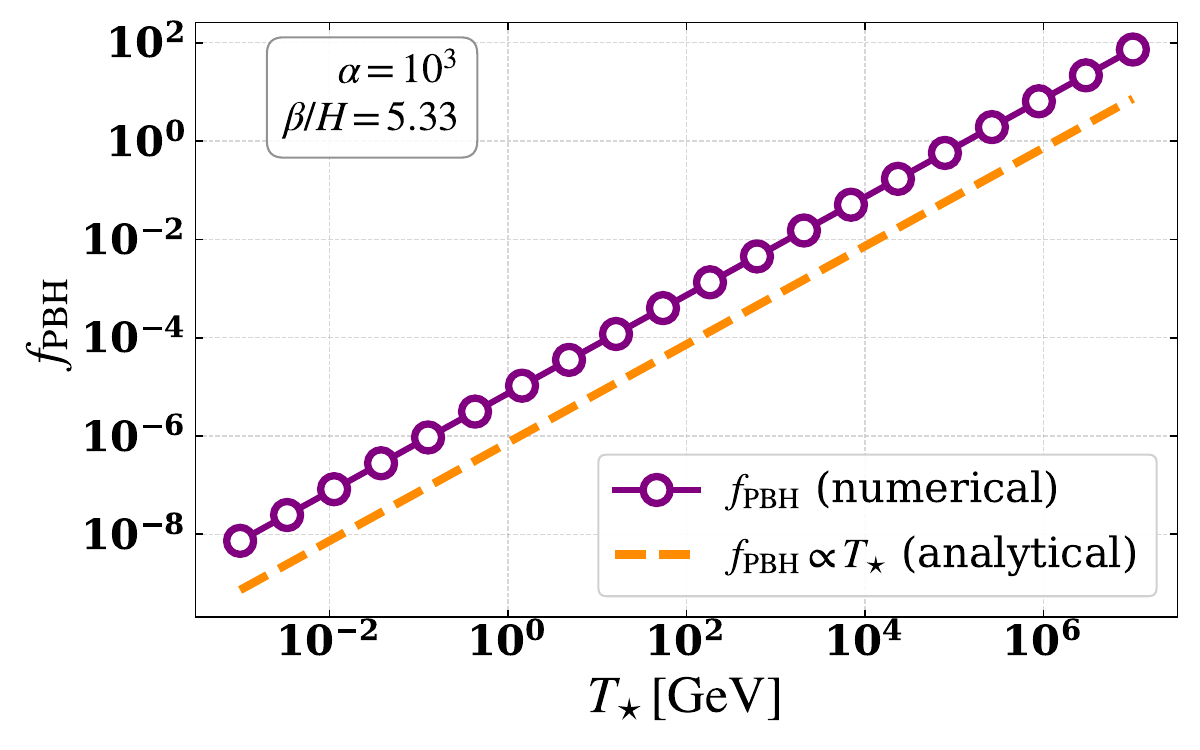}
    \includegraphics[scale = 0.42]{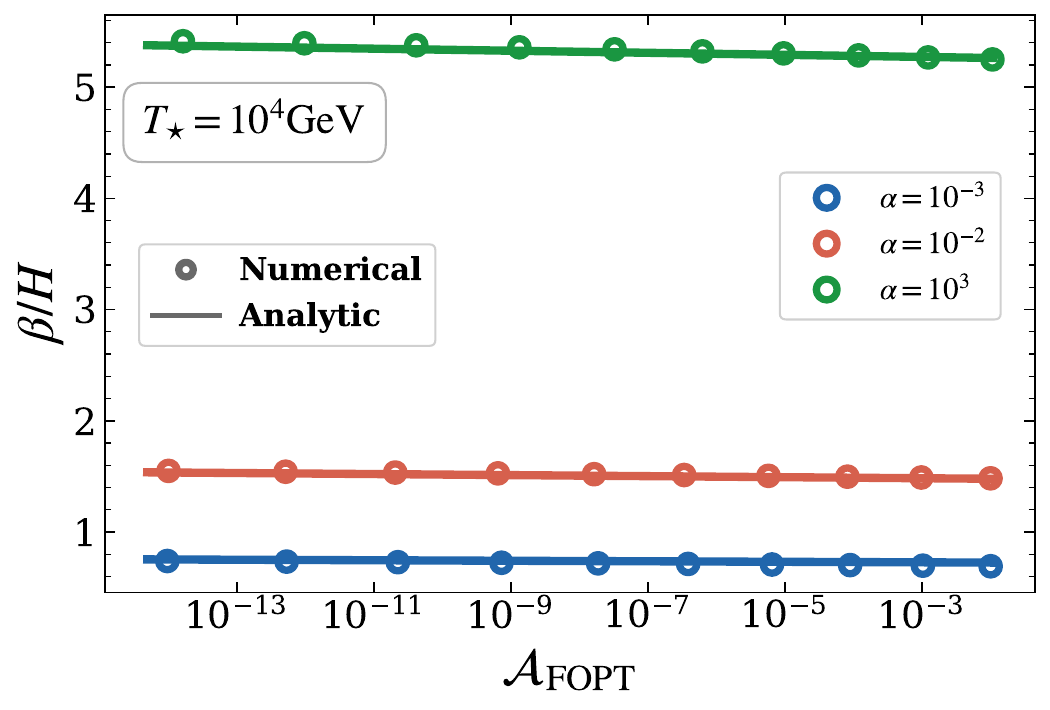}
    \caption{\it Comparison between the numerical and semi-analytical estimates of the fractional PBH abundance in terms of the FOPT parameters, following Eqs.~\eqref{eq:fpbh_fopt} and \eqref{apneq:MPBH_fopt}.}
    \label{fig:FPBH_PT_ana}
\end{figure}

The left panel of \cref{fig:FPBH_PT_ana} shows that $f_{\rm PBH} \propto T_{\star}$. The proportionality constant, i.e., $\mathcal{A}_{\rm FOPT}$, is a function only of the other FOPT parameters $\alpha$ and $\beta / H$. It is tedious to find the closed form for $\mathcal{A}_{\rm FOPT}$, so alternatively we look for allowed values of $\beta / H $ in terms of other model parameters including $\mathcal{A}_{\rm FOPT}$. In order to give the analytical result for $\beta / H$, we employ the symbolic regression tool \textsc{PySR} \cite{cranmer2023interpretable} to obtain:

\begin{equation} 
   \beta / H = \frac{0.58 \left(\log_{10} \alpha + 5.26\right)^{2}}{4.035 + 0.43 \times 2^{\log_{10} \alpha} + 1.2 \times 10^{-2} \left[\frac{\log_{10} \alpha}{\log_{10}(2.82 \alpha)} + \log_{10} \mathcal{A}_{\rm FOPT}\right]} \label{eq:betaH_ana}
\end{equation}
This symbolic formula is valid for $\alpha \in [10^{-3}, 10^{3}]$ and $\mathcal{A}_{\rm FOPT} \in [10^{-15},10^{-2}]$. Remarkably, this fitting formula matches the numerical results to within $\leq 5 \%$. The right panel of \cref{fig:FPBH_PT_ana} shows the comparison of the numerical and analytical $\beta / H$ results. It is also important to note that, for $\alpha \gg 1$, $P_{\rm FOPT}$ becomes independent of $\alpha$, consistent with the behavior of the FOPT GW spectrum.

\begin{figure}[]
    \centering
    \includegraphics[scale=0.41]{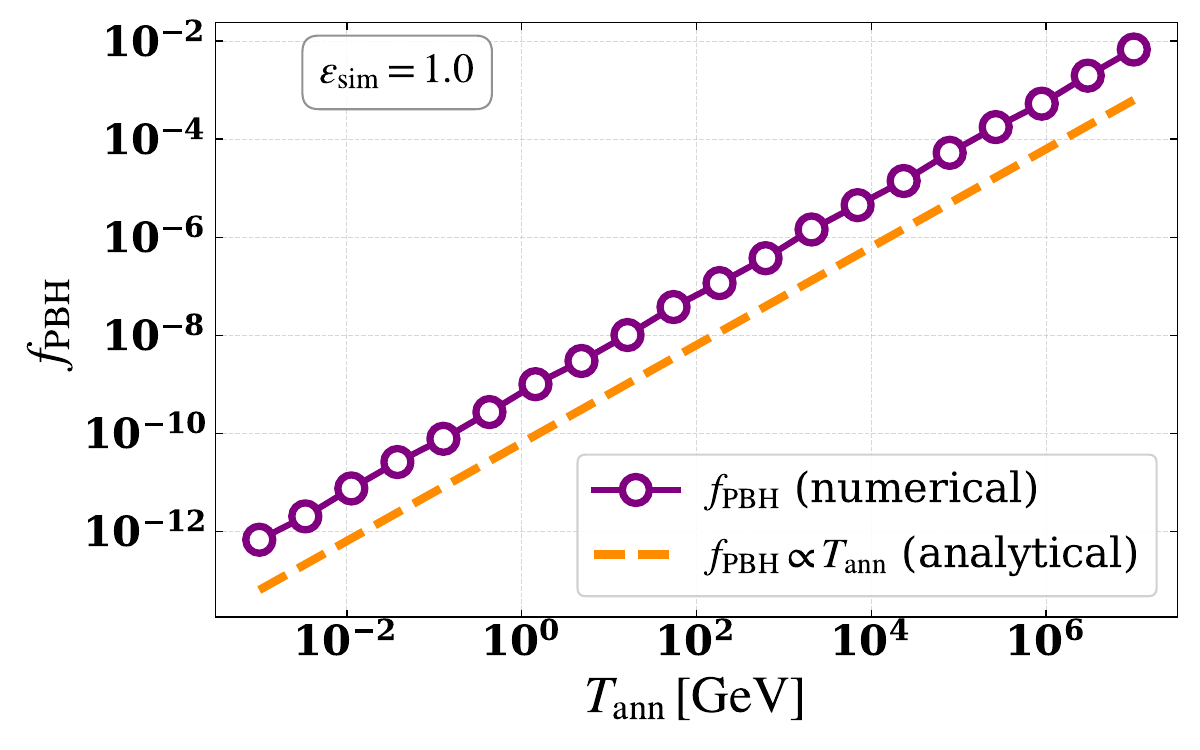}
    \includegraphics[scale=0.42]{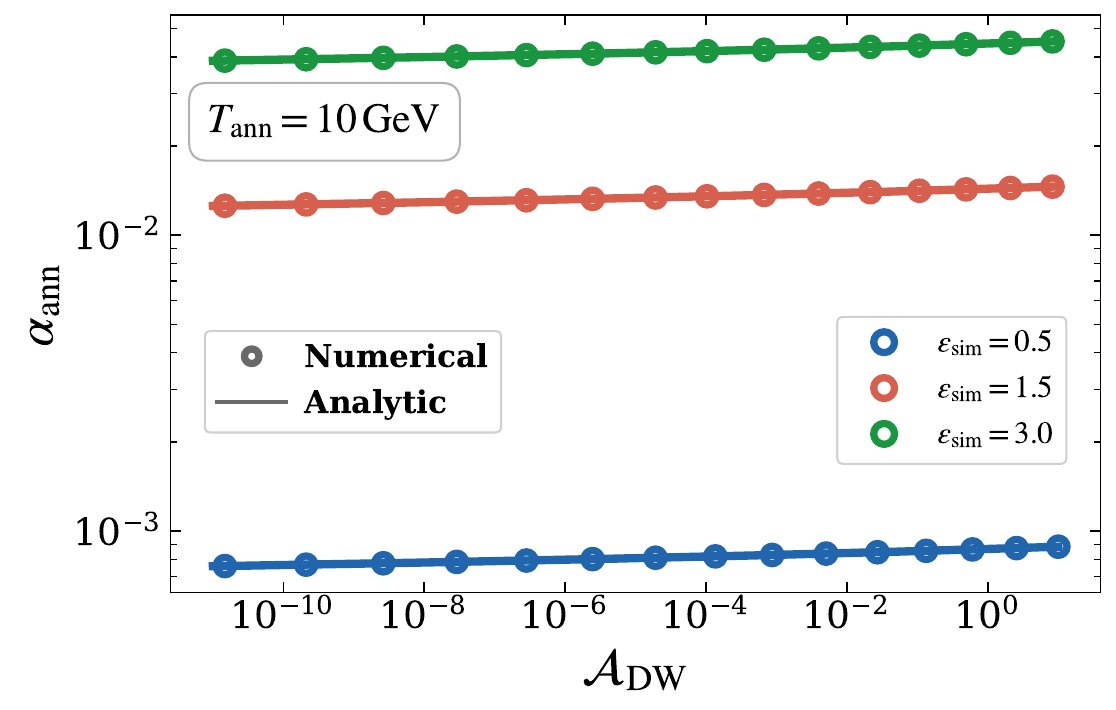}
    \caption{\it Comparison between the numerical and semi-analytical estimates of the fractional PBH abundance in terms of the DW parameters, following Eqs.~\eqref{eq:fpbh_dw} and \eqref{apneq:MPBH_dw}.}
    \label{fig:FPBH_DW_ana}
\end{figure}

Finally, we conclude the discussion of allowed values of $\alpha_{\rm ann}$ by providing the analytical formula for $f_{\rm PBH}$ in terms of DW parameters.
For the domain-wall case as well, we find $f_{\rm PBH} \propto T_{\rm ann}$, as can be seen in the left panel of \cref{fig:FPBH_DW_ana}. We express $f_{\rm PBH}$ today for DWs as:
\begin{equation} 
   f_{\rm PBH} = \mathcal{A}_{\rm DW} \left(\alpha_{\rm ann}, \epsilon_{\rm sim} \right) \, \left(\frac{T_{\rm ann}}{\rm 1 \, GeV} \right) \\
   = 2.20 \times 10^{-2} \mathcal{A}_{\rm DW} \left(\alpha_{\rm ann}, \epsilon_{\rm sim} \right) \, \left(\frac{V_{\rm bias} \left(\rm MeV^4\right)}{\sigma \left(\rm TeV^3\right)}\right)^{1/2}  \label{eq:fpbh_dw}
\end{equation}
where $\mathcal{A}_{\rm DW}$ is the proportionality constant responsible for normalization with respect to $T_{\rm ann}$.
$M_{\rm PBH}$ is given in terms of $T_{\rm ann}$ via:
\begin{equation} 
    M_{\rm PBH}^{\rm FOPT} = (C_{\rm TIS}^{\rm DW})^{-2}T_{\rm ann}^{-2} \simeq 10^{-6}\, M_{\rm eq} \left(\frac{20}{g_{\star}\left(T_{\rm eq}\right)}\right)^{1/2}\, \left(\frac{T_{\rm eq}}{T_{\rm ann}}\right)^{2} \label{apneq:MPBH_dw}
\end{equation}
Finally, we find the analytical fitting form for $\alpha_{\rm ann} \left(\epsilon_{\rm sim}, \mathcal{A}_{\rm DW} \right)$ \cite{cranmer2023interpretable}:

\begin{equation} 
    \log_{10} \alpha_{\rm ann} = 1.25 \times 10^{1} \left(6.8 \times 10^{-3} \left( \epsilon_{\rm sim} + \log_{10} \mathcal{A}_{\rm DW}\right) - 0.9348 + \frac{1.4622}{\log_{10}\left(\epsilon_{\rm sim} + 0.5821 - 1.206 \right)} \right) \label{eq:ana_DW}
\end{equation}

\cref{eq:ana_DW} is accurate to within $5\%$ for $\epsilon_{\rm sim} \in [0.30,3.5]$ and $\mathcal{A}_{\rm DW} \in [10^{-12}, 10^{2}]$, as shown in the right panel of \cref{fig:FPBH_DW_ana}.

We want stress that these formulae are symbolic and only valid in stated range of parameter values. We do not use these relations to obtain the allowed values of parameters for both FOPTs and annihilating DWs.
\section{Inflationary tensor sources}\label{sec:inflation}

This section clarifies how the tensor-induced density formalism developed in the Letter extends to inflationary tensor spectra. The main text focuses on causal, post-inflationary, sub-Hubble sources-first-order phase transitions and annihilating domain walls-because they provide direct particle-physics realizations with sharply identifiable source parameters. Inflationary tensor perturbations are conceptually different: the tensor modes are specified as primordial super-Hubble initial conditions, rather than being matched from an active sub-Hubble stress source after production. Consequently, inflationary tensors must be treated with the $x_{\rm ini}\to0$ kernels in Eqs.~\eqref{eq:anakernela}-\eqref{eq:anakernelb}, not with the finite-$x_{\rm ini}$ kernels used for the FOPT and DW benchmarks. This difference is technical but important: the convolution formula is the same, while the initial condition and the interpretation of the tensor spectrum are different.

For an inflationary tensor spectrum we take the primordial tensor power spectrum ${\cal P}_{\chi,\rm prim}(k)$ as the input. After horizon re-entry in radiation domination the tensor modes evolve with the free transfer function $T_h(k\eta)=\sin(k\eta)/(k\eta)$, but the induced density perturbation is obtained by inserting the primordial spectrum into the super-Hubble version of the tensor-tensor convolution. Thus the appropriate master expression is

\begin{equation}
{\cal P}^{\rm inf}_{\Delta^{(2)}}(k)
=\frac12\int_0^\infty dv\int_{|1-v|}^{1+v}du\,
\frac{F(u,v)}{(uv)^3}\overline{{\cal I}_0^2(u,v)}
{\cal P}_{\chi,\rm prim}(ku){\cal P}_{\chi,\rm prim}(kv),
\label{eq:inflation_master_convolution}
\end{equation}

where $\overline{{\cal I}_0^2}$ denotes the late-time average of the $x_{\rm ini}=0$ kernel constructed from Eqs.~\eqref{eq:anakernela}-\eqref{eq:anakernelb}. This equation is the inflationary analogue of Eq.~\eqref{eq:PSDf}. The finite-source matching relation used for causal sources, Eq.~\eqref{apneq:GW_PT}, is not the starting point here; instead, ${\cal P}_{\chi,\rm prim}$ is specified directly by the inflationary mechanism. This is the first technical point that separates the inflationary validation exercise from the FOPT/DW constraints shown in the main Letter.

As a controlled set of examples, we consider three representative primordial tensor shapes.

\textbf{First, a log-normal tensor spectrum, commonly used as a model-independent proxy for a localized enhancement of primordial inflationary tensor power}\cite{Guzzetti:2016mkm,Dimastrogiovanni:2016fuu,Thorne:2017jft,Namba:2015gja}, is written as

\begin{equation}
{\cal P}^{\rm LN}_{\chi,\rm prim}(k)
=\frac{A_T}{\sqrt{2\pi}\Delta}
\exp\left[-\frac{\ln^2(k/k_p)}{2\Delta^2}\right],
\label{eq:inflation_lognormal}
\end{equation}

where $A_T$ controls the integrated tensor amplitude, $\Delta$ controls the logarithmic width, and $k_p$ fixes the characteristic horizon mass.

\textbf{Second, to test how sharply peaked inflationary tensor spectra behave, we use a broken-power-law FOPT-like tensor }template \cite{An:2023jxf},

\begin{equation}
{\cal P}^{\rm IFT}_{\chi,\rm prim}(k)
=A_T^{\rm IFT}
\frac{(a+b)^c}{\left[b(k/k_p)^{-a/c}+a(k/k_p)^{b/c}\right]^c},
\label{eq:inflation_fopt_template}
\end{equation}
with the same shape parameters $(a,b,c)$ used for the post-inflationary FOPT benchmark, but now interpreted only as a primordial tensor shape.

\textbf{Third, we use a DW-like template, which are nucleated during primordial inflation:} \cite{An:2023idh}
\begin{equation}
{\cal P}^{\rm IDW}_{\chi,\rm prim}(k)
=A_T^{\rm IDW}
\left[\frac14\left(\frac{k_p}{k}\right)^{1/\delta}
+\frac34\left(\frac{k}{k_p}\right)^{1/\delta}\right]^{-\delta},
\label{eq:inflation_dw_template}
\end{equation}

again understood as a primordial-shape test and not as the same post-inflationary annihilation calculation used in the main constraints. These three examples span broad, moderately localized, and sharply structured tensor spectra. They are therefore useful stress tests of the response kernel and of the PBH mass-function pipeline.

The PBH abundance is then computed exactly as in Sec.~\ref{sec:PBH}, with the variance evaluated from the inflationary induced spectrum,

\begin{equation}
\sigma_R^2(\eta_R)=\int d\ln q\,W^2(qR)
{\cal P}^{\rm inf}_{\Delta^{(2)}}(q,\eta_R),
\qquad aH|_{\eta_R}=R^{-1}.
\label{eq:inflation_variance}
\end{equation}

For the Gaussian benchmark one obtains the formation probability from the same threshold prescription as in Eq.~\eqref{eq:GaussianPDF}, and the critical-collapse mass function follows Eq.~\eqref{apneq:massfunction}. The mapping from wavenumber to horizon mass remains

\begin{equation}
M_H(k)\simeq17M_\odot
\left(\frac{k}{10^6\,{\rm Mpc}^{-1}}\right)^{-2}
\left(\frac{g_*}{10.75}\right)^{-1/6},
\qquad k\simeq k_p^{\rm TIS},
\label{eq:inflation_mass_mapping}
\end{equation}

where $k_p^{\rm TIS}$ is found from the peak of Eq.~\eqref{eq:inflation_master_convolution}. The important lesson is that, for inflationary tensors, the PBH mass scale is controlled by the primordial tensor peak $k_p$, while the abundance is exponentially controlled by the tensor amplitude and width through the induced variance. This is the same mass-amplitude logic used for FOPTs and DWs, but with a different initial-condition prescription.

\begin{figure}[t]
    \centering
    \includegraphics[width=0.62\linewidth]{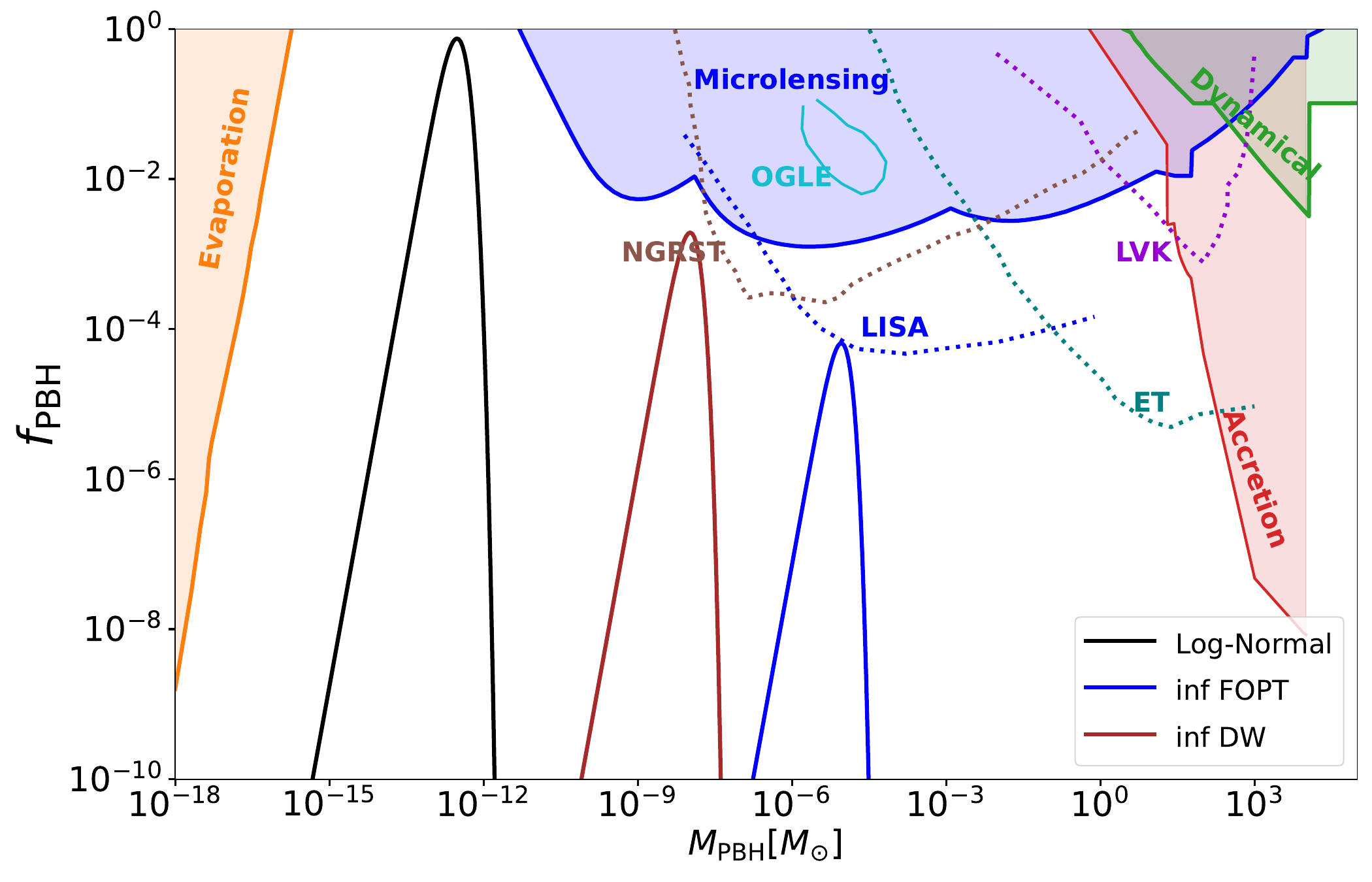}
    \caption{\it PBH mass function $f_{\rm PBH}(M)$ generated by tensor-induced density perturbations sourced by inflationary tensor spectra. The black curve corresponds to the log-normal primordial tensor spectrum in Eq.~\eqref{eq:inflation_lognormal}. The blue curve corresponds to the FOPT-like primordial tensor template in Eq.~\eqref{eq:inflation_fopt_template}. The brown curve corresponds to the DW-like primordial tensor template in Eq.~\eqref{eq:inflation_dw_template}. The figure is not used to set the FOPT/DW constraints in the main Letter; it is a validation and generality check showing that the same tensor-tensor convolution and PBH mass-function pipeline can be applied to super-Hubble inflationary initial conditions after replacing the finite-$x_{\rm ini}$ kernel by the $x_{\rm ini}=0$ kernel.}
    \label{fig:PT_inf}
\end{figure}

Figure~\ref{fig:PT_inf} should be read as a simple measure plot for the super-Hubble limit of the formalism, rather than as an additional constraint plot. The horizontal axis is the PBH mass generated when the scalar fluctuation induced by the primordial tensor spectrum re-enters the horizon, while the vertical axis is the corresponding present-day differential abundance $f_{\rm PBH}(M)$. The black curve shows the response to the log-normal template in Eq.~\eqref{eq:inflation_lognormal}; the blue and brown curves show the responses to the FOPT-like and DW-like primordial tensor templates in Eqs.~\eqref{eq:inflation_fopt_template} and~\eqref{eq:inflation_dw_template}. The location of each curve is controlled mainly by the tensor peak scale $k_p$, through the approximate relation $M_{\rm PBH}\sim M_H(k_p^{\rm TIS})\propto (k_p^{\rm TIS})^{-2}$. The height of each curve is controlled mainly by the induced variance $\sigma_R^2$, and hence by the square of the tensor amplitude entering Eq.~\eqref{eq:inflation_master_convolution}. The shape of each curve is controlled by the convolution of the tensor spectral width with the window function and by critical collapse.

The plot makes three physical points transparent. First, a localized enhancement in primordial tensor power produces a localized PBH mass function. This is because the convolution in Eq.~\eqref{eq:inflation_master_convolution} is dominated by configurations for which both tensor momenta, $ku$ and $kv$, sample the enhanced region of ${\cal P}_{\chi,\rm prim}$. The scalar response therefore peaks at a calculable scale $k_p^{\rm TIS}$ rather than at an arbitrary mass. Second, the width of $f_{\rm PBH}(M)$ is a memory of the tensor source: a broader log-normal spectrum spreads power over a wider interval in $k$, and therefore over a wider interval in $M\propto k^{-2}$, whereas sharper FOPT-like or DW-like shapes generate more localized PBH spectra. Third, the abundance is highly nonlinear in the tensor amplitude. At the level of the variance one has schematically

\begin{equation}
\sigma_{R,p}^2\sim
\int d\ln q\,W^2(qR)
\int dv\,du\,\overline{{\cal I}_0^2(u,v)}
{\cal P}_{\chi,\rm prim}(qu){\cal P}_{\chi,\rm prim}(qv),
\label{eq:inflation_variance_schematic}
\end{equation}

so that for a single-amplitude family ${\cal P}_{\chi,\rm prim}\propto A_T$, the peak variance scales approximately as $\sigma_{R,p}^2\propto A_T^2$. The PBH abundance then behaves as

\begin{equation}
 f_{\rm PBH}(M_p)\propto
 \exp\left[-\frac{\Delta_c^2}{2\sigma_{R,p}^2}\right],
\label{eq:inflation_exponential_sensitivity}
\end{equation}

up to the mass-redshifting and critical-collapse factors discussed in Sec.~\ref{sec:PBH}. This double sensitivity-quadratic in tensor power at the variance level and exponential at the PBH-tail level-is the main reason that Fig.~\ref{fig:PT_inf} is a stringent test of the numerical pipeline.

The figure also clarifies the physical role of the response kernel. In the sub-Hubble FOPT/DW cases, the finite value of $x_{\rm ini}$ encodes when the active source switches off and when the tensor spectrum is matched to free propagation. In the inflationary case, the relevant modes are already present as super-Hubble initial conditions, so the lower integration limit is zero and the kernel is the analytic $x_{\rm ini}=0$ response. Therefore Fig.~\ref{fig:PT_inf} verifies that the same tensor-tensor source term produces sensible PBH mass functions in the opposite initial-condition regime. This directly checks that the formalism is not an artifact of the finite-$x_{\rm ini}$ treatment used for phase transitions and domain walls.

This observation is useful for sharpening the claims of the paper. The core result of the work is not tied to a particular microscopic source; it is the statement that a first-order tensor spectrum can induce a second-order comoving-density spectrum, and that the induced scalar spectrum can seed PBHs after smoothing and thresholding. The source-specific ingredients are: (i) the tensor spectrum itself, (ii) the correct initial-condition kernel, and (iii) the mapping from the source parameters to the tensor amplitude and peak scale. For inflationary tensors, item (ii) is the $x_{\rm ini}=0$ kernel and item (iii) depends on the inflationary model and reheating history. For causal post-inflationary FOPT/DW sources, item (ii) is the finite-$x_{\rm ini}$ kernel and item (iii) is fixed by $(\alpha,\beta/H_\star,T_\star)$ or $(\alpha_{\rm ann},T_{\rm ann},\epsilon_{\rm sim})$. Once these ingredients are specified, the subsequent steps-Eq.~\eqref{eq:inflation_master_convolution}, smoothing, critical collapse, and the extended mass function-are common.

We therefore use Fig.~\ref{fig:PT_inf} as a generality and consistency demonstration, while keeping the main text focused on non-inflationary sources. This choice is deliberate. Post-inflationary FOPT and DW sources have direct particle-physics interpretations and lead to a clean correlation between a stochastic GW signal, a source temperature, and a PBH mass window. Inflationary tensor sources are certainly within the scope of the formalism, but a quantitative inflationary constraint would require additional assumptions about the tensor-production mechanism, scalar perturbations generated by the same sector, tensor-scalar correlations, reheating, and the collapse threshold appropriate for super-Hubble initial conditions. In the letter, we derive and apply a general tensor-induced PBH mechanism, and chose non-inflationary sources as the minimal, model-transparent applications to illustrate the main formalism in simplistic manner.

This also addresses the fact that the absence of inflationary benchmark constraints from the main figures should not be interpreted as a limitation of the tensor-induced formalism. Rather, it reflects a choice to avoid mixing two different classes of initial conditions and two different model-building questions in one set of contours. The inflationary examples in Fig.~\ref{fig:PT_inf} show explicitly how to use the same machinery for primordial super-Hubble tensors: replace the finite-$x_{\rm ini}$ source matching by the $x_{\rm ini}=0$ kernel, specify ${\cal P}_{\chi,\rm prim}$, and then follow the same PBH pipeline. A dedicated inflationary study can therefore be built directly on the present formalism, but is logically separate from the post-inflationary FOPT/DW phenomenology emphasized in the Letter.

\medskip
\section{Viable particle-theory embeddings}
\label{model}

The Letter treats the source parameters $(\alpha,\beta/H_\star,T_\star)$ for FOPTs and $(\alpha_{\rm ann},\sigma,V_{\rm bias})$ for annihilating DWs as macroscopic variables. This section explains how these variables arise in two simple UV-complete particle-physics settings and how they map onto the PBH observables used in the main analysis. The purpose is not to impose a unique microscopic model on the phenomenological scan, but to show that the benchmark points used in the Letter can be interpreted as macroscopic representatives of well-motivated particle theories.

\subsection{General map from microscopic parameters to PBH observables}
\label{sec:model_general_map}

The common structure is
\begin{equation}
\boxed{
\text{particle parameters}
\quad\Longrightarrow\quad
\text{source parameters}
\quad\Longrightarrow\quad
\Omega_{\rm GW}(f)
\quad\Longrightarrow\quad
{\cal P}_{\Delta^{(2)}}(k)
\quad\Longrightarrow\quad
(M_{\rm PBH},f_{\rm PBH}) .}
\label{eq:model_chain}
\end{equation}

The first arrow is model dependent: it requires the finite-temperature potential for FOPTs or the wall tension and bias for DWs. The remaining arrows are the model-independent tensor-induced calculation developed in the Letter. Once the source spectrum is specified, the induced scalar spectrum scales schematically as
\begin{align}
{\cal P}_{\Delta^{(2)},p}^{\rm FOPT}
&\simeq
{\cal N}_{\rm FOPT}
\left(\frac{\alpha}{1+\alpha}\right)^4
\left(\frac{H_\star}{\beta}\right)^{2p_i},
\label{eq:model_scaling_fopt_general}\\
{\cal P}_{\Delta^{(2)},p}^{\rm DW}
&\simeq
{\cal N}_{\rm DW}
\epsilon_{\rm GW}^{2}\alpha_{\rm ann}^{4},
\label{eq:model_scaling_dw_general}
\end{align}

where ${\cal N}_{\rm FOPT,DW}$ include the convolution kernel, spectral shape, finite-$x_{\rm ini}$ matching, and smoothing. The PBH abundance is exponentially sensitive to the variance
\begin{equation}
\sigma_R^2=\int d\ln k\,W^2(kR){\cal P}_{\Delta^{(2)}}(k),
\end{equation}

and in the Gaussian benchmark used in the Letter
\begin{equation}
\beta_{\rm form}(M_H)\simeq
\frac{\sigma_R}{\Delta_c\sqrt{2\pi}}
\exp\left[-\frac{\Delta_c^2}{2\sigma_R^2}\right],
\qquad
f_{\rm PBH}(M)\propto
\left(\frac{M_{\rm eq}}{M_H}\right)^{1/2}\beta_{\rm form}(M_H).
\label{eq:model_betaform}
\end{equation}

Thus particle parameters affect $f_{\rm PBH}$ mainly through the amplitude combination in Eqs.~\eqref{eq:model_scaling_fopt_general}-\eqref{eq:model_scaling_dw_general}, while the PBH mass is fixed mostly by the source temperature and the numerical peak ratio $C_{\rm TIS}=k_p^{\rm TIS}/k_p^{\rm GW}$.

\subsection{Key particle-model parameters and physical meaning}
\label{sec:model_key_parameters}

Before discussing the two examples, it is useful to separate microscopic particle parameters from macroscopic source parameters. The microscopic parameters are Lagrangian quantities, while the macroscopic quantities are the only inputs required by the tensor-induced PBH calculation. The bridge between them is the finite-temperature transition calculation for FOPTs, and the wall-scaling/annihilation calculation for DWs.

\begin{table}[t]
\centering
\scalebox{0.85}{%
\begin{tabular}{|c|c|c|}
\hline
Quantity & Particle-model meaning & Role in the GW-PBH prediction \\
\hline
$v_\phi$ & $U(1)_{B-L}$ breaking scale & Sets the overall high-energy scale of the FOPT \\
$g_{B-L}$ & $B-L$ gauge coupling & Sets $m_{Z'}=2g_{B-L}v_\phi$ and affects the thermal barrier \\
$y_N$ & RH-neutrino Yukawa coupling & Sets $M_N=y_Nv_\phi/\sqrt2$ and links the transition to seesaw/leptogenesis \\
$T_\star$ & transition completion temperature & Fixes the GW peak and, with $C_{\rm TIS}$, the PBH mass scale \\
$\alpha$ & released vacuum energy over radiation energy & Controls the GW amplitude and hence $f_{\rm PBH}$ exponentially \\
$\beta/H_\star$ & inverse transition duration & Smaller values mean slower transitions, larger GW amplitude, and larger PBH abundance \\
$C_{\rm TIS}$ & numerical TIS/GW peak ratio & Converts the GW frequency scale to the PBH mass scale \\
\hline
\end{tabular}
}
\caption{\it Main microscopic and macroscopic parameters for the FOPT example. The first three are Lagrangian or symmetry-breaking parameters; the last four are the macroscopic quantities entering the tensor-induced PBH calculation.}
\label{tab:FOPT_key_parameters}
\end{table}

\begin{table}[t]
\centering
\begin{tabular}{|c|c|c|}
\hline
Quantity & Particle-model meaning & Role in the GW-PBH prediction \\
\hline
$u$ & $Z_2$-breaking singlet VEV & Sets the wall tension through $\sigma\propto u^3$ \\
$\lambda_\varphi$ & singlet quartic coupling & Fixes the wall width and normalization of $\sigma$ \\
$Y_L,Y_R,M_N$ & Dirac-seesaw parameters & Determine $m_\nu\simeq Y_LY_Rvu/(\sqrt2 M_N)$ and leptogenesis scale \\
$\sigma$ & DW surface tension & Controls wall energy density and the annihilation condition \\
$V_{\rm bias}$ & soft $Z_2$-breaking vacuum bias & Triggers annihilation and fixes $T_{\rm ann}$ through $H_{\rm ann}\sim V_{\rm bias}/\sigma$ \\
$\alpha_{\rm ann}$ & wall energy fraction at annihilation & Controls the GW amplitude and hence $f_{\rm PBH}$ exponentially \\
$T_{\rm ann}$ & annihilation temperature & Fixes the GW peak and the PBH mass scale \\
$\epsilon_{\rm sim}$ & finite-$x_{\rm ini}$ matching parameter & Shifts the kernel response and the numerical value of $C_{\rm TIS}^{\rm DW}$ \\
\hline
\end{tabular}
\caption{\it Main microscopic and macroscopic parameters for the biased-DW example. The microscopic parameters $(u,\lambda_\varphi,Y_L,Y_R,M_N,V_{\rm bias})$ determine the macroscopic inputs $(\sigma,V_{\rm bias},\alpha_{\rm ann},T_{\rm ann})$ used in the main analysis.}
\label{tab:DW_key_parameters}
\end{table}

The physical interpretation is then direct. The mass $M_{\rm PBH}$ is primarily a clock: it is set by the cosmic time of the FOPT or DW annihilation, equivalently by $T_\star$ or $T_{\rm ann}$. The abundance $f_{\rm PBH}$ is primarily an amplitude measure: it is controlled by the source strength through $\alpha$ and $\beta/H_\star$ for FOPTs, and by $\alpha_{\rm ann}$ for DWs. This is why the same microscopic event can lead simultaneously to a stochastic GW signal and to a sharply correlated PBH mass window.

\subsection{FOPT example: classically scale-invariant \texorpdfstring{$U(1)_{B-L}$}{U(1) B-L} model}
\label{sec:model_fopt_bl}

A minimal UV completion producing strong FOPTs is the classically scale-invariant $U(1)_{B-L}$ extension of the Standard Model. This model contains a complex scalar field $\Phi$ whose vacuum expectation value $v_\phi$ breaks $U(1)_{B-L}$, a gauge boson $Z'$, and three right-handed neutrinos. The same symmetry breaking generates Majorana masses for the right-handed neutrinos and can support resonant leptogenesis. It has been shown that the $B-L$ transition can be strongly first order and can source observable GWs, including in the context of testable leptogenesis~\cite{Ellis:2019oqb,Ellis:2020nnr,Dasgupta:2022isg}.

The $U(1)_{B-L}$ example is only one representative realization. Strong FOPTs with observable GWs also occur in a broad class of BSM settings, including singlet-extended sectors, conformal or nearly conformal gauge extensions, axion/Peccei-Quinn sectors, neutrino-mass models, and models motivated by baryogenesis or collider complementarity~\cite{DelleRose:2019pgi,VonHarling:2019rgb,Ghoshal:2020vud,Jaeckel:2016jlh,Jinno:2016knw,Marzola:2017jzl,Iso:2017uuu,Chao:2017ilw,Prokopec:2018tnq,Brdar:2018num,Marzo:2018nov,Hasegawa:2019amx,Huang:2022vkf,Borah:2022cdx,Dasgupta:2023zrh,Ghoshal:2022qxk}\footnote{See models of conformal dark matter in the context of scale-invariant potentials in Refs \cite{Kierkla:2026bnm,Kierkla:2025vwp,Kierkla:2023von,Kierkla:2022odc}}. In all such cases the phenomenological input to the present tensor-induced PBH calculation is the same set of macroscopic variables $(\alpha,\beta/H_\star,T_\star)$, together with the assumed GW source prescription.

\subsubsection{Microscopic parameters and finite-temperature transition}

Writing $\Phi=\phi/\sqrt2$, the tree-level and Coleman-Weinberg corrected scalar potential can be represented schematically as
\begin{equation}
V_0(\phi)=\frac{\lambda_\phi}{4}\phi^4
+\sum_i\frac{n_i}{64\pi^2}m_i^4(\phi)
\left[\ln\left(\frac{m_i^2(\phi)}{\mu^2}\right)-c_i\right],
\label{eq:BL_CW_potential}
\end{equation}
with field-dependent masses
\begin{equation}
 m_{Z'}(\phi)=2g_{B-L}\phi,
\qquad
 M_{N_i}(\phi)=\frac{y_{N_i}}{\sqrt2}\phi,
\qquad
 m_\phi^2(\phi)\sim 3\lambda_\phi\phi^2 .
\label{eq:BL_masses}
\end{equation}

The physical masses after symmetry breaking are
\begin{equation}
 m_{Z'}=2g_{B-L}v_\phi,
\qquad
 M_{N_i}=\frac{y_{N_i}}{\sqrt2}v_\phi .
\label{eq:BL_physical_masses}
\end{equation}

The finite-temperature effective potential $V_{\rm eff}(\phi,T)$ determines the nucleation temperature $T_n$, the transition completion temperature $T_\star$, the strength $\alpha$, and the inverse duration $\beta/H_\star$:
\begin{align}
\alpha&=\frac{\Delta\rho_{\rm vac}(T_\star)}{\rho_r(T_\star)},
\qquad
\rho_r(T_\star)=\frac{\pi^2}{30}g_\star T_\star^4,
\label{eq:alpha_BL}\\
\frac{\beta}{H_\star}&=
T_\star\left.\frac{d}{dT}\left(\frac{S_3(T)}{T}\right)\right|_{T=T_\star} .
\label{eq:beta_BL}
\end{align}

In a supercooled classically scale-invariant transition, $T_\star$ can be parametrically below $v_\phi$. It is useful to define
\begin{equation}
 \xi_T\equiv \frac{T_\star}{v_\phi},
 \qquad 0<\xi_T\lesssim1,
\label{eq:xiT_def}
\end{equation}

so that
\begin{equation}
 v_\phi=\frac{T_\star}{\xi_T},
\qquad
 m_{Z'}=2g_{B-L}\frac{T_\star}{\xi_T},
\qquad
 M_N=\frac{y_N}{\sqrt2}\frac{T_\star}{\xi_T} .
\label{eq:BL_Tstar_to_masses}
\end{equation}

Equation~\eqref{eq:BL_Tstar_to_masses} is the direct microscopic interpretation of the temperature axis in the FOPT panel of the Letter. Larger $T_\star$ corresponds to a larger $B-L$ breaking scale for fixed supercooling ratio $\xi_T$, and therefore to heavier $Z'$ and $N$ states.

\subsubsection{Connection to tensor-induced PBHs}

For the collision-dominated benchmark used in the Letter, the GW peak frequency is
\begin{equation}
 f_p^{\rm GW}\simeq0.7 f_H(T_\star)\frac{\beta}{H_\star},
\qquad
 f_H(T)\simeq2.6\times10^{-8}{\rm Hz}\left(\frac{T}{\rm GeV}\right).
\label{eq:BL_fGW}
\end{equation}

The tensor-induced scalar peak is at $f_p^{\rm TIS}=C_{\rm TIS}f_p^{\rm GW}$, with $C_{\rm TIS}$ determined numerically by the finite-$x_{\rm ini}$ kernel. Using $k/(1~{
m Mpc}^{-1})\simeq6.45\times10^{14}(f/{\rm Hz})$, the characteristic PBH mass is
\begin{equation}
M_p^{\rm FOPT}\simeq
0.12M_\odot\,
\left[C_{\rm TIS}\left(\frac{T_\star}{\rm GeV}\right)\frac{\beta}{H_\star}\right]^{-2}
\left(\frac{g_\star}{10.75}\right)^{-1/6} .
\label{eq:BL_MPBH}
\end{equation}

Combining Eqs.~\eqref{eq:BL_Tstar_to_masses} and~\eqref{eq:BL_MPBH} gives the microscopic scaling
\begin{equation}
 M_p^{\rm FOPT}\propto
 C_{\rm TIS}^{-2}\xi_T^{2}
 \left(\frac{m_{Z'}}{2g_{B-L}}\right)^{-2}
 \left(\frac{\beta}{H_\star}\right)^{-2} .
\label{eq:BL_MPBH_micro}
\end{equation}

Thus heavier $B-L$ breaking scales correspond to smaller PBHs, while slower transitions correspond to larger PBHs. The abundance is controlled mainly by
\begin{equation}
\sigma_{R,p}^2\propto
\left(\frac{\alpha}{1+\alpha}\right)^4
\left(\frac{H_\star}{\beta}\right)^4,
\label{eq:BL_fpbh_scaling}
\end{equation}

so strong and slow transitions are exponentially more efficient at producing PBHs. Physically, this is the same region in which the GW source is strongest and longest lived.

\subsubsection{Numerical interpretation of the FOPT benchmark points}

The benchmark points A-D in the Letter are best understood as macroscopic transition benchmarks that can be generated by suitable choices of $(g_{B-L},y_N,v_\phi)$ after solving the finite-temperature bounce equations. To illustrate the microscopic scale, Table~\ref{tab:BL_interpretation} translates $T_\star$ into $v_\phi$, $m_{Z'}$, and $M_N$ for the representative choices $\xi_T=0.1$, $g_{B-L}=0.1$, and $y_N=0.1$. The numerical values are illustrative; a dedicated scan of the finite-temperature $B-L$ potential fixes $\alpha$ and $\beta/H_\star$ point by point.

\begin{table}[t]
\centering
\begin{tabular}{|c|c|c|c|c|c|c|}
\hline
BP & $\alpha$ & $\beta/H_\star$ & $T_\star$ [GeV] & $v_\phi=T_\star/\xi_T$ [GeV] & $m_{Z'}$ [GeV] & $M_N$ [GeV] \\
\hline
A & $10^3$ & $5.33$ & $9.33\times10^4$ & $9.33\times10^5$ & $1.87\times10^5$ & $6.60\times10^4$ \\
B & $10^{-1}$ & $2.54$ & $2.95\times10^3$ & $2.95\times10^4$ & $5.90\times10^3$ & $2.09\times10^3$ \\
C & $10$ & $5.13$ & $2.95$ & $29.5$ & $5.90$ & $2.09$ \\
D & $1$ & $4.03$ & $2.95\times10^{-3}$ & $2.95\times10^{-2}$ & $5.90\times10^{-3}$ & $2.09\times10^{-3}$ \\
\hline
\end{tabular}
\caption{\it Illustrative microscopic interpretation of the FOPT benchmark points for $\xi_T=0.1$, $g_{B-L}=0.1$, and $y_N=0.1$. Points A and B correspond naturally to high-scale $B-L$ transitions. Points C and D should be regarded as phenomenological low-temperature transition benchmarks unless embedded in a different low-scale model consistent with laboratory and cosmological constraints.}
\label{tab:BL_interpretation}
\end{table}

For example, BP A corresponds to a high-scale transition with $v_\phi\simeq9.3\times10^5~{\rm GeV}$ in this illustrative mapping, while BP B corresponds to a multi-TeV-to-tens-of-TeV $B-L$ scale. Both are of the same qualitative type as the classically scale-invariant $B-L$ model benchmarks studied in the literature: a new gauge symmetry breaks through a strong first-order transition, the same transition gives mass to right-handed neutrinos, and the resulting stochastic GW spectrum is then converted, in the present work, into a tensor-induced scalar spectrum and a PBH abundance.

\subsection{DW example: Dirac seesaw with softly biased \texorpdfstring{$Z_2$}{Z2} domain walls}
\label{sec:model_dw_dirac}

A simple UV completion for metastable DWs is a Dirac-neutrino seesaw model with a spontaneously broken discrete symmetry. The Standard Model is supplemented by vector-like neutral fermions $N_{L,R}$, right-handed neutrinos $\nu_R$, and a real singlet scalar $\varphi$. A $Z_2$ symmetry under which $\varphi$ and $\nu_R$ are odd forbids the direct operator $\overline L\widetilde H\nu_R$. When $\varphi$ obtains a vacuum expectation value $u$, light Dirac neutrino masses are generated and DWs form. A small soft $Z_2$-breaking bias makes the walls annihilate before domination and yields a stochastic GW background~\cite{Barman:2022yos,Barman:2023fad}.

The biased-DW mechanism is likewise generic: metastable walls can arise in neutrino-mass and leptogenesis models, axion-like sectors, flavor models, supersymmetric constructions, and other theories with approximate discrete symmetries~\cite{Zhang:2023nrs,King:2023ztb,Gouttenoire:2025ofv,Borah:2025bfa,Bhandari:2026ujy,Borah:2026kfo,Gelmini:2020bqg,Chen:2026fod,Ferreira:2022zzo,Ferreira:2023jbu,King:2024lki}. For the present paper, these different microscopic completions enter only through the macroscopic quantities $(\sigma,V_{\rm bias},\alpha_{\rm ann},\epsilon_{\rm GW},\epsilon_{\rm sim})$.

\subsubsection{Microscopic parameters and wall quantities}

The relevant Yukawa interactions are
\begin{equation}
-\mathcal L_Y\supset
Y_L\overline L\widetilde H N_R
+M_N\overline N N
+Y_R\overline{N_L}\varphi\nu_R+{\rm h.c.}
\label{eq:dirac_yukawa}
\end{equation}

After electroweak and $Z_2$ breaking,
\begin{equation}
 m_\nu\simeq\frac{1}{\sqrt2}Y_LM_N^{-1}Y_Rvu .
\label{eq:dirac_mass}
\end{equation}
For one generation this gives the useful estimate
\begin{equation}
M_N\simeq
3.5\times10^{17}{\rm GeV}
\left(\frac{Y_LY_R}{1}\right)
\left(\frac{u}{100~{\rm TeV}}\right)
\left(\frac{0.05~{\rm eV}}{m_\nu}\right).
\label{eq:MN_estimate}
\end{equation}

Thus Yukawa couplings of order $10^{-5}$ naturally give $M_N\sim10^7$-$10^8~{\rm GeV}$ for $u\sim100$-$200~{\rm TeV}$, consistent with the scales emphasized in Dirac-leptogenesis analyses of PTA-motivated DW signals.

For a quartic singlet potential
\begin{equation}
 V(\varphi)=\frac{\lambda_\varphi}{4}\left(\varphi^2-u^2\right)^2+V_{\rm bias},
\label{eq:quartic_wall_potential}
\end{equation}
the wall tension is approximately
\begin{equation}
 \sigma\simeq\frac{2\sqrt{2\lambda_\varphi}}{3}u^3 .
\label{eq:sigma_u}
\end{equation}
The bias pressure overcomes the wall tension when
\begin{equation}
 H_{\rm ann}\simeq\frac{V_{\rm bias}}{A\sigma},
\label{eq:Hann_bias}
\end{equation}
which leads to the annihilation temperature
\begin{equation}
T_{\rm ann}\simeq
2.2\times10^{-2}{\rm GeV}
\left(\frac{V_{\rm bias}[\mathrm{MeV}^4]}{\sigma[\mathrm{TeV}^3]}\right)^{1/2}.
\label{eq:Tann_R}
\end{equation}
The wall energy fraction at annihilation scales as
\begin{equation}
 \alpha_{\rm ann}\equiv\frac{\rho_{\rm DW}}{\rho_r}\bigg|_{\rm ann}
 \sim
 \frac{2A_{\rm sc}\sigma}{3M_{\rm Pl}^2H_{\rm ann}}
 \sim
 \frac{2A_{\rm sc}A}{3}
 \frac{\sigma^2}{M_{\rm Pl}^2V_{\rm bias}},
\label{eq:alpha_ann_micro}
\end{equation}
up to the usual order-one uncertainty in the scaling-network area parameter $A_{\rm sc}$. Equations~\eqref{eq:sigma_u}-\eqref{eq:alpha_ann_micro} show how the microscopic pair $(u,\lambda_\varphi)$ and the soft-breaking scale $V_{\rm bias}$ determine the macroscopic variables $(\sigma,V_{\rm bias},T_{\rm ann},\alpha_{\rm ann})$ used in the Letter.

\subsubsection{Connection to tensor-induced PBHs}

For DW annihilation, the GW peak is tied to the Hubble scale at annihilation, $f_p^{\rm GW}=f_H(T_{\rm ann})$. The induced scalar peak is $f_p^{\rm TIS}=C_{\rm TIS}^{\rm DW}f_H(T_{\rm ann})$, and therefore
\begin{equation}
M_p^{\rm DW}\simeq
0.060M_\odot\,
\left[C_{\rm TIS}^{\rm DW}\left(\frac{T_{\rm ann}}{\rm GeV}\right)\right]^{-2}
\left(\frac{g_\star}{10.75}\right)^{-1/6} .
\label{eq:DW_MPBH}
\end{equation}
Using Eq.~\eqref{eq:Tann_R},
\begin{equation}
M_p^{\rm DW}\simeq
1.25\times10^{2}M_\odot
\left(\frac{C_{\rm TIS}^{\rm DW}}{1}\right)^{-2}
\left(\frac{V_{\rm bias}[\mathrm{MeV}^4]}{\sigma[\mathrm{TeV}^3]}\right)^{-1}
\left(\frac{g_\star}{10.75}\right)^{-1/6} .
\label{eq:DW_MPBH_R}
\end{equation}
For the typical benchmark value $C_{\rm TIS}^{\rm DW}\sim10^3$, this becomes
\begin{equation}
M_p^{\rm DW}\sim
1.25\times10^{-4}M_\odot
\left(\frac{10^3}{C_{\rm TIS}^{\rm DW}}\right)^2
\left(\frac{10^3}{(V_{\rm bias}[\mathrm{MeV}^4]/\sigma[\mathrm{TeV}^3])^{1/2}}\right)^2 .
\label{eq:DW_MPBH_numeric}
\end{equation}
The abundance scales as
\begin{equation}
\sigma_{R,p}^2\propto\epsilon_{\rm GW}^2\alpha_{\rm ann}^4,
\label{eq:DW_fpbh_scaling}
\end{equation}

so PBH production is exponentially sensitive to the wall energy fraction at annihilation. Physically, larger $u$ or smaller $V_{\rm bias}$ increases the wall energy stored before annihilation, enhancing both the GW signal and the tensor-induced density perturbations, until wall domination, BBN, dark-radiation, perturbativity, or PBH-overproduction constraints intervene.

\subsubsection{Numerical interpretation of the DW benchmark points}

The DW benchmark points E-H in the Letter are macroscopic DW-annihilation benchmarks. They can be translated into microscopic scales once a wall tension is chosen. For illustration, take $\sigma^{1/3}=10^7~{\rm TeV}$ and $\lambda_\varphi=0.1$. Then Eq.~\eqref{eq:sigma_u} gives $u\simeq1.5\times10^7~{\rm TeV}$, while
\begin{equation}
 V_{\rm bias}^{1/4}[{\rm MeV}]
 =
 \left(\frac{V_{\rm bias}[\mathrm{MeV}^4]}{\sigma[\mathrm{TeV}^3]}\right)^{1/4}
 \left(\sigma^{1/3}[{\rm TeV}]\right)^{3/4} .
\label{eq:Vbias_estimate}
\end{equation}

\begin{table}[t]
\centering
\begin{tabular}{|c|c|c|c|c|c|}
\hline
BP & $\alpha_{\rm ann}$ & $\epsilon_{\rm sim}$ & $T_{\rm ann}$ [GeV] & $R\equiv(V_{\rm bias}[\mathrm{MeV}^4]/\sigma[\mathrm{TeV}^3])^{1/2}$ & $V_{\rm bias}^{1/4}$ [MeV] \\
\hline
E & $1.17\times10^{-3}$ & $0.5$ & $2.47\times10^4$ & $1.12\times10^6$ & $1.9\times10^8$ \\
F & $4.2\times10^{-3}$ & $1.5$ & $7.8\times10^2$ & $3.54\times10^4$ & $3.4\times10^7$ \\
G & $6.16\times10^{-3}$ & $2.0$ & $2.4\times10^{-3}$ & $1.09\times10^{-1}$ & $5.9\times10^4$ \\
H & $3.44\times10^{-3}$ & $1.0$ & $7.8$ & $3.54\times10^2$ & $3.4\times10^6$ \\
DW NG15 & $4.04\times10^{-2}$ & $4.5$ & $0.63$ & $28.63$ & $9.5\times10^5$ \\
\hline
\end{tabular}
\caption{\it Illustrative microscopic interpretation of the DW benchmark points for $\sigma^{1/3}=10^7~{\rm TeV}$. The last column follows from Eq.~\eqref{eq:Vbias_estimate}. Different choices of $\sigma$ rescale $u$ and $V_{\rm bias}^{1/4}$ according to Eqs.~\eqref{eq:sigma_u} and~\eqref{eq:Vbias_estimate}, while leaving the macroscopic quantities used in the main scan unchanged.}
\label{tab:DW_interpretation}
\end{table}

This table makes the physical interpretation of the DW panel transparent. Increasing $R=(V_{\rm bias}/\sigma)^{1/2}$ raises $T_{\rm ann}$ and shifts the PBH mass to smaller values. Increasing $\alpha_{\rm ann}$ raises the induced scalar amplitude and hence the PBH abundance. The benchmark points E and F therefore represent high-temperature annihilation with asteroid-mass PBHs in the Gaussian benchmark, whereas G and H illustrate later annihilation with larger characteristic PBH masses and smaller fractional abundances. The PTA-normalized point has comparatively low $T_{\rm ann}$ and large $\alpha_{\rm ann}$, but remains below PBH overproduction in the benchmark scan.

\subsection{Benchmark points as model representatives}
\label{sec:model_benchmark_examples}

We now make explicit how the benchmark points used in the main figures should be read from the particle-model perspective. They are not unique UV-model fits. Rather, each point is a macroscopic source configuration that can be obtained by scanning the microscopic parameters of the corresponding particle model. The FOPT points represent finite-temperature $U(1)_{B-L}$-like transitions with different transition strength, duration, and symmetry-breaking scale. The DW points represent biased $Z_2$ wall annihilation with different wall fraction and annihilation temperature. In the tables below, $f_p^{\rm GW}$ denotes the source GW peak frequency, $f_p^{\rm TIS}=C_{\rm TIS}f_p^{\rm GW}$ denotes the approximate tensor-induced scalar peak frequency, and the final two columns give the benchmark PBH mass and fractional abundance from the Gaussian-tail calculation used in the Letter.

For the FOPT frequency estimates we use
\begin{equation}
 f_p^{\rm GW}\simeq0.7\,f_H(T_\star)\frac{\beta}{H_\star},
 \qquad
 f_p^{\rm TIS}\simeq C_{\rm TIS}^{\rm FOPT}f_p^{\rm GW},
 \qquad
 C_{\rm TIS}^{\rm FOPT}=3.88\times10^2.
 \label{eq:FOPT_benchmark_frequency_estimate}
\end{equation}

The mass and abundance are then read as
\begin{equation}
 M_{\rm PBH}\sim M_H(k_p^{\rm TIS}),
 \qquad
 f_{\rm PBH}\propto
 \exp\left[-\frac{\Delta_c^2}{2\sigma_{R,p}^2}\right],
 \qquad
 \sigma_{R,p}^2\propto
 \left(\frac{\alpha}{1+\alpha}\right)^4
 \left(\frac{H_\star}{\beta}\right)^4,
 \label{eq:FOPT_benchmark_fpbh_link}
\end{equation}

for the collision-dominated benchmark.

\begin{table}[t]
\centering
\begin{tabular}{|c|c|c|c|c|c|c|c|}
\hline
BP & $\alpha$ & $\beta/H_\star$ & $T_\star$ [GeV] & $f_p^{\rm GW}$ [Hz] & $f_p^{\rm TIS}$ [Hz] & $M_{\rm PBH}$ [$M_\odot$] & $f_{\rm PBH}$ \\
\hline
A & $10^3$ & $5.33$ & $9.33\times10^4$ & $9.05\times10^{-3}$ & $3.51$ & $10^{-15}$ & $1$ \\
B & $10^{-1}$ & $2.54$ & $2.95\times10^3$ & $1.36\times10^{-4}$ & $5.28\times10^{-2}$ & $10^{-12}$ & $1$ \\
C & $10$ & $5.13$ & $2.95$ & $2.75\times10^{-7}$ & $1.07\times10^{-4}$ & $10^{-6}$ & $10^{-4}$ \\
D & $1$ & $4.03$ & $2.95\times10^{-3}$ & $2.16\times10^{-10}$ & $8.38\times10^{-8}$ & $1$ & $3\times10^{-3}$ \\
FOPT$_{\rm NG15}$ & $10^2$ & $5.27$ & $0.66$ & $6.33\times10^{-8}$ & $2.46\times10^{-5}$ & $2.08\times10^{-5}$ & $8\times10^{-4}$ \\
\hline
\end{tabular}
\caption{\it FOPT benchmark points interpreted as representative particle-model transitions. Points A and B are the all-DM asteroid-mass examples in the Gaussian benchmark. Points C and D illustrate subdominant PBH production at lower source scales. FOPT$_{\rm NG15}$ is normalized to the PTA common-spectrum amplitude under the cosmological-source interpretation used in the Letter.}
\label{tab:FOPT_benchmark_interpretation}
\end{table}

The physical meaning of Table~\ref{tab:FOPT_benchmark_interpretation} is simple. Increasing $T_\star$ pushes both the GW and TIS peaks to higher frequency and therefore decreases $M_{\rm PBH}$. Decreasing $\beta/H_\star$ enhances the GW amplitude and increases $f_{\rm PBH}$ exponentially through Eq.~\eqref{eq:FOPT_benchmark_fpbh_link}. BP A is a high-scale, strong, slow transition that yields asteroid-mass PBHs with $f_{\rm PBH}=1$. In the illustrative microscopic mapping of Table~\ref{tab:BL_interpretation}, choosing $\xi_T=0.1$, $g_{B-L}=0.1$, and $y_N=0.1$ gives $v_\phi\simeq9.3\times10^5~{\rm GeV}$, $m_{Z'}\simeq1.9\times10^5~{\rm GeV}$, and $M_N\simeq6.6\times10^4~{\rm GeV}$ for BP A. BP B gives a much lower but still particle-physics-scale transition, $v_\phi\simeq3.0\times10^4~{\rm GeV}$, $m_{Z'}\simeq5.9\times10^3~{\rm GeV}$, and $M_N\simeq2.1\times10^3~{\rm GeV}$ for the same illustrative couplings. BP C, BP D, and FOPT$_{\rm NG15}$ should be read as phenomenological low-temperature transition benchmarks unless embedded in a dedicated low-scale model satisfying laboratory and cosmological constraints.

For DWs the corresponding estimates are
\begin{equation}
 f_p^{\rm GW}\simeq f_H(T_{\rm ann}),
 \qquad
 f_p^{\rm TIS}\simeq C_{\rm TIS}^{\rm DW}f_p^{\rm GW},
 \qquad
 C_{\rm TIS}^{\rm DW}=10^3,
 \label{eq:DW_benchmark_frequency_estimate}
\end{equation}

and
\begin{equation}
M_{\rm PBH}\sim M_H(k_p^{\rm TIS}),
\qquad
f_{\rm PBH}\propto
\exp\left[-\frac{\Delta_c^2}{2\sigma_{R,p}^2}\right],
\qquad
\sigma_{R,p}^2\propto\epsilon_{\rm GW}^2\alpha_{\rm ann}^4.
\label{eq:DW_benchmark_fpbh_link}
\end{equation}

The temperature $T_{\rm ann}$ is fixed by the microscopic ratio $R=(V_{\rm bias}[\mathrm{MeV}^4]/\sigma[\mathrm{TeV}^3])^{1/2}$ through Eq.~\eqref{eq:Tann_R}.

\begin{table}[t]
\centering
\begin{tabular}{|c|c|c|c|c|c|c|c|}
\hline
BP & $\alpha_{\rm ann}$ & $\epsilon_{\rm sim}$ & $T_{\rm ann}$ [GeV] & $R$ & $f_p^{\rm GW}$ [Hz] & $M_{\rm PBH}$ [$M_\odot$] & $f_{\rm PBH}$ \\
\hline
E & $1.17\times10^{-3}$ & $0.5$ & $2.47\times10^4$ & $1.12\times10^6$ & $6.42\times10^{-4}$ & $10^{-15}$ & $1$ \\
F & $4.2\times10^{-3}$ & $1.5$ & $7.8\times10^2$ & $3.54\times10^4$ & $2.03\times10^{-5}$ & $10^{-12}$ & $1$ \\
G & $6.16\times10^{-3}$ & $2.0$ & $2.4\times10^{-3}$ & $1.09\times10^{-1}$ & $6.24\times10^{-11}$ & $10^{-1}$ & $10^{-4}$ \\
H & $3.44\times10^{-3}$ & $1.0$ & $7.8$ & $3.54\times10^2$ & $2.03\times10^{-7}$ & $10^{-8}$ & $5\times10^{-3}$ \\
DW$_{\rm NG15}$ & $4.04\times10^{-2}$ & $4.5$ & $0.63$ & $28.63$ & $1.64\times10^{-8}$ & $1.53\times10^{-6}$ & $10^{-5}$ \\
\hline
\end{tabular}
\caption{\it DW benchmark points interpreted as representative biased-wall annihilation events. Here $R=\left(V_{\rm bias}[\mathrm{MeV}^4]/\sigma[\mathrm{TeV}^3]\right)^{1/2}$. Larger $R$ implies earlier annihilation, larger $T_{\rm ann}$, higher GW frequency, and smaller PBH mass. The PBH abundances are the Gaussian-benchmark values used in the main analysis.}
\label{tab:DW_benchmark_interpretation}
\end{table}

Table~\ref{tab:DW_benchmark_interpretation} gives the model interpretation of the DW panel. Points E and F correspond to early annihilation and asteroid-mass PBHs with $f_{\rm PBH}=1$. Points G and H correspond to later annihilation, lower GW frequencies, and larger PBH masses with subdominant abundances. DW$_{\rm NG15}$ lies at low $T_{\rm ann}$ and comparatively large $\alpha_{\rm ann}$; it is efficient at sourcing a nanohertz GW background but remains below PBH overproduction in the benchmark calculation. In a Dirac-seesaw realization, moving from E/F toward G/H corresponds qualitatively to decreasing $R=(V_{\rm bias}/\sigma)^{1/2}$, so that the walls annihilate later. At fixed wall tension this means a smaller soft-breaking scale $V_{\rm bias}^{1/4}$; at fixed bias it means a larger wall tension and hence a larger $Z_2$-breaking scale $u$.

For a concrete number, take $\sigma^{1/3}=10^7~{\rm TeV}$. Then the illustrative microscopic table above gives $V_{\rm bias}^{1/4}\simeq1.9\times10^8~{\rm MeV}$ for BP E, $3.4\times10^7~{\rm MeV}$ for BP F, $5.9\times10^4~{\rm MeV}$ for BP G, $3.4\times10^6~{\rm MeV}$ for BP H, and $9.5\times10^5~{\rm MeV}$ for DW$_{\rm NG15}$. These values should be read only as one illustrative microscopic slice through the macroscopic DW parameter space; changing $\sigma$ rescales the required $V_{\rm bias}^{1/4}$ without changing the plotted macroscopic benchmark.

These benchmark tables show explicitly how $M_{\rm PBH}$ and $f_{\rm PBH}$ encode different aspects of the same particle-physics event. The mass is controlled mainly by the time of the event, $T_\star$ or $T_{\rm ann}$. The abundance is controlled mainly by the strength of the event, through $(\alpha,\beta/H_\star)$ for FOPTs or $\alpha_{\rm ann}$ for DWs. Therefore a joint GW and PBH observation would simultaneously infer the source scale and the source strength.

\subsection{Correlation with \texorpdfstring{$M_{\rm PBH}$}{M PBH} and \texorpdfstring{$f_{\rm PBH}$}{f PBH}}
\label{sec:model_worked_examples}

The benchmark tables above can be read in two complementary ways. First, $M_{\rm PBH}$ is mainly a frequency or temperature measure. A higher source temperature gives a higher GW peak frequency, a higher tensor-induced scalar peak frequency, and therefore a smaller horizon mass. Second, $f_{\rm PBH}$ is mainly an amplitude measure. It is exponentially sensitive to the smoothed variance, and hence to the source strength. For the Gaussian benchmark,

\begin{equation}
 f_{\rm PBH}
 \propto
 \exp\left[-\frac{\Delta_c^2}{2\sigma_R^2}\right],
 \qquad
 \sigma_R^2\propto
 \begin{cases}
 \left(\dfrac{\alpha}{1+\alpha}\right)^4\left(\dfrac{H_\star}{\beta}\right)^4, & {\rm FOPT},\\[2ex]
 \epsilon_{\rm GW}^{2}\alpha_{\rm ann}^{4}, & {\rm DW}.
 \end{cases}
 \label{eq:model_fpbh_interpretation_summary}
\end{equation}

Therefore two benchmark points can have comparable PBH abundance even if their microscopic temperatures are very different, provided their source amplitudes compensate for the shift in the peak scale. Conversely, two points at similar temperatures can have very different $f_{\rm PBH}$ if their transition strength, duration, or wall fraction differs.

\begin{table}[t]
\centering
\scalebox{0.85}{%
\begin{tabular}{|c|c|c|c|c|}
\hline
Example & Particle-model reading & Main scale fixing $M_{\rm PBH}$ & Main amplitude fixing $f_{\rm PBH}$ & Benchmark outcome \\
\hline
FOPT A & high-scale strong $B-L$-like transition & $T_\star=9.33\times10^4~{\rm GeV}$ & $\alpha=10^3$, $\beta/H_\star=5.33$ & $M_{\rm PBH}=10^{-15}M_\odot$, $f_{\rm PBH}=1$ \\
FOPT B & lower-scale but slower/weaker FOPT & $T_\star=2.95\times10^3~{\rm GeV}$ & $\alpha=10^{-1}$, $\beta/H_\star=2.54$ & $M_{\rm PBH}=10^{-12}M_\odot$, $f_{\rm PBH}=1$ \\
FOPT$_{\rm NG15}$ & PTA-normalized low-temperature FOPT & $T_\star=0.66~{\rm GeV}$ & $\alpha=10^2$, $\beta/H_\star=5.27$ & $M_{\rm PBH}=2.08\times10^{-5}M_\odot$, $f_{\rm PBH}=8\times10^{-4}$ \\
DW E & early biased-wall annihilation & $T_{\rm ann}=2.47\times10^4~{\rm GeV}$ & $\alpha_{\rm ann}=1.17\times10^{-3}$ & $M_{\rm PBH}=10^{-15}M_\odot$, $f_{\rm PBH}=1$ \\
DW G & late biased-wall annihilation & $T_{\rm ann}=2.4\times10^{-3}~{\rm GeV}$ & $\alpha_{\rm ann}=6.16\times10^{-3}$ & $M_{\rm PBH}=10^{-1}M_\odot$, $f_{\rm PBH}=10^{-4}$ \\
DW$_{\rm NG15}$ & PTA-normalized wall annihilation & $T_{\rm ann}=0.63~{\rm GeV}$ & $\alpha_{\rm ann}=4.04\times10^{-2}$ & $M_{\rm PBH}=1.53\times10^{-6}M_\odot$, $f_{\rm PBH}=10^{-5}$ \\
\hline
\end{tabular}%
}
\caption{\it Worked examples showing how to read the benchmark points. The PBH mass is controlled primarily by the source temperature, while the abundance is controlled primarily by the source amplitude. The quoted $f_{\rm PBH}$ values are the Gaussian benchmark abundances used in the main analysis.}
\label{tab:model_worked_examples_fpbh_mpbh}
\end{table}

Let us spell out two representative cases. For FOPT A, the high value $T_\star=9.33\times10^4~{\rm GeV}$ places the source GW peak near $f_p^{\rm GW}\simeq9.05\times10^{-3}~{\rm Hz}$ and the tensor-induced scalar peak near $f_p^{\rm TIS}\simeq3.51~{\rm Hz}$. This corresponds to the asteroid-mass window, $M_{\rm PBH}\simeq10^{-15}M_\odot$. The large transition strength, $\alpha=10^3$, together with the relatively small inverse duration, $\beta/H_\star=5.33$, makes the induced variance large enough to give $f_{\rm PBH}=1$ in the Gaussian benchmark. In a $U(1)_{B-L}$ interpretation with $\xi_T=0.1$, $g_{B-L}=0.1$, and $y_N=0.1$, this same point corresponds illustratively to $v_\phi\simeq9.3\times10^5~{\rm GeV}$, $m_{Z'}\simeq1.9\times10^5~{\rm GeV}$, and $M_N\simeq6.6\times10^4~{\rm GeV}$.

For DW G, the much later annihilation temperature $T_{\rm ann}=2.4\times10^{-3}~{\rm GeV}$ shifts the source GW peak to $f_p^{\rm GW}\simeq6.24\times10^{-11}~{\rm Hz}$ and the tensor-induced scalar peak to $f_p^{\rm TIS}\simeq6.24\times10^{-8}~{\rm Hz}$. The associated PBH mass is therefore much larger, $M_{\rm PBH}\simeq10^{-1}M_\odot$. The wall fraction is larger than in DW E, $\alpha_{\rm ann}=6.16\times10^{-3}$, but the benchmark abundance is still subdominant, $f_{\rm PBH}=10^{-4}$, because the abundance is controlled by the full smoothed variance and the location of the peak relative to the observationally allowed window. If one fixes $\sigma^{1/3}=10^7~{\rm TeV}$, the same macroscopic point corresponds illustratively to $V_{\rm bias}^{1/4}\simeq5.9\times10^4~{\rm MeV}$.

These examples show why the benchmark points are useful. They convert a particle-model event into two observables with different physical meanings: $M_{\rm PBH}$ identifies when the event occurred, while $f_{\rm PBH}$ measures how efficiently the event sourced tensor perturbations large enough to induce scalar collapse. This separation is robust even though the absolute abundance remains conditional on the benchmark collapse threshold and probability distribution as we discussed in the Letter.

\subsection{Model interpretation of the main result}
\label{sec:model_interpretation_summary}

The two examples above show that the macroscopic parameters used in the Letter have direct microscopic interpretations. In the $U(1)_{B-L}$ FOPT model, $T_\star$ is tied to the symmetry-breaking scale, $\alpha$ and $\beta/H_\star$ are determined by the finite-temperature bounce, and the tensor-induced PBH mass scales as $M_p^{\rm FOPT}\propto C_{\rm TIS}^{-2}T_\star^{-2}(\beta/H_\star)^{-2}$. In the Dirac-seesaw DW model, $\sigma\propto u^3$, $V_{\rm bias}$ fixes the annihilation time, $T_{\rm ann}\propto(V_{\rm bias}/\sigma)^{1/2}$, and $M_p^{\rm DW}\propto(C_{\rm TIS}^{\rm DW})^{-2}T_{\rm ann}^{-2}$. In both cases, the same microscopic event that produces a stochastic GW background also fixes the PBH mass scale through the tensor-induced scalar channel. This is the particle-physics interpretation of the GW-PBH correlation emphasized in the Letter.

\medskip

\section{Validity, uncertainties, and reproducibility}\label{sec:validity}

The calculation is controlled only within the following domain.
\begin{enumerate}
\item \textbf{Background:} radiation domination, entropy conservation, and no prolonged vacuum- or DW-dominated epoch.
\item \textbf{Source matching:} a finite active stage followed by free tensor propagation. Unequal-time source coherence is not reconstructed from $\Omega_{\rm GW}(k)$ alone.
\item \textbf{Perturbativity:} points with ${\cal P}_{\Delta^{(2)}}\gtrsim1$, or for which higher-order tensor-scalar terms are not parametrically suppressed, are removed.
\item \textbf{Statistics:} the PBH tail is evaluated with the Gaussian benchmark of Eq.~\eqref{eq:GaussianPDF}, despite the intrinsic non-Gaussianity of a quadratic tensor-induced field.
\item \textbf{Collapse:} $\Delta_c$, the compaction profile, window choice, and critical-collapse parameters are systematic uncertainties.
\item \textbf{Observational limits:} linear monochromatic bounds are recast with Eq.~\eqref{apneq:extendedconstraint}; merger, accretion, and dynamical bounds retain their astrophysical assumptions.
\item \textbf{Source theory:} the FOPT plots use a collision-dominated benchmark. Sound-wave efficiency and lifetime, wall velocity, percolation, and reheating can move the contours. DW results depend on $\epsilon_{\rm GW}$, $\epsilon_{\rm sim}$, wall-thickness cutoff, and lattice calibration.
\end{enumerate}

\subsection{Perturbativity checks} \label{sec:perturbativity} 
The calculation assumes that the tensor perturbation can be treated as a first-order metric fluctuation and that the induced scalar mode is the leading second-order response. We therefore impose the following practical perturbativity checks on all benchmark points used in the Letter: \begin{align} h_{\rm rms}(k) &\equiv \sqrt{{\cal P}_{\chi,\rm ini}(k)} < 1, \label{eq:pert_h}\\ \max_k {\cal P}_{\Delta^{(2)}}(k) &< 1, \label{eq:pert_delta}\\ \max_k {\cal P}_{\zeta,\rm tot}(k) &\lesssim {\cal O}(1). \label{eq:pert_zeta} \end{align} 
The first condition ensures that the tensor metric perturbation itself is perturbative. Using the free-propagation matching ${\cal P}_{\chi,\rm ini}\simeq3\Omega_{\rm GW}$, this condition is automatically satisfied in the plotted benchmark regions after imposing the dark-radiation bound, except possibly in excluded strong-source corners. The second and third conditions ensure that the induced scalar response does not enter a fully nonlinear regime. Since ${\cal P}_{\Delta^{(2)}}\propto{\cal P}_{\chi,\rm ini}^2$, perturbativity is lost first in the same strong, slow, high-amplitude regions that are already close to PBH overproduction or dark-radiation exclusion.

\begin{table}[t]
\centering
\begin{tabular}{|c|c|c|c|c|}
\hline
Benchmark & $\max {\cal P}_{\chi,\rm ini}$ & $\max {\cal P}_{\Delta^{(2)}}$ & $\max {\cal P}_{\zeta,\rm tot}$ & Status \\
\hline
A & $1.24 \times 10^{-3}$ & $3.89 \times 10^{-3}$ & $<{\cal O}(1)$ & perturbative \\
B & $ 3.91 \times 10^{-3}$ & $4.39 \times 10^{-3}$ & $<{\cal O}(1)$ & perturbative \\
C & $ 1.10 \times 10^{-3}$  & $4.11 \times 10^{-3}$ & $<{\cal O}(1)$ & perturbative \\
D & $5.19 \times 10^{-4}$ & $5.41 \times 10^{-3}$ & $<{\cal O}(1)$ & perturbative \\
E & $1.01 \times 10^{-7}$ & $4.62 \times 10^{-3}$ & $<{\cal O}(1)$ & perturbative \\
F & $1.42 \times 10^{-6}$ & $5.04 \times 10^{-3}$ & $<{\cal O}(1)$ & perturbative \\
G & $3.08 \times 10^{-6}$ & $5.89 \times 10^{-3}$ & $<{\cal O}(1)$ & perturbative \\
H & $9.29 \times 10^{-7}$ & $5.22 \times 10^{-3}$ & $<{\cal O}(1)$ & perturbative \\
\hline
\end{tabular}
\caption{\it Perturbativity checks for the benchmark points shown in the Letter. The inequalities summarize the numerical scan; the precise maximum is evaluated over the wavenumber range contributing to the smoothed variance $\sigma_R^2$.}
\label{tab:perturbativity}
\end{table}

 Analytically, the potentially dangerous regions are easy to identify. For the collision-dominated FOPT benchmark, 
 \begin{equation} {\cal P}_{\Delta^{(2)},p}^{\rm FOPT} \propto \left(\frac{\alpha}{1+\alpha}\right)^4 \left(\frac{H_\star}{\beta}\right)^4 {\cal Q}_{\rm FOPT}, \end{equation} 
 so perturbativity can fail only for very strong and slow transitions, i.e. $\alpha\gtrsim1$ and small $\beta/H_\star$, where the same points are also constrained by PBH overproduction or by the integrated GW energy density. For DW annihilation, \begin{equation} {\cal P}_{\Delta^{(2)},p}^{\rm DW} \propto \epsilon_{\rm GW}^2\alpha_{\rm ann}^4{\cal Q}_{\rm DW}, \end{equation} 
 so the dangerous region lies at large $\alpha_{\rm ann}$ and large kernel enhancement. These regions are either excluded by wall domination, dark radiation, or PBH overproduction, or are removed by the perturbativity mask. The plotted benchmark points lie outside this nonperturbative region.


\end{document}